\documentclass[british]{extarticle}
\usepackage[T1]{fontenc}
\usepackage[utf8]{inputenc}
\usepackage{xcolor}
\usepackage{textcomp}
\usepackage{amsmath}
\usepackage{esint}
\usepackage[authoryear]{natbib}

\makeatletter
\newcommand{\lyxaddress}[1]{
	\par {\raggedright #1
	\vspace{1.4em}
	\noindent\par}
}

\usepackage{tikz}
\usetikzlibrary{shapes,snakes}
\usepackage{pgfplots}
\usepackage{hyperref}
\usepackage{breakurl}
\usepackage{latexsym}
\usepackage{stmaryrd}
\newtheorem{principle}{Principle}

\makeatother

\usepackage{babel}
\begin{document}
\selectlanguage{british}
\title{The construction of Electromagnetism}
\author{M.A. Natiello$^{\dagger}$and H.G. Solari$^{\ddagger}$}
\maketitle

\lyxaddress{$^{\dagger}$Centre for Mathematical Sciences, Lund University;}

\lyxaddress{$^{\ddagger}$Departamento de Física, FCEN-UBA and IFIBA-CONICET}
\begin{abstract}
We examine the construction of electromagnetism in its current form,
and in an alternative form, from a point of view that combines a minimal
realism with strict rational demands. We begin by discussing the requests
of reason when constructing a theory and next, we follow the historical
development as presented in the record of original publications, the
underlying epistemology (often explained by the authors) and the mathematical
constructions. The historical construction develops along socio-political
disputes (mainly, the reunification of Germany and the second industrial
revolution), epistemic disputes (at least two demarcations of science
in conflict) and several theories of electromagnetism. Such disputes
resulted in the militant adoption of the ether by some, a position
that expanded in parallel with the expansion of Prussia. This way
of thinking was facilitated by the earlier adoption of a standpoint
that required, as a condition for understanding, the use of physical
hypothesis in the form of analogies; an attitude that is  antithetic
to Newton's ``\emph{hypotheses non fingo}''. While the material
ether was finally abandoned, the epistemology survived in the form
of ``substantialism'' and a metaphysical ether: the space. The militants
of the ether attributed certainties regarding the ether to Faraday
and Maxwell, when they only expressed doubts and curiosity. Thus,
the official story is not the real history. This was achieved by the
operation of detaching Maxwell's electromagnetism from its construction
and introducing a new game of formulae and interpretations. Large
and important parts of Maxwell work are today not known, as for example,
the rules for the transformation of the electromagnetic potentials
between moving systems. When experiments showed that all the theories
based in the material ether were incorrect, a new interpretation was
offered: Special Relativity (SR). At the end of the transformation
period a pragmatic view of science, well adapted to the industrial
society, had emerged, as well as a new protagonist: the theoretical
physicist. The rival theory of delayed action at distance initiated
under the influence of Gauss was forgotten in the midst of the intellectual
warfare. The theory is indistinguishable in formulae from Maxwell's
and its earlier versions are the departing point of Maxwell for the
construction of his equations. We show in a mathematical appendix
that such (relational) theory can incorporate Lorentz' contributions
as well as Maxwell's transformations and C. Neumann's action, without
resource to the ether. Demarcation criteria was further changed at
the end of the period making room for habits and intuitions. When
these intuited criteria are examined by critical reason (seeking for
the fundaments) they can be sharpened with the use of the Non Arbitrariness
Principle, which throws light over the arbitrariness in the construction
of SR. Under a fully rational view SR is not acceptable, it requires
to adopt a less demanding epistemology that detaches the concept from
the conception, such as Einstein's own view in this respect, inherited
from Hertz. In conclusion: we have shown in this relevant exercise
how the reality we accept depends on earlier, irrational, decisions
that are not offered for examination but rather are inherited from
the culture.

\centerline{Draft, v0.0 of 25/09/19;v0.1 10/04/19; v0.3 of 22/01/20}


\textbf{ORCID numbers:}M. A. Natiello: 0000-0002-9481-7454 H. G. Solari:
0000-0003-4287-1878
\end{abstract}

\section{Introduction}

At the beginning of the XIX century, physics was dominated by ideas
that evolved from Galileo, Descartes, Newton and Leibniz, and a good
number of distinguished mathematicians such as Laplace, Lagrange,
Legendre, Poisson, Ampère, Gauss and Hamilton that expanded mathematics
and physics at the same time. In continental European physics the
space was relational (following Leibniz) although Descartes' filled
space persisted as a \textsf{luminiferous}\textsf{\emph{ }}\textsf{ether}
in the theory of light and had supporters in electromagnetism (EM)
\citep{cane80}. In this scheme, the only well known force, gravitational
force, was described as an instantaneous action at distance producing
accelerations in the direction of the line connecting interacting
bodies. The tradition in Great Britain (insular Europe) was of a mixed
type. In part, it followed the absolute space tradition of Newton,
in the vulgar form of a frame fixed to the distant stars. At the same
time, in the philosophical side, it followed the empiricist tradition
of Hume. Although Hume speaks about causes and not actions (much less
interactions) it appears to us that in the following paragraph causes
can be exchanged by actions:
\begin{quote}
The idea, then, of causation must be deriv’d from some relation among
objects; and that relation we must now endeavour to discover. I find
in the first place, that whatever objects are consider’d as causes
or effects, are contiguous; and that nothing can operate in a time
or place, which is ever so little remov’d from those of its existence.
{[}...{]} We may therefore consider the relation of contiguity as
essential to that of causation; at least may suppose it such, according
to the general opinion, till we can find a more proper occasion to
clear up this matter, by examining what objects are or are not susceptible
of juxtaposition and conjunction.\\
The second relation I shall observe as essential to causes and effects,
is not so universally acknowledg’d, but is liable to some controversy.
’Tis that of priority of time in the cause before the effect. Some
pretend that ’tis not absolutely necessary a cause shou’d precede
its effect; but that any object or action, in the very first moment
of its existence, may exert its productive quality, and give rise
to another object or action, perfectly co-temporary with itself. {[}...{]}
The consequence of this wou’d be no less than the destruction of that
succession of causes, which we observe in the world; and indeed, the
utter annihilation of time. For if one cause were co-temporary with
its effect, and this effect with its effect, and so on, ’tis plain
there wou’d be no such thing as succession, and all objects must be
co-existent.\citep{hume96}
\end{quote}
Action at distance and even worse, instantaneous action at distance
would have been felt inappropriate to the empiricist tradition since
it would imply a body causing an effect on another which is non continuous.
Lord Kelvin put it in simple form:
\begin{quote}
The idea that the Sun pulls Jupiter, and Jupiter pulls back against
the sun with equal force, and that the sun, earth, moon, and planets
all act on one another with mutual attractions seemed to violate the
supposed philosophic principle that matter cannot act where it is
not. \citep[Kelvin's preface to][]{hert93}
\end{quote}
It is interesting to observe a sort of pendulum-like movement of ideas
between the continent and Great Britain. Kelvin refers in \citep[Kelvin's preface to][]{hert93}
to an observation made by Voltaire in 1727 when he wrote:
\begin{quote}
A Frenchman who arrives in London finds a great alteration in philosophy,
as in other things. He left the world full, he finds it empty. At
Paris you see the Universe composed by vortices of subtile matter;
at London we see nothing of the kind...
\end{quote}
However, by the end of the XVIII century action at a distance was
the dominant theory \citep[Kelvin's preface of][]{hert93}. The resurgence
of ether is credited to Faraday.\emph{ }Kelvin writes:
\begin{quote}
...before his death, in 1867, he had succeeded in inspiring the raising
generation of the scientific world with something approaching to faith
that electric force is transmitted by a medium called ether, of which,
as it had been believed by the whole scientific world for forty years,
light and radiant heat are transversal vibrations.
\end{quote}
And he continues:
\begin{quote}
...for electricity and magnetism Faraday's anticipations and Clerk-Maxwell's
splendidly developed theory has been established on the sure basis
of experiments by Hertz' work...
\end{quote}
In this form goes the standard story, but, is it correct? Is it faithful
to history? We will show that there are missing parts in this story
probably because Hertz gave the name \textsf{Maxwell's theory} to
all theories that agreed with Maxwell's in the final equations to
be tested \citep{hert93}. Hertz mentions Helmholtz theory as well
and he indicates the existence of others. The omitted theories were
those of the Göttingen group that gathered around the figure of Gauss
and included rivals of Maxwell as Lorenz, Riemann, Newmann and especially
Weber whose differences with Hertz' mentor, Helmholtz, are well known
\citep{assi94}. Thus, personal rivalries and other social phenomena,
as we will later see, might have played a role in the final (social)
outcome.\footnote{Helmholtz and Hertz developments occur at a particular time of German
history. Helmholtz becomes head of the Berlin school of physics soon
after the beginning of the first Reich (the \emph{Kaiserreich}) \citep{hoff98}.
In that period there was a battle against Hegel's philosophy. Helmholtz
is among those struggling to abandon Hegel's idealism and methods
\citep{dago04}.}

Towards the mid of the XIX century, the force of the Enlightenment
was declining in Europe while the force of the British Empire and
the second industrial revolution was emerging along with the struggle
for the unification of Germany. Political events seldom have an influence
on the development of science; rather, they influence indirectly through
the ``ethos of the times“. However, in the construction of Electromagnetism
they influenced in both forms as we will see.

Let us turn back to philosophical issues. The construction of EM must
confront difficulties not present in the construction of mechanics.
In particular, most EM phenomena do not appear as such, i.e., as an
occurrence, perceptible to the senses, since we have constructed two
covering concepts, matter and light, that naturalise most of our experience
with electricity. In that time, matter and light were conceived as
“external“ entities on which EM phenomena occur, rather than entities
whose properties are almost entirely of electromagnetic nature. Thus,
the study of EM proceeds with little sensorial input in the form of
experimental outcomes. Our possibilities of using our intuition and
direct experience is then scarce and we must rely on reason and the
process of \textsf{abduction} \citep{peir55}. In turn, abduction
--the process of adopting (testable) explanatory hypotheses-- is
often aided by analogies and habits as a form of producing the hypothesis
for consideration. However, the value of the abduced hypothesis is
not to be judged by such auxiliary methods but rather for its explanatory
power over the facts, that must necessarily be a larger set than those
that motivated the hypothesis. Peirce prevents us:
\begin{quote}
observed facts relate exclusively to the particular circumstances
that happened to exist when they were observed. They do not relate
to any future occasions upon which we may be in doubt how we ought
to act. They, therefore, do not, in themselves, contain any practical
knowledge. Such knowledge must involve additions to the facts observed.
The making of those additions is an operation which we can control;
and it is evidently a process during which error is liable to creep
in.\citep[p. 150]{peir55}
\end{quote}
Did these recommendations guided the construction of EM? We anticipate
that the answer is no. On the contrary, we will show that analogical
thinking operates as a restriction upon what can be constructed.

By the beginning of the XX century, a complete turn around of ideas
in physics had been achieved. Those ideas became the consensus form
of physics some years later. The changes that were operated far exceeded
a scientific revolution. We intend to show in this work the epistemic
transformation of physics produced by the lectors of Maxwell (rather
than by Maxwell himself). The transformation of physics was not limited
to the incorporation of a new domain of understanding. Rather, it
was a complex development producing a new type of scientist, the \textsf{theoretical
physicist}, that took the duties of constructing the theories, which
was previously performed by mathematicians \citep[p. 6]{jung17}.
The new field of studies, the new social subject, came equipped with
its own epistemology, a radical change of which the practitioners
were probably not aware, except perhaps (and partially) two of the
main protagonists: Poincaré \citep{poin13foundations} and Einstein
\citep{eins36,eins40}.

The plan for this work is to critically follow the transformation
of the ideas in Electromagnetism in the period going from Ampère \citep{ampe25,assi15}
to the emergence of special relativity (SR) \citep{eins05,schw77,schw77b,schw77c}
focusing in the epistemic changes produced. We will begin by focusing
in Ampère-Weber's relational approach and its social decline (a decline
without refutation). Next, we move to the epistemological contributions
of Maxwell that anticipate key elements in Husserl's phenomenology
and the controversy regarding physical hypotheses (such as the electric
fluid and the ether). The next stage consists in the abandonment of
Maxwell philosophical attitude by his followers: the beginning of
the new era. This crucial step explains (or perhaps just describes)
how philosophy was left behind and how a new form of construction
of science was born, with new elements such as interpretations and
analogies being central to it. It is the \textsf{epistemological change}
what converts a phenomenological theory as Maxwell's (in his own words)
into a fundamental theory, first of the material ether and later of
the philosophical ether: the space. The lax rules of the new epistemology
allowed the new scientists to patch their theories, producing an era
of ``continuous progress``, greatly facilitated by the abandonment
of the idea of refutation, of the search for the fundamentals (i.e.,
critical thought, philosophically understood) and of the unity of
reason of Kantian idealism. This is, the Gordian-knot was repeatedly
cut rather than untied, thus resulting in a conceptual change of what
it is meant by \textsf{science}: the adoption of pragmatic-realism
as already described by other authors \citep{torr00} and substantially
in agreement with the ethos of the time.

While the new scientist developed and occupied the social niche of
producer of theory, the old epistemological approach did not get extinct.
From time to time, challenges to the \textsf{main stream} have been
raised by, or re-emerged in, various scientific groups having a closer
affinity with the Galilean-Newtonian heritage. Among the challenges,
those emerged in the electrical engineering domain are the most interesting
since this scientific community has considered discriminating experiments
between a relational electromagnetism and the relativistic version
of electromagnetism. Curiously enough, the situations they considered
by 1956, motivated by the initial explorations of outer space have
emerged in later years as an ``anomaly'' of the main stream electromagnetism.
But what would have been trumpeted as an extraordinary predictive
success of science if achieved by the dominant conceptions has gained
little or no transcendence: social forces guard us from heretic knowledge
as Bourdieu \citep{bour99} has taught us.

It is important to notice that reviewing the emergence of a new field
of study can be done from two (extreme) sides; one side being the
praiseful form and the other its opposite, the derogatory form. We
began our search trying to find what at that moment we thought was
a small missing link between a relativistic and a relational theory
of EM. Along the way we realised that the differences cannot be bridged
since there is no common epistemological ground to both approaches.
A similar transition can be found in Dingle \citep{ding60}.

The Maxwell-Lorentz theory of electromagnetism has not evolved for
at least 100 years. Yet, inconsistencies on its formulation (or at
least in its most common interpretation) have emerged in recent years.
Inconsistencies in the standard use of fields \citep{lazarovici2018against}
and in considering the Maxwell-Lorentz equations as an initial value
problem \citep{deckert2016initial} have been found. The philosophical
consequences have been recently treated in \citep{hart20}. However,
the analysis has been performed within the current epistemology of
theoretical physics. 

Using a constructivist approach we have sought to transcend the foundations
of mechanics \citep{sola18b} unearthing and giving a mathematical
form to a principle of reason: the no-arbitrariness principle (NAP).
It has been the philosophical guidance provided by NAP what made us
to regress in time until an epoch where physics was compatible with
it. We were not able to locate a period in time which was free from
arbitrariness. There is e.g., arbitrariness in the Ampère-Weber theory,
although it can easily be removed. In so doing, it becomes even closer
to Maxwell's theory. Maxwell himself practices the \emph{epojé} \footnote{Scepticism. The act of refraining from any conclusion for or against
anything as the decisive step for the attainment of truth.} hence arbitrariness is put in parenthesis/suspended. This attitude
differs drastically with the attitude of his followers which is the
entry-point through which a full flared arbitrariness is introduced.
Thus, as we adopted as starting point a commitment to reason, and
a method for preserving reason, departures from the rational ideal
will be highlighted. We begin by presenting requisites of reason that
must be satisfied in the construction of a mathematical theory (Section
\ref{sec:The-scientific-attitude}), this is, we introduce ``rational
realism''. Next we review the main line of development of Electromagnetism
up to and including Maxwell (Section \ref{sec:Matter-and-electricity}).
Among other historical information we identify in Maxwell's writings
the transformations undergone by the potentials upon a Galilean change
of coordinates. The ether era is discussed in Section \ref{sec:The-ether}.
We analyse the epistemological changes that made it possible (actually,
almost mandatory) to believe in the ether. The two epistemological
venues are then contrasted. Subsection \ref{subsec:DAD} and Appendices
\ref{sec:Delayed} and \ref{sec:lagrangian-appendix} explore the
electromagnetism that results from applying the constructive rational
rules of Section \ref{sec:The-scientific-attitude}. Subsection \ref{subsec:Albert}
is dedicated to the contribution by Einstein, the final point of the
evolution of current electromagnetism. We show that this view is not
possible under the ``rational realism'', complementary, in Appendix
\ref{subsec:blindness} we offer a relevant and example on how epistemological
blindness operates. The final sections correspond to discussions and
conclusions.

\section{The scientific attitude and the inference of a theory\label{sec:The-scientific-attitude}}

Scientists conceive the world as a cosmos, a harmonious totality.
For them, there is nothing as fascinating as discovering this harmony.
For this task they are equipped basically with two tools: reason and
experience. What they call understanding is the result of the interplay
of the two, for experience does not constitute knowledge if not for
the intervention of reason. These ideas (and some words) are taken
from Kant \citep{kant87} and Peirce \citep{peir55} and do not change
in their strength if, rather than considering the world as a cosmos,
we change the proposition to:\textsf{ the goal of the scientist is
to articulate a harmonic vision of the world, to make a cosmos out
of the sensorial input she/he receives.}

\subsection{Some rules for the construction of a cosmos\label{subsec:Real+NAP}}

The task of understanding the construction of EM requires some precision
on what we mean by reason and the requisites for inference.

\paragraph*{The principle of reality}

In the first place, we must indicate that the attempt of constructing
a cosmos out of sensorial input implies the assumption that there
is something real that reaches us through the senses, this is to say,
that there is subject and object. While the truth of this statement
is debatable, we can consider the dangers involved in accepting or
rejecting it. Little damage is done if accepting reality were an error
and it turns to be that everything is part of a unique encompassing
(solipsistic) being. On the contrary, if we were in error when rejecting
reality, we would become completely dysfunctional and miss one of
the greatest opportunities in life. The principle is summarised in:
\begin{quote}
``Such is the method of science. Its fundamental hypothesis, restated
in more familiar language, is this: There are Real things, whose characters
are entirely independent of our opinions about them; those Reals affect
our senses according to regular laws, and, though our sensations are
as different as are our relations to the objects, yet, by taking advantage
of the laws of perception, we can ascertain by reasoning how things
really and truly are; and any man, if he have sufficient experience
and he reason enough about it, will be led to the one True conclusion.
The new conception here involved is that of Reality.''\citep[p. 18]{peir55}
\end{quote}
Hence, we adopt realism as a starting point and reject conspiratorial
theories, this is, we reject hypotheses which cannot be put to experimental
test. We state this starting point in the form of a \textsf{principle}
\citep{sola18b}:

\begin{principle} There is a material world that we perceive with our senses (including experiments).
\end{principle}

\paragraph{The no arbitrariness principle}

In a recent work, we have shown that if we introduce some arbitrary
decisions in the scientific discourse (be it for the sake of the argument
or with the aim of facilitating an explanation), the set of possible
arbitrary elements must have the internal structure of a group\footnote{For example: We can say that the relations in the invariant relational
space are lifted into relations in the subjective spaces by arbitrary
decisions, but since the subjective statements must remain equivalent,
there must be a group of transformations, ${\cal G}$, that allows
us to move from one presentation to the other. If we conceive now
a theory as a space of statements, $E$, relating different concepts
belonging to our subjective presentation, what is real in them is
only the core that remains when we remove (mod out) the arbitrariness,
$T=E/{\cal G}$, which is the result of identifying statements that
only differ by the introduced arbitrariness. Thus, $T$ is invariant
while $E$ is equivariant with respect to ${\cal G}$.}, being then the set of all possible presentations of the argument
a representation of the group and as such equivalents \citep{sola18b}.
Further, we have shown that the facilitation of the relational concept
of space due to Leibniz produced by the introduction of a privileged
observer introduces a (useful) subjective element, the \textsf{subjective
space} (the space of all elementary physics texts) along with a series
of properties of this space as well as conditions that the statements
regarding physical laws must satisfy if they are going to remain rational.

The set of arbitrary decisions deserves some further consideration.
It has been indicated \citep{marg57}, in consideration of their own
versions of NAP, that choosing different arbitrary sets where the
statement (observation) should identically hold might lead to different
theories. Therefore, a clarification of the concept of arbitrariness
in this context is needed. It is important to indicate that any difference
that is dictated by experimental and observational methods should
be considered \textsf{not} to be arbitrary, but there is more to it.
For example, the class of (idealised) isolated systems, when we disregard
their internal structure, admits a group of arbitrariness. We call
this class of systems \textsf{inertial}\texttt{\textsc{ }}\citep{sola18b}.
They are to be distinguished from those systems that indicate as a
necessity the presence of a companion one. But since isolated systems
are an idealisation, the same can be said for inertial systems, thus,
approximately-inertial systems must exist. In practice, then, if a
system can be considered inertial or not, depends on the extent that
the presence of other matter not being accounted for can sensibly
modify the experimental outcome. This view has been held by experimentalists
such as Michelson \citep{mich04}.

This view of inertia must be contrasted with the subjective view.
In the subjective view the subject tries to draw a definition from
her/his own experience. She/he imagines being in a cage from which
all input from the exterior has been cut, and extrapolates experiences
within the cage to: ``If the cage moves with constant velocity (in
some subjective space she/he does not dare to mention) I won't be
able to grasp the velocity with which I am moving. On the other hand,
if the cage changes direction, it comes to a stop or sets to move
faster, I will know the cage is accelerated and then non-inertial,
for I am likely to bump on the cage borders''. Thus, the non-inertial
system is revealed as not being isotropic nor homogeneous, which are
precisely the symmetries required to suppress the arbitrariness of
the subject. In case our subject finds her/himself in a non-inertial
system, she/he will try to find out which are the particular properties
of this space, something that has to be constructed by comparison
with the expectations with respect to inertial space. Such space will
now be loaded with characteristics potentially varying from point
to point, this is, described by a mathematical field. This is about
how far the construction of reality initiated by the self-centered-child
can be taken, for any step further would lead to the recognition of
the observer as arbitrary (one out of many possible observers), rather
than observing through the eyes of God, that sees reality the way
reality is. Such a surpassing step would force our observer to leave
behind some of her/his most cherished toys, such as space, or kinetic
energy (in as much as it is not relational).

When constructing a theory we have to make an early decision: are
we going to introduce arbitrariness or not? The decision has not much
relevance if we keep track of the introduced arbitrariness, and acknowledge
the necessity of (and the methods for) removing it. However, if we
lose consciousness of our constructive effort, we might inadvertently
enter into the realm of arbitrariness. No amount of mathematics will
take us ever out of the subjetivist cage, since the necessary step
is not an analytic/deductive judgement but rather a synthetic/critical
one. This is, we need to understand not what the consequences of our
beliefs are, but rather which is the foundation of our beliefs.

If arbitrariness is the absence of reason, the double negation in
no-arbitrariness is equivalent to reason. This is, the rejection of
arbitrariness is a condition put on every rational construction \citep{sola18b}.

\begin{principle} {\bf [ No Arbitrariness Principle (NAP)]} No knowledge of Nature depends on arbitrary decisions. 
\end{principle}

\paragraph{The imagination as limit}

The introduction of explanatory hypothesis, the process of abduction,
is subject to the control of rationality and to the condition that
the newly introduced hypothesis explains a class of problems larger
that the one that motivated it, this is, that the hypothesis bears
some of the main ingredients of \textsf{cognitive surpass} \citep{piag89}
and offers itself more openly to refutation. However, the requisites
for the acceptance of explanatory hypotheses (i.e., to be able to
stand in front of refutation attempts) says nothing about the method
of production. There are no conditions for the process of creation
of hypotheses to be tested. The most natural source would be the intuition
of the sensible, \emph{empiria}, but in the case of EM data is scarce
and manifests itself as e.g., the deviation of a needle or a change
in the equilibrium point of a scale, that only indirectly represent
the phenomena. Hypotheses are then constructed supplementing data
with imagination. Since entertaining and testing hypotheses requires
considerable effort, there is a preselection of promising hypotheses.
In most (all?) cases the chosen hypotheses are those produced by analogy
and/or habit. Thus, physical hypothesis in the form of mechanical
systems (particles, fluids, springs, ...) enter the electromagnetic
scene and are erroneously placed as empirical input. Under the name
of ``physical reasoning'', physics instructors will train their
students\footnote{Anthropologist Sharon Traweek observes about high-energy physicists
that ``Undergraduate physics students, to be successful, must display
a high degree of intellectual skill, particularly in analogical thinking.
The students learn from textbooks whose interpretation of physics
is not to be challenged; in fact, it is not to be seen as interpretation.``
\citep[p. 74, ][]{traw92}. Also (p. 77): ``Teachers show students
how to recognize that a new problem is like this or that familiar
problem; in this introduction to the repertoire of soluble problems
to be memorized, the student is taught not induction or deduction
but analogic thinking.''} to seek answers in these terms. In so doing, they advance irrational
constructions in substitution of the far more difficult rational ones.
The increment in the use of analogies during the second part of the
XIX century is well documented in \citep{jung17}. But these irrational
constructs play a fundamental role when an observational problem has
to be cast into formulae and vice versa. The physical hypothesis and
analogies (a constitutive part of theories seldom recognised as such
since only formulae appear to have the attributes necessary for recognition)
are the main nexus with the observable reality.

What we can construct as images and memories comes from what is perceived
by our senses and since human beings are largely visual animals, the
root ``image'' in ``imagination'' must be taken almost at a literal
value. However, if we limit our explanatory hypotheses to those that
can be produced by these methods, we will soon produce a sensible
analogy among matters that, precisely, are of a different (immaterial)
kind. For example interactions are inferred elements and do not have
a material form, but our method of generating hypotheses will assign
to them the characteristics of bodies, thus, they will be placed in
space by analogy with waves in a material media or particles travelling
from here to there. Even Lorenz, that strongly objected the ether
\citep{lore67} found it necessary to have a material medium \citep{lore1863}.
The following text of Lord Kelvin illustrates analogy and imagination
as a limit:
\begin{quote}
I never satisfy myself until I can make a mechanical model of a thing.
If I can make a mechanical model I can understand it. As long as I
cannot make a mechanical model all the way through I cannot understand;
and that is why I cannot get {[}this is probably the reporter's Americanism
for the word ``accept''{]} the electromagnetic theory.\citep[p.235]{thompson2011life}
\end{quote}

\paragraph{The continuity principle (reduction to the obvious/evident)}

Argumentations are constructed in such a way that they rest upon small
units we consider evident or obvious. Yet, what is obvious or evident
for some, may not be so for others. One of the forms in which we usually
identify potentially irrational arguments is by detecting hiatus or
\emph{lacunæ} in the argumentation. The request “please, fill in the
gap“ quite often reveals a belief that cannot be supported while being
essential to the argument. On the contrary, the rational argumentation
proceeds to fill the gaps by explaining how they consist of the concatenation
of smaller pieces, iterating the process until the pieces are accepted
as evident or obvious. This self-similar form corresponds to what
in mathematics is called\texttt{ }\textsf{continuity}.

\paragraph{The mediation principle and the dialectical openings}

We do not usually accept as reasonable that which appears out of nothingness
as self-evident assertions. We normally request a new rational-belief
to be derived (mediated) by acceptable argumentation from accepted
beliefs. This recurrent form of reasoning cannot be pursued indefinitely.
It comes to an end when we reach a point in which beliefs can no longer
be derived from other accepted beliefs. At this point there seems
to be only one option: Either we make explicit a layer of arbitrary
assumptions (axioms) which is the opaque end that reason lets us see,
or we find a set of opposing concepts and ideas that in their interplay
constitute the foundation of our discussion; the dialectical openings\citep{sola18b}.
The participating concepts or ideas support each other by oppositions
and they exist only as dialectically opposing elements, while it is
this opposition what we consider as perceived, as real. In case that
we decide to introduce arbitrary hypothesis and yet remain rational,
we would be forced to mod them out as explained in NAP, thus something
not arbitrary would have to survive. Hence, the alternative is: irrational
or dialectic.

\paragraph{Logical action in front of contradictions}

Whenever a chain of reasoning arrives to a contradiction, the whole
chain is rejected. When the contradiction results from comparing theoretical
prediction and experimental reality we speak of an experimental refutation.
The logical scheme can be depicted as $A\Rightarrow B\Rightarrow\left[\mathrm{further\ consequences}\right]\Rightarrow\mathrm{False}$.
No matter how pleasant $B$ and other intermediate consequences are,
there is no support for them. The most evident example is the hypothesis
of the ether which is fundamental for the proposition of Maxwell's
displacement field. Discarding the existence of the ether (following
empirical evidence) under the present principle would mean the refutation
of the hypothesis as well as all of its consequences. The theory of
EM would have to be retracted to the point at which it incorporated
light and then reconstructed again under new hypotheses. Sustaining
a thesis despite that its argumentation has been shown to be false
is the case of\emph{ }\textsf{persistence}\emph{ }in Peirce's classification
of the methods of fixating beliefs \citep[ch. 2]{peir55}. It is,
as such, irrational.

There is another instance of the same logical scheme which is not
usually considered, namely when the contradiction stems from the logical
structure of the theory (e.g., inconsistent postulates). Assume $A$
is True, then $-A$ is False. The construction $A\Rightarrow B\Rightarrow\left[\mathrm{further\ consequences}\right]\Rightarrow(-A)$
discloses an internal contradiction of the theory and, as above, it
forces us to reject the full chain. It is observed that some authors
are tempted to change the sequence into: Axiom $B$ is true, $B\Rightarrow\left[\mathrm{further\ consequences}\right]\Rightarrow(-A)$
detaching the statements from its production process keeping (without
an alternative argumentation) the desired result. Informally, we call
the process “cutting the branch we are sitting on“, we fall in this
case into a violation of the continuity principle and the mediation
principle. This situation relates directly to observations made by
\citet{ding60} in which the author finds that in the construction
of Special Relativity the formulae are detached of their physical
meaning.

\paragraph{On reason}

What is reason then? How can we reason about what reason is? Reason
can only be conceived in front of no-reason, of arbitrariness. But
arbitrary is the statement that introduces a belief without admitting
doubt about it (perhaps by conviction/persistence, or authority).
Hence, to discuss reason and arbitrariness we need to discuss belief
and doubt, the forms of fixation of beliefs, and the relations between
these elements. Further, since what is imposed one way or another
(as in arbitrariness) cannot be said to be harmonious, we find that
reason is directly related to harmony. Perhaps at this point it is
better to consider our fundamental insight: we feel in peace only
with a rational world, a world with harmony, a cosmos, and not a world
that is the result of power and arbitrariness. Hence, our choice of
reason for the organisation of the world emerges directly from our
feelings. These feelings can be made universal, for it is possible
to conceive a humanity of free persons that consistently entertain
doubts and peacefully cooperate to progress in their common beliefs,
while it is not possible to conceive a peaceful coexistence of arbitrary
beliefs, except in the final state of uniformity (death). Rather than
coexisting peacefully, arbitrary beliefs tend to replace each other
by forceful imposition. Thus, our choice for rationality is not rational,
it emerges from our feelings, and it is perhaps because of them that
reason will ever re-emerge.

\subsection{Vulgar pragmatism}

Would a scientist agree on the need of being reasonable and on the
meaning of reason? We expect not to find a scientist rejecting the
idea of being reasonable. Furthermore, we cannot conceive a scientist
admitting not being reasonable without at the same time withdrawing
her/his argument. But self-evaluation of our own reasonability is
usually poor in self-criticism, for reasoning is often \textsf{this
thing that I am doing}, while I have no access to the totality of
the processes of reasoning in others. Some schools of thought (we
will call them \textsf{sophists}) regard reasoning as just one form
(among several others) of disputing and convincing. Philosophers have
in turn indicated that reasoning is not always at the core of scientific
research \citep{feye87} and that only instrumental reason is actually
being used \citep{hork47}.

One of the cornerstones of reasonability is to reject logical inconsistency.
The attitude we take in front of error is decisive: the scientific
person yields to reason and to empirical evidence. If a theory is
found to be inconsistent, and also if it produces predictions against
empirical evidence, its reasonability collapses, we are no longer
satisfied by its explanatory power and it is therefore rejected, as
Peirce explicitly requested\footnote{See e.g., ``It is a great mistake to suppose that the mind of the
active scientist is filled with propositions which, if not proved
beyond all reasonable cavil, are at least extremely probable. On the
contrary, he entertains hypotheses which are almost wildly incredible,
and treats them with respect for the time being. Why does he do this?
Simply because any scientific proposition whatever is always liable
to be refuted and dropped at short notice.“ (p.36) “But the scientific
spirit requires a man to be at all times ready to dump his whole cartload
of beliefs, the moment experience is against them.“ (p.46) \citep{peir55}} and later Popper will make central to his philosophy of science \citep{popp59}.

At first sight, this vision is probably shared by most scientists.
On the contrary, the attitude of persistence in beliefs (tenacity
\citep[ch. 2, ][]{peir55}, in Peirce's denomination) despite empirical
evidence against them or despite logical inconsistencies, is one form
of being \textsf{non}-scientific. Note that our objection is not to
the results or predictions that a non-scientific theory could produce,
but rather to its construction. Considering success in prediction
ability as the only measure of scientific correctness amounts to replacing
the attitude of the scientist with some form of\emph{ }\textsf{vulgar
pragmatism}. Peirce \citep[Fixation of belief,][]{peir55} regards
human actions as the outcome of the struggle between doubt and belief.
The former provokes tension and the urge for resolution, the latter
is a state of peace of mind towards which we strive. He further identifies
reason as the only way of fixating beliefs that is free of arbitrariness,
quite in line with the concept of \textsf{critical pragmatism}\emph{
}\citep{pennycook1997vulgar} or --as we prefer to call it-- \textsf{scientific
pragmatism}. Any method of fixating beliefs other than reason, e.g.,
the argument by confirmation or verification that a belief is \textsf{useful}
for any purpose other than reasonability (personal benefit, success,
the imposition of an idea, etc.) will be called \textsf{vulgar pragmatism},
in line with \citep{pennycook1997vulgar}.

The problems presented by the incorporation of light into EM had become
so frustrating by the beginning of the XX century that Poincaré was
willing to accept a radical change in the scientific attitude, this
is, the acceptance of “persistence in beliefs“, the unwillingness
of dropping our theories when they are contradicted by experiments
(a form of vulgar pragmatism already identified by Peirce \citep[ch. 2]{peir55}
by 1887). But perhaps most remarkable, is the degree of collective
blindness that developed, since shortly after Maxwell's landmark paper,
read in December 1864 \citep{maxw64}, Lorenz presented his extension
of Weber's theory incorporating light without ether\emph{ }\citep{lore67},
a relational theory that as such predicted way in advance the outcome
of the famous Michelson-Morley experiment and that was found to be
identical to Maxwell's theory when later applied to Hertz' experiments.
We look to this situation with perplexity. By 1881 there were two
theories (Lorenz' and Maxwell's) and one discriminant experiment (Michelson's).
Despite the correspondence of the experimental results with only one
of the theories, we persist in patching the failed theory using conspiratorial
thoughts?

This was not an isolated fact. By the turn of the century the change
in epistemological attitude away from the scientific approach and
towards verificationism became increasingly evident. Let us consider
the proposition:
\begin{quote}
Supposons que, dans une nuit, toutes les dimensions de l'univers deviennent
mille fois plus grandes: le monde sera resté semblable à lui-même,
en donnant au mot de similitude le même sens qu'au 3° livre de géométrie.
Seulement, ce qui avait un mètre de long mesurera désormais un kilomètre,
ce qui était long d'un millimètre deviendra long d'un mètre. Le lit
où je suis couché et mon corps lui-même se seront agrandis dans la
même proportion. Quand je me réveillerai le lendemain matin, quel
sentiment éprouverai-je en présence d'une aussi étonnante transformation?
Eh bien, je ne m'aper\-cevrai de rien du tout. Les mesures le plus
précises seront incapables de me rien révéler de cet immense bouleversement
puisque les mètres dont je me ser virai auront varié précisément dans
les mêmes proportions que les objets que je chercherai à mesurer...
D'après Lorentz et Fitzgerald tous les corps entraînés dans le mouvement
de la Terre subis sent une déformation. Cette déformation est à la
vérité très faible, puisque toutes les dimensions parallèles au mouvement
de la Terre diminueraient d'un cent millionième, tandis que les dimensions
perpendiculaires à ce mouvement ne seraient pas altérées. \citep{poin06}
\end{quote}
This argumentation has several problems. The proposed contraction
is imperceptible, nothing changes in our relation to the world if
we accept it or reject it except probably the issue that it intends
to explain. As an explanatory hypothesis it does not meet the requisite
of explaining more than the observation that suggested it. The explanation
proceeds through something otherwise unverifiable and of no practical
consequences whatsoever (as the excerpt claims). There are no grounds
to prefer this explanation to other alternatives. It only serves the
purpose of fixating our belief that there is no mistake in the considerations
that led us to the conclusion that the speed of light should depend
on the velocity of Earth relative to the ether\emph{.} The idea of
having to accept such a conspiratorial hypothesis to save a theory
of its wreckage is indignant. With higher dignity, Kelvin rejects
the vibrations of the ether as an explanation of light since: “I firmly
believe in an electromagnetic theory of light, and that when we understand
electricity and magnetism and light we shall see them all together
as parts of a whole. But I want to understand light as well as I can,
without introducing things that we understand even less of.“ \citep[pp. 835--836]{thompson2011life}

It should be noticed on passing that Poincaré's attitudes towards
Lorentz' theories had been fluctuating. In a homage to Lorentz, Poincaré
writes:
\begin{quote}
It would no doubt seem strange that in a monument raised to the glory
of Lorentz I would review the considerations which I presented previously
as an objection to his theory. I could say that the pages which follow
are rather in the nature of an attenuation rather than a magnification
of that objection. But I disdain that excuse, because I have one which
is 100 times better: Good theories are flexible. Those which have
a rigid form and which can not change that form without collapsing
really have too little vitality. But if a theory is solid, then it
can be cast in diverse forms, it resists all attacks, and its essential
meaning remains unaffected. \citep{poin00}
\end{quote}
It appears from this text that a theory is not “vital“ enough if it
“collapses“ in front of contradictory empirical evidence.

What is, then, the goal of an explanatory hypothesis? In the views
of scientific pragmatism it is “to lead to the avoidance of all surprise
and to the establishment of an habit of positive expectation that
shall not be disappointed“ \citep[p. 267,][]{peir55}, through subjection
to the test of experiment. In the absence of any special reasons for
the contrary, any hypothesis, therefore, may be admissible provided
it being capable of experimental verification, and only in so far
as it is capable of such verification. Peirce again: “An explanatory
hypothesis, that is to say, a conception which does not limit its
purpose to enabling the mind to grasp into one a variety of facts,
but \textsf{which seeks to connect those facts with our general conceptions
of the universe}, ought, in one sense, to be verifiable; that is to
say, it ought to be little more than a ligament of numberless possible
predictions concerning future experience, so that if they fail, it
fails“ \citep[p. 267, ][(our emphasis)]{peir55}.

Other indications of the advance of different forms of vulgar pragmatism
arose along the XX century:
\begin{quote}
Science is the attempt to make the chaotic diversity of our sense
experience a logically uniform system of thought.\citep{eins40}
\end{quote}
The word “uniform“ has taken the place of “harmony“. But harmony is
the quality of producing a single and pleasant totality, while uniformity
implies the suppression of differences, something that many of us
find utterly unpleasant since we regard life as a harmonious diversity
and death as uniformity (remaining the same in all cases and at all
times). The search for the unity of reason is then replaced by the
search of a uniform symbolic manipulation of all forces, see for example
\citep{smol08}, being the later a goal akin to industrialisation
rather than to Enlightenment. The unity of reason requires to transcend
the oppositions/differences (at the same time preserving them) or
unifying them in their differences by surpassing them \citep{piag89},
achieving a more abstract and encompassing vision.

According to Einstein:
\begin{quote}
“Physics constitutes a logical system of thought which is in a state
of evolution, and whose basis cannot be obtained through distillation
by any inductive method from the experiences lived through, but which
can only be attained by free invention. The justification (truth content)
of the system rests in the proof of usefulness of the resulting theorems
on the basis of sense experiences, where the relations of the latter
to the former can only be comprehended intuitively. Evolution is going
on in the direction of increasing simplicity of the logical basis“.
\citep{eins40}
\end{quote}
It appears from this statement that physics has no roots, it floats
in the air as a free invention without ties. Since the ties that enter
the construction of a theory rule the use of the theory, the reverse
link between theoretical results and sensorial experiences rests upon
some sort of “intuition“. It is not intuition with its usual philosophical
meaning (a central idea in \citep{huss83} where intuition mediates
in the process of ideation, the construction of the empirical input
into ideas and later theories). The word “intuition“ in the above
quotation must be understood as involving elements removed from rational
control. The “physics“ is thus bracketed between nothingness and irrationality
since the ties of philosophical intuition as well as the rational
ties of the continuity and mediation principles have been removed,
constructing then a new epistemology that resembles vulgar pragmatism:
it is true because it is useful.

\subsection{Maxwell's epistemic position\label{subsec:Maxwell's-epistemic}}

James Clerk Maxwell's contributions to electromagnetism culminated
in his \textsf{Treatise} \citep{maxw73} where he extends his foundational
paper \citep{maxw64} into a comprehensive work encompassing old and
new mathematics and physics developed along several decades. In both
works, Maxwell introduces considerations that correspond better to
epistemology than to mathematics or physics.

From the beginning, his goal is to describe electromagnetic phenomena
by means of local interactions with the surrounding medium\emph{ }(the\emph{
}ether), as opposed to the action at a distance theories, more developed
outside England. Maxwell begins by celebrating the continental tradition
of electromagnetism \citep{maxw64}:
\begin{quote}
The most obvious mechanical phenomenon in electrical and magnetical
experiments is the mutual action by which bodies in certain states
set each other in motion while still at a sensible distance from each
other. The first step, therefore, in reducing this phenomena into
scientific form is to ascertain the magnitude and direction of the
force acting between the bodies {[}...{]}

In this way, mathematical theories of statical electricity, of magnetism,
of the mechanical action between conductors carrying currents and
of the induction of currents have been formed {[}...{]}

These theories assume, more or less explicitly, the existence of substances
the particles of which have the property of acting on one another
at a distance by attraction or repulsion. The most complete development
of a theory of this kind is that of M. W. \citep{webe46} who made
the same theory include both electrostatic and electromagnetic phenomena.

In doing so, however, he has found it necessary to assume that the
force between two electric particles depend on their relative velocity,
as well as on their distance.

This theory, as developed by M. W. Weber and C. Neumann, is exceedingly
ingenious, and wonderfully comprehensive in its application to phenomena
of statical electricity, electromagnetic attractions, induction of
currents and diamagnetic phenomena; and it comes to us with the more
authority as it has served to guide the speculations of one who has
made so great an advance in the practical part of electric science,
both by introducing a consistent systems of units in electrical measurement,
and by actually determining electrical quantities with an accuracy
hitherto unknown\footnote{Maxwell tries to confine Weber within the status of somebody concerned
with the “practical part“, since the great advance indicated is precisely
Weber's work.}.
\end{quote}
In other parts of the Treatise, Maxwell points out that his theory
and Weber's arrive to the same descriptions of different electromagnetic
phenomena. See e.g., \citep[{[644], [855]}][]{maxw73} among similar
comments scattered along the work. However, he finds Weber's approach
too difficult \citep{maxw64}, arguing as follows for his point of
view:
\begin{quote}
(2) The mechanical difficulties, however, which are involved in the
assumption of particles acting at distance with forces that depend
on their velocities are much as to prevent me from considering this
theory as an ultimate one, thought it might have been and may yet
be useful in leading to the coordination of phenomena.

I have therefore preferred to seek an explanation of the fact in another
direction by supposing them to be produced by actions which go on
the surrounding medium as well as in the excited bodies, and endeavouring
to explain the action between distant bodies without assuming the
existence of forces capable of acting directly at sensible distances.

(3) The theory I propose may be therefore be called a theory of the
\textsf{Electromagnetic Field}, because it has to do with the space
in the neighbourhood of the electric or magnetic bodies, and may be
called a \textsf{Dynamical} Theory, because it assumes that in that
space there is matter in motion, by which the observed electromagnetic
phenomena are produced.
\end{quote}
In his Treatise, Maxwell explains further \citep[{[529]}][]{maxw73}:
\begin{quote}
We are accustomed to consider the universe as made up of parts, and
mathematicians usually begin by considering a single particle, and
then conceiving its relation to another particle and so on. This has
generally been supposed the most natural method. To conceive of a
particle, however, requires a process of abstraction, since all our
perceptions are related to extended bodies, so that the idea of the
\textsf{all} that is in our consciousness at a given instant is perhaps
as primitive an idea as that of any individual thing. Hence there
may be a mathematical method in which we proceed from the whole to
the parts instead of from the parts to the whole. For example, Euclid,
in his first book conceives a line as traced out by a point, a surface
as swept out by a line and a solid as generated by a surface. But
he also defines a surface as the boundary of a solid, a line as the
edge of a surface and a point as the extremity of a line.
\end{quote}
Both in the main text of these works and also in the Preface of the
Treatise, Maxwell claims that his views are based on the approach
of Faraday to electromagnetic phenomena. However, while Faraday entertained
doubts about the necessity of the ether and indeed advanced an alternative
explanation, Maxwell misinterprets Faraday writings (in text and spirit)
as already pointed out clearly in \citet[p. 250-52]{ding60}. In the
work cited by Maxwell, Faraday writes, in the manuscript entitled
“Thoughts on ray vibrations“ \citep[p. 447]{fara55},
\begin{quote}
The point intended to be set forth for consideration of the hearers
was, whether it was not possible that the vibrations that in certain
theory are assumed to account for radiation and radiant phænomena
cannot occur in the lines of force which connect particles, and consequently
masses of matter together; a motion that, as far as it is admitted,
will dispense with the æther, which in another view is supposed to
be the medium in which this vibrations take place.

You are aware of the speculation which I some times since uttered
respecting that view of the nature of matter that considers its ultimate
atoms as centres of force, and not as so many little bodies surrounded
by forces, the bodies being considered in the abstract as independent
of the forces and capable of existing without them. In the later view,
this little atoms have a definite form and a certain limited size;
in the former view such is not the case, for that which represents
size may be considered as extending to any distance to which the lines
of force of the particle extends: the particle indeed is supposed
to exist only by these forces, and where they are it is.
\end{quote}
Faraday's speculation about matter is “A speculation touching Electric
Conduction and the Nature of Matter“ \citep[p. 284]{fara44} and rests
upon ideas of Boscovich. The speculation is not an opinion, Faraday
philosophically proceeds to show that current ideas on matter (of
that time, 1844) were incompatible with electricity before adopting
Boscovich's view. The view basically consists in a duality\footnote{We will return to this issue in Section \ref{sec:Matter-and-electricity}}.
We only know about matter because of its actions (forces) and we locate
matter where the centre of this action lies. If the action extends
through the universe, the atoms are “where they act“, this is in all
the universe. Matter and force cannot be separated, they are two aspects
of an unity. Maxwell actually resents such ideas and tries to amend
Faraday:
\begin{quote}
{[}speaking of Faraday{]} ...He even speaks of the lines of force
belonging to a body as in some sense part of itself, so that in its
action to distant bodies it cannot be said to act where it is not.
This, however, is not a dominant idea with Faraday. I think he would
rather have said that the field of space is full of lines of force,
whose arrangement depends on that of the bodies in the field, and
that the mechanical and electrical action on each body is determined
by the lines which abut on it \citep[{[529]}][]{maxw73}.
\end{quote}
From a rational perspective, Maxwell's argumentation is disappointing.
To sustain that Faraday should have said what he did not say amounts
to the vulgar strategy of substituting adverse observations for desired
observations. If the idea was dominant or not with Faraday is beside
the point. The beliefs of Faraday are of no importance, what matters
are his arguments and Maxwell chooses not to confront them but to
hold to his own beliefs. This way of avoiding adverse ideas contrasts
with the otherwise careful and open-minded attitude of Maxwell both
as a philosopher and even more as a mathematician \footnote{Recall for example the discussion on Ampère's force \citep[{[509],[510], [526]}][]{maxw73},
the derivation of the electromagnetic force inside a conductor \citep[{[598]}][]{maxw73}
or his remarks about the limits of the fluid analogy \citep[{[574]}][]{maxw73}.}. We will return to Maxwell's philosophical views in Section \ref{subsec:Hertz},
contrasting them with Hertz' ideas.

The tradition of natural philosophy to which Faraday adhered comes
to an abrupt end in Maxwell. For the sake of completeness it must
be noted that Faraday never criticised nor put down his idea on ray
vibrations. It is advanced as a hypothesis open to criticism, and
there it stands, still today. It appears that he never returned to
the subject.

The foundation of Maxwell's beliefs can be found in \citep[{[866]}][]{maxw73}
where he discusses action at distance and writes
\begin{quote}
Now we are unable to conceive propagation in time, except either as
the flight of a material substance through space, or as the propagation
of a condition of motion or stress in a medium already existing in
space.{[}...{]} If something is transmitted from one particle to another
at a distance, what is its condition after it has left the one particle
and before it has reached the other?
\end{quote}
The belief (provisional in Maxwell) in the ether is clearly stated
as the result of a limitation by reduction to ideas that are part
of our experience with matter, but applied to inferred ideas such
as (inter)actions. It represents a fear of abstraction (at the end
a fear to reason) which is present for example in \citep[10 intro]{berk17}.
This is the main hidden lemma of theoretical physics which is of epistemic
or philosophical character and consists in the rejection, \emph{in
limine}, of ideas that can not be imagined (constructed with mental
images of material experiences). Faraday's “vibrating rays“ could
be easily ridiculed if taken literally. The rays are only a form of
illustration of some aspects of an interaction; imagining a vibrating
illustration is a valid pedagogical tool, but it can hardly be considered
part of the Real (except as illustration). But the image facilitates
communication (and then acceptance) reaching those, as Kelvin for
example, that need a material support for their thoughts. The obstacle
is not Physics, or the Universe or the Cosmos, the obstacles are our
own capabilities for reasoning and our confidence in reason. Such
a problem is not present in the construction of classical mechanics.

Last, but not least, Maxwell's introduces a new form of exploring
possible organisations of the Cosmos. The introduction of a Lagrangian
for electromagnetic phenomena (the starting point of Maxwell's research)
is compared \citet[{[554]}][]{maxw73} with the use of the same mathematical
object by Lagrange “the end of Lagrange was to bring dynamics under
the power of the calculus{[}...{]} our aim, on the other hand, is
to cultivate our dynamical ideas“. Thus, Maxwell introduces mathematical
analogies. If the idea of current, as explained by Faraday, implies
the movement of electricity, and we have an expression (due to F.
Neumann \citep[{[542-543]}][]{maxw73}) for the energy involved in
the interaction between two circuits with electrical currents, then
we can construct the associated momentum (the \textsf{electro-tonic
state} of Faraday \citep[Par. 60][]{fara39}) of the current as the
derivative of the “kinetic energy“ (the energy involved in the movement)
with respect to the current. This fundamental quantity is what in
today's language is known as the vector potential, $A$, and correspondingly,
its time derivative is a “force“, the electromagnetic force\footnote{Note that the time derivative here is a convective, or “total“, derivative
and not a partial derivative.}, $fem=-\frac{dA}{dt}$. We will come back to the important details
of this matter.

\subsection{Poincaré's proposed reorganisation of science\label{subsec:Poincar=0000E9}}

Poincaré's attitude towards Electrodynamics contrasts with his important
contributions to Mechanics. He argues for the ether hypothesis from
a conservative standpoint (without it the description of phenomena
would be more complicated than what we are familiar with). Also, his
mathematical arguments are incorrect in several points:
\begin{quote}
And does our ether really exist? We know the origin of our belief
in the ether. If light reaches us from a distant star, during several
years it was no longer on the star and not yet on the earth; it must
then be somewhere and sustained, so to speak, by some material support.

The same idea may be expressed under a more mathematical and more
abstract form. What we ascertain are the changes undergone by material
molecules; we see, for instance, that our photographic plate feels
the consequences of phenomena of which the incandescent mass of the
star was the theater several years before. Now, in ordinary mechanics
the state of the system studied depends only on its state at an instant
immediately anterior; therefore the system satisfies differential
equations. On the contrary, if we should not believe in the ether,
the state of the material universe would depend not only on the state
immediately preceding, but on states much older; the system would
satisfy equations of finite differences. It is to escape this derogation
of the general laws of mechanics that we have invented the ether.

That would still only oblige us to fill up, with the ether, the interplanetary
void, but not to make it penetrate the bosom of the material media
themselves. Fizeau's experiment goes further. By the interference
of rays which have traversed air or water in motion, it seems to show
us two different media interpenetrating and yet changing place one
with regard to the other.

We seem to touch the ether with the finger.

Yet experiments may be conceived which would make us touch it still
more nearly. Suppose Newton's principle, of the equality of action
and reaction, no longer true if applied to matter alone, and that
we have established it. The geometric sum of all the forces applied
to all the material molecules would no longer be null. It would be
necessary then, if we did not wish to change all mechanics, to introduce
the ether, in order that this action which matter appeared to experience
should be counterbalanced by the reaction of matter on something.

Or again, suppose we discover that optical and electrical phenomena
are influenced by the motion of the earth. We should be led to conclude
that these phenomena might reveal to us not only the relative motions
of material bodies, but what would seem to be their absolute motions.
Again, an ether would be necessary, that these so-called absolute
motions should not be their displacements with regard to a void space,
but their displacements with regard to something concrete.\citep[p. 147]{poin13foundations}\footnote{\citet{torr07} quotes his own translation of the first two paragraphs
and comments that the mathematical insight of the second paragraph
is wrong. We add that it was known to be wrong to Gauss and the school
of Göttingen as well as to Maxwell.}
\end{quote}
But Poincaré realises that the ether is an intuition: “We seem to
touch the ether with the finger“ is the English translation of the
French expression: “On croit toucher l'éther du doigt“ which meas
to understand intuitively.\footnote{There are at least two digital versions of this book. We quote the
translation by Halsted which is faithful to the French original version
in this point as well as to printed versions. There is another version
translated by W.J.G. \citep{poin13E} which reads: “The ether is all
but in our grasp“}

The first paragraph almost reproduces Maxwell's argument (quoted above).
It is interesting to realise that when the ether was at last derogated,
the operation was performed by switching to the other option given
by Maxwell in his alternative (the flight of a material substance),
this is, by creating the photon that, as Poincaré understands, must
make some elements of the old physics (as Poincaré calls it) persistent.

\citep[pp. 300--301]{poin13foundations} proposes a reorganisation
of physics in terms of “principles“ and he lists them:
\begin{enumerate}
\item Conservation of energy (Mayer's principle)
\item Degradation of energy (Carnot's)
\item Equality of action and reaction (Newton's)
\item Relativity, according to which the laws of physical phenomena must
be the same for an stationary observer or for an observer carried
along in a uniform form of translation; so that we have not and cannot
have any means of discerning whether or not we are carried along in
such motion.
\item Conservation of mass (Lavoisier's)
\item (I will add) the principle of least action.
\end{enumerate}
\begin{quote}
The application of these five or six general principles to the different
physical phenomena is sufficient for our learning of them all that
we could reasonably hope to know of them. The most remarkable example
of this new mathematical physics is, beyond question, Maxwell's electromagnetic
theory of light.{[}...{]}

This principles are results of experiments boldly generalized; but
they seem to derive from their very generality a high degree of certainty.
In fact, the more general they are, the more frequent are the opportunities
to check them, and the verifications multiplying, taking the most
varied, the most unexpected forms, end by no longer leaving place
for doubt. \citep[Original in French of 1905]{poin13foundations}
\end{quote}
Poincaré establishes his principles (beliefs) from experience as he
says. But the experiences here implied are of two different kinds.
On one side 1, 2, 5 and perhaps 3 can be the subject of experimental
tests, they come from \emph{empiria}. Actually, Lavoisier's law was
a falling belief at the time of his writings. In contrast, the relativity
and the least action principles come from the experience of our practices,
they are constructive principles \citep{marg57} that come from \emph{habitus}.
Poincaré does not seem to be aware of the difference between the empirical
input (the observation) and our ideation.

\subsubsection{The principle of relativity}

In their epistemological appraisal of special relativity, Margeneau
and Mould \citep{marg57} analyse the principle of relativity from
a philosophical perspective. For these authors it is a constructive
principle, and we agree. They introduce two versions of the principle:
the new and the old. Referring to the old they say: “Historically,
relativity is associated with problems arising out of the need to
provide a reference for particle motion. The question which philosophers
have asked is: Should quantities like particle position, velocity
and acceleration be referred to an absolute, primitively given spatial
frame of reference; or should the kinematic notions which fix the
state of the particle have meaning relative only to other particles?“
and later, they state “The modern form of the principle is not concerned
with the status (primitive or defined) of the concepts it treats,
but rather with the extensibility of the axioms over the range of
individuals included in the axiomatic structure. It requires the elimination
of special or preferred individuals. 'Individual' here refers to membership
in any given or generated collection of constructs of a specified
kind.“

The principle of relativity appears in \citep{poin13foundations,marg57,ding60}
as emerging from irrationality, this is, it is not mediated by other
beliefs but it appears as true to the elite that has been previously
trained in Newtonian mechanics, for it extrapolates the \emph{habitus}
of the Newtonian construction. It is only very recently that we have
surpassed the direct irrational acceptance showing that it is a requirement
of reason and of the principle of reality (a bolder belief). If there
exists something real which is the concern of science, the real cannot
depend on our decisions, as stated in Peirce's paragraph quoted above.
By considering that the objective must be \textsf{intersubjective}
we arrived to NAP \citep{sola18b}. But NAP requires a mathematical
structure relating arbitrariness, the structure of a mathematical
group.\footnote{Margenau and Mould use the word group \citet{marg57} in at least
two forms: as equivalent to set (as in “this group includes many teachers“)
and with the mathematical meaning used in NAP \citep{sola18b} (as
in “transformation group“ or the non-existent but frequently mentioned
“Lorentz group“).}

Although NAP was thought in terms of the description of the real with
empirical foundation, its significance overpasses it. If we consider
a fundamental theory as real (although not directly related to observations)
the theory must be free of arbitrariness as well, and since the consecutive
applications of arbitrary actions is an arbitrariness as well, the
set of arbitrary elements must be related by operations pertaining
to a group. Poincaré's relativity principle can be viewed then as
the requirement that the relation between symbols be invariant (indirectly,
that the concepts ideated from the empiria and mediated by the \textsf{rules
of correspondence} have invariant relations).

Margenau and Mould seek to escape from the problems of Special relativity
by enlarging the set of arbitrariness the way Einstein did in the
General Theory of relativity \citet{marg57}. The problem is then:
Do (approximately) inertial systems exist? If the idea of inertial
systems were an arbitrariness, then we should go along with Einstein.
Otherwise, we should restrain to label as arbitrary what is in fact
an idealisation emerging from an empirical result. If we conceive
the motion of a reference system in relation to absolute space or
the ether, it is clear that the notion falls when we reject both ideas.
But as we have shown \citep{sola18b} the idea of inertial motion
emerges from the concept of isolation, and absolute isolation is an
idealisation of empirical observations accessible to trained and untrained
eyes. Non-inertial motion can be measured. A key experiment related
to the present issues was performed using a ring interferometer and
a rotating table \citep{sagn13}. Michelson showed the effect of rotation
of the Earth in the propagation of light \citep{mich25,mich25b} and
discussed why the experiment could not account for the movement of
Earth around the sun \citep{mich04}. Sagnac (ring) interferometers
are used regularly as inerciometers in aerial navigation \citep{post67}.\footnote{Ironically, Sagnac's experiment is used by relativists (for example
\citep{maly00}) as well as by defenders of the ether (e.g., \citep{silv89})
to confirm their theories, and there is no contradiction in that because
only pre-existing beliefs can be confirmed. The fact is that the formula
produced by “extending“ special relativity to slowly rotating systems
and by retaining the ether are just the same expression.}Hence, approximately inertial is measurable and inertial systems are
an idealised category. Margenau and Mould have no right to mix inertial
frames and non-inertial frames in the same arbitrariness class, a
tradition that goes back (at least) to Mach \citet[(II.VI.5, p. 232)]{mach19}
and was criticised by Poincaré \citep[Ch. VII]{poin13foundations}
using precisely the idea of isolation.

\section{Matter and electricity\label{sec:Matter-and-electricity}}

From the earlier investigations on static electricity it was clear
that there were two kinds of materials: conductors and insulators.
While in conductors the static electricity was able to move, no such
macroscopic displacement was possible in insulators. Electricity was
therefore conceived as matter inside matter. This is, either as a
fluid (or two fluids) or as particles moving inside matter. In the
second conception, the particles were positive and negative, and it
was later assumed that they would move in opposite directions with
equal velocities, an image known as \textsf{Fechner hypothesis} \citep[p. 52]{assi94}.
These early images still lurk behind some debates that oppose Lorentz'
force to Ampère's force, despite their difference being clarified
by Maxwell (see below). Since electricity was bound to matter, the
force exerted within the electrical substance would be transmitted
to the wire that was constraining it. The concepts arising from this
matter-inside-matter picture led to the distinctions \textsf{ponderable
matter} vs. \textsf{imponderable matter} and \textsf{mechanical force}
vs. \textsf{electromotive force}.

However, the movement of electricity (so conceived) within some materials
challenged the existing ideas of matter. It was Faraday who, at an
early stage, clearly understood the problem:
\begin{quote}
The view of the atomic constitution of matter which I think is most
prevalent, is that which considers the atom as a something material
having a certain volume, upon which those powers were impressed at
the creation, which have given it, from that time to the present,
the capability of constituting, when many atoms are congregated together
into groups, the different substances whose effects and properties
we observe. These, though grouped and held together by their powers,
do not touch each other, but have intervening space, otherwise pressure
or cold could not make a body contract into a smaller bulk, nor heat
or tension make it larger; in liquids these atoms or particles are
free to move about one another, and in vapours or gases they are also
present, but removed very much further apart, though still related
to each other by their powers. {[}...{]}

\textbf{But it is always safe and philosophic to distinguish, as much
as is in our power, fact from theory; the experience of past ages
is sufficient to show us the wisdom of such a course; and considering
the constant tendency of the mind to rest on an assumption, and, when
it answers every present purpose, to forget that it is an assumption,
we ought to remember that it, in such cases, becomes a prejudice,
and inevitably interferes, more or less, with a clear-sighted judgment
{[}...{]}}

If the view of the constitution of matter already referred to be assumed
to be correct, and I may be allowed to speak of the particles of matter
and of the space between them (in water, or in the vapour of water
for instance) as two different things, then space must be taken as
the only continuous part, for the particles are considered as separated
by space from each other. Space will permeate all masses of matter
in every direction like a net, except that in place of meshes it will
form cells, isolating each atom from its neighbours, and itself only
being continuous \citep[p. 284--286]{fara44}(emphasis added).
\end{quote}
Faraday proceeds then to show that space cannot be conceived neither
as an insulator nor a conductor, hence the conception of matter was
incompatible with experimental phenomena. He then continues his exposition:
\begin{quote}
If we must assume at all, as indeed in a branch of knowledge like
the present we can hardly help it, then the safest course appears
to be to assume as little as possible, and in that respect the atoms
of Boscovich appear to me to have a great advantage over the more
usual notion. His atoms, if I understand aright, are mere centres
of forces or powers, not particles of matter, in which the powers
themselves reside. If, in the ordinary view of atoms, we call the
particle of matter away from the powers \emph{a}, and the system of
powers or forces in and around it \emph{m}, then in Boscovich's theory
\emph{a} disappears, or is a mere mathematical point, whilst in the
usual notion it is a little unchangeable, impenetrable piece of matter,
and \emph{m} is an atmosphere of force grouped around it. \citep[p. 289--290]{fara44}
\end{quote}
Notice that the already quoted article by Maxwell \citep[{[529]}][]{maxw73}
precisely tries to amend Faraday adhering to the criticised theory.
Faraday proceeds further with this idea:
\begin{quote}
To my mind, therefore, the \emph{a} or nucleus vanishes, and the substance
consists of the powers or \emph{m}; and indeed what notion can we
form of the nucleus independent of its powers? all our perception
and knowledge of the atom, and even our fancy, is limited to ideas
of its powers: what thought remains on which to hang the imagination
of an \emph{a} independent of the acknowledged forces? A mind just
entering on the subject may consider it difficult to think of the
powers of matter independent of a separate something to be called
\emph{the matter}, but it is certainly far more difficult, and indeed
impossible, to think of or imagine that \emph{matter} independent
of the powers. \textbf{Now the powers we know and recognize in every
phenomenon of the creation, the abstract matter in none; why then
assume the existence of that of which we are ignorant, which we cannot
conceive, and for which there is no philosophical necessity? }{[}...{]}
(emphasis added)

In the view of matter now sustained as the lesser assumption, matter
and the atoms of matter would be mutually penetrable. As regards the
mutual penetrability of matter, one would think that the facts respecting
potassium and its compounds, already described, would be enough to
prove that point to a mind which accepts a fact for a fact, and is
not obstructed in its judgement by preconceived notions. With respect
to the mutual penetrability of the atoms, it seems to me to present
in many points of view a more beautiful, yet equally probable and
philosophic idea of the constitution of bodies than the other hypotheses,
especially in the case of chemical combination. If we suppose an atom
of oxygen and an atom of potassium about to combine and produce potash,
the hypothesis of solid unchangeable impenetrable atoms places these
two particles side by side in a position easily, because mechanically,
imagined, and not infrequently represented; but if these two atoms
be centres of power they will mutually penetrate to the very centres,
thus forming one atom or molecule with powers, either uniformly around
it or arranged as the resultant of the powers of the two constituent
atoms; and the manner in which two or many centres of force may in
this way combine, {[}...{]} \citep[p. 290--293]{fara44}(emphasis
added).
\end{quote}
At a moment in time when one or two fluids and imponderable matter
were being discussed, Faraday's views are marvellously bold. He regards
as \textsf{accidental} such concepts as shape, extension, etc., pertaining
to the traditional view. For him, the identity of matter relates instead
to its “powers“, i.e., the different possibilities of interaction
and their associated fields. As stated before (see \citep[p. 447]{fara55}
above), Faraday soon realised that this view allowed to get rid of
the ether.

Faraday's views developed over a long period of time. By 1821 he had
written:
\begin{quote}
Those who consider electricity as a fluid, or as two fluids, conceive
that a current or currents of electricity are passing through the
wire during the whole time it forms the connection between the poles
of an active {[}voltaic{]} apparatus. There are many arguments in
favour of the materiality of electricity, and but few against it;
but still it is only a supposition; and it will be as well to remember,
while pursuing the subject of electro-magnetism, that we have no proof
of the materiality of electricity, or of the existence of any current
through the wire. \citep[p. 212--213][quote from Faraday]{assi15}
\end{quote}
We cannot avoid to underline the philosophical attitude of Faraday,
who is making all possible efforts to avoid prejudice and to preserve
understanding. Faraday is a philosopher and refers to himself as such.
To our knowledge, he is the last natural scientist to regard himself
as a philosopher.

\subsection{Following Ampère's footprints\label{subsec:Amp}}

While the work by Coulomb considering electrostatic forces did not
rise controversies and was later incorporated in all theories of EM,
the work by Ampère did raise controversies, while it also was fundamental
in the progression of EM theories. A recent work \citep{assi15} informs
us how research and controversies about currents and their forces
developed around 1820-1830, involving colleagues/adversaries (producers
of alternative views) all across Europe. Scientists such as Ørsted,
Faraday, Biot-Savart and Grassmann criticised Ampère's force between
\textsf{current elements}. Most of the criticisms revolve around different
analogies with material entities (fluids or particles) that were favoured
by one or the other scientist. Grassmann pointed out some degree of
arbitrariness in Ampère's assumption that forces between current elements
were \textsf{central} (along the line joining the elements). Most
noticeably, Faraday only presented words of caution, since he wisely
entertained doubts.

Finally, Maxwell \citep[{[518-527]}][]{maxw73} showed that a complete
experimental deduction of the force between current elements starting
from the forces between closed circuits was not possible. We add that
the current element is a mental segmentation of real current-carrying
wires, and it is closer to metaphysics than to physics since it cannot
be physically realised. According to Assis' research, Ampère was criticised
both for \textsf{not} considering electric particles moving inside
the conductor (Ørsted) and for considering them (Biot-Savart). Actually,
the core of the discussion was about which is the “correct“ analogy
with matter, be it moving charges, one or two magnetic fluids, atomic
magnets and so on. It is remarkable though, that the subtitle of Ampère's
contribution was “Theory of Electrodynamic Phenomena, Uniquely Deduced
from Experience“ an expression that illustrates how our hypotheses
are hidden for our own observation.

Despite the controversy with respect to the force between current
elements, the integrated force between closed circuits that carry
currents presented by Ampère still stands without objections. Indeed,
the other alternatives to this force (Grassman, Biot-Savart) yield
the same result as Ampère's \citep[{[518-527]}][]{maxw73}.

\subsection{The relational point of view\label{relational}}

Work on the relational view was centred in Göttingen and followed
the lead of Carl F. Gauss who left his guidance in two short communications
\citep[bd.5 p. 602-626 and 627-629,][]{gaus70}. This view rests on
the assumption that electromagnetism is an interaction analogous to
e.g., gravity, in that it can be fully described as the mutual influence
between charged particles, depending on intrinsic and relative properties.
This description is expressed by a ``force'' (although of a different
nature than e.g., the gravitational force) between pairs of particles,
depending on the individual charges, relative distance, relative velocity
and relative acceleration. In particular, the dependence on velocity
is discussed in Maxwell's Treatise, where it says in {[}851{]} that
Gauss in July 1835 interpreted as a fundamental law of electrical
action, that: ``Two elements of electricity in a state of relative
motion attract or repel one another, but not in the same way as if
they are in a state of relative rest''. Gauss suspended his work
on EM by 1836 (his own account in \citep[bd.5 p. 627-629,][]{gaus70})\footnote{Soon later, Wilhelm Weber, his friend and main experimental mind in
the research, was expelled out of Göttingen following the protests
in favour of the (old) liberal constitution. Weber was one of the
“Seven of Göttingen“. He lost then his laboratory and the daily contact
with Gauss.}. Ten years later Gauss will express the suspicion that the interaction
does not reflect instantaneous action at a distance but rather a delayed
action at a distance \citep[bd.5 p. 627-629,][]{gaus70}.

Some of these relational efforts are reviewed by Maxwell in \citep[v.2 Ch.XXIII][]{maxw73}
where forces advanced by Weber, Gauss, Riemann, Clausius and Betti
are discussed).

\subsubsection{Weber's force}

Among the attempts just mentioned, Weber's force $F_{21}$ between
charged particles \citet{webe46} is the one that has deserved most
attention, being still discussed in our days \citet{assi94}:
\[
F_{21}=\frac{q_{1}q_{2}}{4\pi\epsilon_{0}}\frac{\overrightarrow{r}_{\!\!12}}{r_{12}^{3}}\left(1-\frac{\dot{r}_{12}^{2}}{2C^{2}}+\frac{r_{12}\,\ddot{r}_{12}}{C^{2}}\right)
\]
Here, $F_{21}$ is the force on particle $2$ due to particle $1$,
$q_{1},$$q_{2}$ are the electric charges,$\overrightarrow{r}_{\!\!12}=\overrightarrow{r}_{\!\!2}-\overrightarrow{r}_{\!\!1}$
is the relative position vector between the particles, with length
$r_{12}$. The force involves the derivatives of this length. Weber's
work settled the choice of units by relating this force to other known
forces also measurable with a dynamometer. The constant $\epsilon_{0}$
relates to static electricity (where the derivatives of the distance
are zero) and the constant $\mu_{0}$ relates to magnetic forces.
The ratio of these two forces relates to the quantity $C^{2}=\left(\mu_{0}\epsilon_{0}\right)^{-1}$,
later known as the speed of light and omnipresent in electromagnetism.

Maxwell has mixed opinions about Weber's approach. On one hand, we
will see in Section \ref{subsec:Maxwell} that he criticises instantaneous
action at a distance in favour of the ``propagating medium'' theory
(we defer the discussion of this to Section \ref{sec:The-ether},
advancing only that the criticism is not conclusive but expresses
just that two possible approaches could be pursued, and Maxwell pursues
one of them). On the other hand, Maxwell recognises in {[}552{]} that
this approach leads to conceptions that are ``as beautiful as they
are bold``, while in {[}856{]} he notes that ``Weber's law, with
the various assumptions about the nature of electric currents which
it involves, leads by mathematical transformations to the formula
of Ampère. {[}...{]} Weber's law is also consistent with the principle
of the conservation of energy'' and ``Weber's law will explain the
induction of electric currents.``

Let us consider in detail the insights behind Weber's force. The force
expresses instantaneous action at a distance. Further, it is a central
force obeying Newton's principle of action and reaction. These two
assumptions were current in those times (although Gauss, Riemann,
Betti and Lorenz suspected otherwise) and were not put to test. Also,
for zero relative velocity Weber force reduces to Coulomb's force
between charged particles.

One of the goals of Weber was to incorporate the forces between currents
as studied by Ampère\footnote{Actually, action and reaction could not be put to test, since it cannot
be deduced from Ampère's force, as noted by Maxwell in the Treatise,
{[}527{]}.} and Faraday. Weber assumed that the current within a conductor responds
to the fact that positively and negatively charged (point-like) particles
move within the conductor with certain relative velocity. Originally,
positive and negative charges were supposed to have opposite velocities
of equal size --what is called the Fechner hypothesis \citep[p. 52,][]{assi94}--,
but Weber's framework does not depend on this hypothesis \citep[Ch. 4.2,][]{assi94}
to describe Ampère's force or Faraday's induction law. In short, Weber's
force is the simplest relational central force of instantaneous action
that satisfies Coulomb's, Ampère's and Faraday's laws under the assumption
that current consists of electrically charged (point like) particles
in relative motion\footnote{The idea of conduction current as the movement of charged particles
will return in Section \ref{subsec:Lorentz}, where we discuss that
it is more restricted than the view of Maxwell and it enters in conflict
with experimental data.}. Weber's electrical particles were supposed to exist within matter
and to be imponderable (weightless), as opposed to usual (ponderable)
matter.

Weber's mechanicist view of electricity was the current theory around
1850. It came to coexist with Maxwell's theory some 15 years later,
while after 1885 Maxwell's theory (as understood by his followers)
was dominating. A serious difficulty with Weber's approach was the
impossibility to unify light and electricity, something that Maxwell's
theory was able to achieve \citep[v.2 Ch.XX,][]{maxw73}. However,
electrical waves were first advanced within Weber's framework.

\subsubsection{Franz Neumann's contribution\label{subsec:FNeumann}}

Ampère's force was a fundamental expression for further developments
by the EM community. In its most symmetric form it reads
\begin{eqnarray*}
F & = & -\frac{\mu_{0}}{4\pi}I_{1}I_{2}\oint_{1}\oint_{2}\frac{x_{1}(l_{1})-x_{2}(l_{2})}{|x_{1}(l_{1})-x_{2}(l_{2})|^{3}}\left(dl_{1}\cdot dl_{2}\right)\\
 & = & \frac{\mu_{0}}{4\pi}I_{1}I_{2}\oint_{1}\oint_{2}\left(\nabla_{x}\frac{1}{|x|}\right)_{x=x_{1}(l_{1})-x_{2}(l_{2})}\left(dl_{1}\cdot dl_{2}\right)
\end{eqnarray*}
where $I_{1},$$I_{2}$ are the currents through conductors $1$ and
$2$. The second line suggests that there is an energy involved in
producing the configuration of
\begin{equation}
P=\frac{\mu_{0}}{4\pi}I_{1}I_{2}\oint_{1}\oint_{2}\frac{1}{|x_{1}(l_{1})-x_{2}(l_{2})|}\left(dl_{1}\cdot dl_{2}\right)\label{eq:potencial}
\end{equation}
This form, introduced by F. Neumann \citep{neum1846-induction}, was
one of the starting points in Maxwell's theory. In contrast to Weber,
Neumann did not rest on a material image, but rather abducted the
expression because it served to organise and integrate Faraday's induction
with Ampère force. The integrated expression, eq.(\ref{eq:potencial}),
is a safer starting point than Weber's force since it requires a lesser
number of hypothesis (usually analogies) about the unknown.

Eq.(\ref{eq:potencial}) admits at least two readings. The energy
can be regarded as the potential energy that results from starting
in a situation where the circuits do not interact and subsequently
bringing one circuit to the proximity of the other, preserving the
currents in each of them (in other words it is, a mechanical energy).
Also, the expression can be considered to be the kinetic energy associated
to the electrical currents. In formulas,
\begin{eqnarray}
P & = & \oint_{2}I_{2}A(x_{2}(l_{2}))\cdot dl_{2}\nonumber \\
A(x_{2}) & = & \frac{\mu_{0}}{4\pi}I_{1}\oint_{1}\frac{1}{|x_{1}(l_{1})-x_{2}|}dl_{1}\label{eq:P,A}
\end{eqnarray}
F. Neumann is credited with the first introduction of what today is
called the \textsf{vector potential}, $A$.

It is worth to keep in mind that these developments use the space
as an auxiliary element since all expressions are relational. The
same can be said of Weber's force.

\subsubsection{Kirchhoff and electrical waves}

In 1857 Kirchhoff\citep{kirchhoff1857liv} advanced a model for the
propagation of electricity in wires, combining existing and new elements.
He started by computing the electromotive force associated to charges
and currents within a wire inspired in Weber's theories. Further,
he combined these results with \textsf{Ohm's law} (namely that current
in a wire is proportional to electromotive force) and the \textsf{continuity
equation} (i.e., that spatial variation in current corresponds to
time-variation in charge, also called the conservation of charge),
thus obtaining a wave-like equation for the current (again, current
here is identified with charges in motion). Apparently, Weber was
simultaneously working on the same lines at the time \citep{assi00,assi03},
although his results were published some years later. The resulting
wave-like equation came to be called the ``Telegraph equation''.

As much as this wave behaviour was encouraging, its differences with
light were only too large. A wire is filled with electrically active
particles (this is the current view about wires still today), both
positive and negative and in roughly the same amount if the wire is
to be electrically neutral as a whole, thus offering a tangible electrical
medium for the waves to propagate (more or less like the pressure
waves originated in a string musical instrument). Light, on the other
hand, was expected to exist in vacuum where there is no tangible medium
at all. A possible escape way at the time was to endow the vacuum
with electrical properties, by resorting to the ether (see Section
\ref{sec:The-ether}), an idea that was already circulating in different
forms.

\subsubsection{Lorenz and Delayed Action at a Distance (DAD)}

The danish physicist Ludvig Lorenz presented a series of works on
light propagation \citep{lore61,lore1863,lore67} in the 1860's. Especially
in his memory of 1867 he expresses discomfort with the ether hypothesis,
which had only been useful to ``furnish a basis for our imagination''
(p. 287). He sets up to develop an ether-free theory of light, something
that he achieves by introducing delays. Electrodynamics could be an
action at a distance theory, but not an instantaneous one. This was
a dramatic novelty, that is still not completely grasped. Retarded
action was being discussed at the time. Indeed, in Maxwell's Treatise
(Ch. XXIII) the theories of C. Neumann, Betti and Riemann (all published
in 1867-68) are briefly discussed, along with a criticism by Clausius.
Curiously, Lorenz far-reaching contribution is not mentioned in that
part of the Treatise. It appears in Art. {[}805{]} as a novel theory
of light propagation, without mention of the delayed action at a distance.
Maxwell ends his recollection by noting that Lorenz' conclusions are
``similar to those of this chapter, though obtained by an entirely
different method'' (Maxwell also points out that his own theory was
published earlier, in 1865).

The proposal of delayed action opens up for different possibilities
regarding how this delay could take place. Within an emission theory,
light generates at the source, it travels through space and reaches
the detector, somewhat like sound waves, or even a projectile. This
idea goes back at least to Huygens and Newton and it was supported
by Maxwell, Clausius and almost all of Maxwell's followers. It rests
strongly in the assumption of a propagating medium. Consequently,
this view attempts to describe delays with information about the location
of the source at the departing time and that of the detector at the
(later) detection time. However, this quantity is not universal, it
takes different values for different observers and it cannot be used
``as is'' to gain understanding about light propagation (see below).
Outside the emission theory, other possibilities open up. The Göttingen
school advocated a different view, better adapted to Faraday's intuitions
\citep[p. 447]{fara55} based on an objective measure of the delay
involved in the interaction, as it will be presently exemplified.
A more detailed discussion about delayed action at a distance is given
in Appendix \ref{sec:Delayed}.

\subsubsection{C. Neumann, Betti, Riemann and Clausius' criticism\label{subsec:Neumann-Betti-Riemann}}

Three more attempts were made to incorporate delayed action at a distance
by Gauss' followers Betti\citep{bett67}, Riemann\citep{riem67} and
Carl Neumann \citep{neum68}. These scientists tried to justify in
different forms the use of an objective form of the delay, aiming
to formulate the propagation law guessed by Gauss (the wave equation)
and mathematically enunciated by Maxwell (1865) and Lorenz (1867).
\footnote{Lorenz' and Maxwell's propagation of electromagnetic disturbance arrive
to the same final equation, the difference being in the foundations.
Maxwell uses the displacement current based on the ether to obtain
what is now called Ampère-Maxwell's law, ${\displaystyle \Delta A=-\mu_{0}j+\frac{1}{C^{2}}\frac{\partial^{2}A}{\partial t^{2}}}$
while Lorenz obtains ${\displaystyle \Delta A-\frac{1}{C^{2}}\frac{\partial^{2}A}{\partial t^{2}}=-\mu_{0}j}$
using the delays, explicitly avoiding to invoke any ether ($j$ is
here the current density, i.e., the current per unit volume).}

The fate of Neumann's theory is better known as a consequence of the
polemic with Clausius discussed by Archibald and Assis \citep{arch86,assi94}.
Archibald observes that ``These exchanges illuminate some of the
points that are at issue when an instantaneous action at a distance
model is replaced by one where the action is propagated with finite
velocity. In so doing, they illustrate two views of the role of mathematics
in physical science, showing that the twentieth century is not the
first time that mathematical model offering good results have posed
problems for those who seek a more intuitive picture''.

Neumann explicitly proposes a potential ``travelling'' from source
to detector. The locations and times of the interaction must satisfy
the relation
\begin{equation}
|x_{d}(t)-x_{s}(t)|=C(t-t_{0})\label{eq:univ}
\end{equation}
($d$ for detector and $s$ for source) being $t_{0}$ the time for
the electromagnetic disturbance in the source. This relation, common
to Lorenz and the whole Göttingen school, is universal (independent
of observers or choices of reference frames); hence, all observers
compute the same ``velocity'', ${\displaystyle C=\frac{1}{(\mu_{0}\epsilon_{0})}}$,
which is Weber's measured relation between static and dynamic electricity,
a quantity for which it makes no sense to consider a frame of measurement.
However, Neumann's focus is in the action of a delayed potential.
He insists that the travel is similar but not identical to that of
light \citep{neum69}.

Clausius \citep{clau69} criticism of Neumann (and also of Riemann
in this particular topic) is based on an emission theory. Clausius
claims that the ``distance'' between source and detector must satisfy
\[
|x_{d}(t_{d})-x_{s}(t_{s})|=C(t_{d}-t_{s}).
\]
This expression does not have the universal character of eq.\eqref{eq:univ}.
In the terms of NAP, Clausius proposal equates a subjective quantity
$|x_{d}(t_{d})-x_{s}(t_{s})|$ to an objective one \citep{sola18b}.
The laws of physics are thus made to depend on the arbitrariness of
the selection of a reference frame since for an arbitrary change from
a frame to another moving with relative velocity $v$ we would have
$|x_{d}(t_{d})-x_{s}(t_{s})|\mapsto|x_{d}(t_{d})-x_{s}(t_{s})+v(t_{d}-t_{s})|$.
This view cannot be sustained unless an absolute reference frame is
introduced: the ether or absolute space. The substantialist view\footnote{Substantialism: The doctrine that behind phenomena there are substantial
realities (Oxford dictionary). In Newton the space is the place of
objects. Now it is taken further since it becomes the place of impending
actions as well. See for example https://plato.stanford.edu/entries/spacetime-theories/\#5.2} at the basis of the emission theory of light has never been questioned,
but it persists beyond any doubts about its appropriateness.

Further, it was later realised that this view forces to deform our
conceptions of space and time in order to compensate for the arbitrariness.
Clausius' criticism does not go further than that: Neumann's programme
is regarded as inadequate since it does not conform to Clausius' views.
It suggests without proof that the ether-based emission theory would
be appropriate and final. Rather than a search for the foundations
and an attempt to refine or improve the criticised work, the criticism
dismissed the proposal on improper grounds.

Maxwell presented his views on C. Neumann's theory in articles {[}863{]}
and {[}866{]} of the Treatise:
\begin{quote}
Besides this, the velocity of transmission of the potential is not,
like that of light, constant relative to the aether or to space, but
rather like that of a projectile, constant relative to the velocity
of the emitting particle at the instant of emission.

It appears, therefore, that in order to understand the theory of Neumann,
we must form a very different representation of the process of the
transmission of potential from that to which we have been accustomed
in considering the propagation of light. Whether it can ever be accepted
as the 'construirbar Vorstellung'\footnote{Maxwell quotes from Gauss' letter to Weber \citep[bd.5 p. 627-629,][]{gaus70}
an expression of difficult translation: ``construirbare Vorstellung''
(possibly meaning ``constructible representation'', or simply ``a
representation that can be carried out'').} of the process of transmission, which appeared necessary to Gauss,
I cannot say, but I have not myself been able to construct a consistent
mental representation of Neumann's ' theory. (art {[}863{]})

In the theory of Neumann, the mathematical conception called Potential,
which we are unable to conceive as a material substance, is supposed
to be projected from one particle to another, in a manner which is
quite independent of a medium, and which, as Neumann has himself pointed
out, is extremely different from that of the propagation of light.
In the theories of Riemann and Betti it would appear that the action
is supposed to be propagated in a manner somewhat more similar to
that of light. (art {[}866{]})
\end{quote}
Apparently, Maxwell tries to regard C. Neumann's potential as a material
substance, without success. As we have previously seen (Section \ref{subsec:Maxwell's-epistemic}),
Maxwell's epistemic view blinds him when considering Faraday's ray
theory which is precisely what Neumann's has built in formulae.

Riemann was closer to Lorenz in that he considered only necessary
to introduce the wave-propagation\footnote{To be fair, we should note that the proposal is earlier than Maxwell's,
although it was published at a later time. In the Treatise, article
{[}862{]}, Maxwell dismissed Riemann's approach referring to Clausius'
criticism.} without resting on analogies or images of matter or the space. Remarkably,
he offered some possibilities in which his hypothesis could be thought
in material terms
\begin{quote}
This supposition can be fulfilled in various ways. Let us assume,
for instance, that the conductors are crystalline in their smallest
particles, so that the same relative distribution of the electricity
is periodically repeated at definite distances which are infinitely
small compared with the dimensions of the conductors
\end{quote}
Unlike Lorenz', Riemann's theory did not incorporate light polarisation.

Clausius criticises Riemann in two points, the first of them being
the same as his critic of Neumann. The other is a procedural issue.
Riemann's calculations hold only if the delay is characterised by
a relation such as eq.(\ref{eq:univ}), but it is unclear what Riemann
meant since he uses various notations to describe distances along
the work. Betti's work attempted to provide a different framework
for Riemann's ideas. Clausius' criticism of Betti's work focuses on
general formal properties of series expansions, but he did not discuss
if such criticism was relevant for the issue in question.

The work of Lorenz mentioned above also considers a delay compatible
with eq.(\ref{eq:univ}). Remarkably, his work appeared contiguously
after Riemann's work in the same issue of the \emph{Poggendorff Annalen}
(1867)\footnote{Riemann's manuscript was presented in 1858 but published in 1867 (after
his death). Betti comments on this as ``\textit{...supponendo che
la corrente consista nel movimento delle due elettricità positiva
e negativa che vanno contemporaneamente nel filo in direzioni opposte,
{[}...{]} Questo concetto della corrente elettrica tutto ideale è
poco in armonia con ciò che si conosce di essa, e pare che Riemann
non ne fosse soddisfatto, avendo ritirato l'articolo dalla Segreteria
dell'Accademia, ed essendosi astenuto dal pubblicarlo}'', namely
that Riemann possibly was unsatisfied with his hypothesis about the
nature of electric current. Clausius (referring to the alleged ``error''
he discussed) conveys his own version as ``I believe that Riemann
subsequently convinced himself of this error, and that this was the
reason he withdrew his paper'', thus ignoring both Lorenz' work (sustaining
basically the same result than the criticised works) and the reasons
offered by Betti (in the same paper that Clausius also criticises),
friend and collaborator of Riemann in this matter.} but it was not criticised at that time, nor afterwards. It is surprising
that neither Lorenz' nor other non canonical view about light propagation
has ever been considered as an alternative.

The odd circumstance of three papers around the same idea and a weak
attempt of refutation coming soon after, move us to consider the historic
environment of these events. Carl Neumann was the son of Franz Neumann
and had the greatest admiration for Riemann \citep{jung17}. In turn,
Riemann became a friend of Betti during a visit of the latter to Göttingen.
Later Riemann visited Betti in Italy. Riemann was an assistant to
Weber during 18 months around 1849 and Weber is one of his recognised
influences \footnote{See https://www-history.mcs.st-andrews.ac.uk/Biographies/Riemann.html}.
By the year 1866, Hannover and Prussia (``Göttingen and Berlin'')
enter in war. Riemann fled Göttingen for Italy because of the war,
and he died there the same year. The war lasted until 1868 when Hanover
was annexed as a province of the new empire. Clausius was born in
Prussia, studied in Berlin, taught physics at the Royal Artillery
and Engineering School in Berlin and was Privatdozent in Berlin University.
In 1870 he organised an ambulance corps during the Franco-Prussian
war and was wounded in battle. It can be said that the battle of ideas
was held under the emotional atmosphere of the battles for the reunification
of Germany. By 1871 the German Empire was proclaimed.

\subsubsection{Revisions}

The activity on a relationist view of electromagnetism ceased by 1868
except for a reprint of C. Neumann's work and the immediate reaction
of Clausius. More than a half century later, Moore and Spencer (and
collaborators) researched the subject in a series or articles \citep{moon54,moon56,moon60,moon1989binary,moon1989universal,moon1991validity,moon1994electrodynamics})
in which they considered the main experiments that support the currently
accepted theory. Their 1960 book \citep{moon60} presents the systematic
revision and comparison. The result of the revision is that, at the
time (1960), there was no experimental evidence to prefer the Maxwell-Lorentz-Einstein
views to the Delayed Action at Distance theory.In the same terms of
Dingle \citep{ding60} they find that the constancy of the speed of
light remains a conjecture since the moment it was proposed by Einstein
\citep[p. 257, ][]{moon60}.

In front of possible criticism from an intuitive point of view they
state: ``But we can hardly expect intuitive ideas to hold for light.
Light is not a wave in a medium and is not a particle: it is a unique
phenomena unlike anything else in nature. To expect to visualise this
unique phenomenon in terms of mechanistic pictures of water waves
and bullets is indeed naive'' (p.256).

In page 251 Moon and Spencer write an example of synchronisation of
clocks using light assuming the Galilean formulation. They write:
``Thus we have an operational method of establishing an universal
time, subject of course to the assumptions that the space is euclidean
and that 'velocity of light' is a meaningful expression. In the advent
of space ships and the establishment of colonies in other planets,
such a synchronisation of clocks would be highly desirable'' they
proceed then to consider the differences between reflected an re-emitted
light. They also easily show that, within their ``postulational basis''
for the discussion it is not possible to synchronise clocks in terms
of special relativity.

It is then relevant to notice the so called ``flyby anomaly'' (of
the Pioneer and other spacecrafts) \citep{bilb14,bilbao2016}, where
two tracking systems were used: one with a re-emitting device and
the other with a reflector. The systems disagree in values associated
to the relativistic correction of time. These observations are anomalies
for the standard electromagnetism but regularities (actually, predictions)
for the relational electromagnetism. Certainly, there are other \emph{a
posteriori} explanations for the anomalies, but the \emph{ad hoc}
explanations cannot account for the fine detail as the relational
view does. Needless to say, these articles have received little or
no attention from the physics community.

\subsection{Maxwell\label{subsec:Maxwell}}

In Maxwell, as well as in all other developers of EM, a basic assumption
about the behaviour of electrodynamics systems is that the electrodynamic
forces exerted on matter behave in the same way as the previously
known mechanical forces. In particular, whatever external force that
is required to keep a system at rest is taken to be equal and of opposite
sign to the internal force generated by electrodynamic interactions
(and that, if no compensated, would alter the static equilibrium).
In this way, by looking at external forces only, it is possible to
learn something about the electrodynamic system.

Today it is believed that “external“ forces such as mechanical resistance
to deformations or frictional forces in the “contact“\footnote{These quoted words from the usual language hide assumptions that we
are barely aware of. We speak freely of contact between bodies, we
may detect a difference between contact and no contact, but still
today it is not a trivial question to relate the detected difference
to the atomic constituents of each body and their interactions.} between two bodies also have an electromagnetic origin, but this
topic was less clear two centuries ago. In any case, it was expected
that isolated electrodynamic systems should have some sort of conserved
electrodynamic energy and electrodynamic forces should be associated
to the time-variation of some “momentum“.

We may identify a series of abductions that increasingly organised
the existent knowledge, starting from F. Neumann who realised that
the force between closed circuits had an associated energy. Helmholtz
and Thomson \citep{thom51} associated this energy to the mechanical
energy (the ability of generating mechanical work). Indeed, Helmholtz
states,
\begin{quotation}
The entire electromotive force of the induced current, generated by
a change of \textbf{position of a magnet relative to a closed conductor},
is equal to the change which thereby takes place in the potential
of the magnet towards the conductor, when the latter is traversed
by the current \citep[p. 157,][]{hemholtz1853}(emphasis added)
\end{quotation}
In formulae, the electromotive force, ${\cal E},$is
\begin{eqnarray}
{\cal E} & =-\frac{d}{dt} & \left(\frac{\mu_{0}}{4\pi}I_{1}\oint_{1}\frac{1}{|x_{1}(l_{1},t)-x_{2}(l_{2},t)|}dl_{1}\right)\label{eq:fem}
\end{eqnarray}

This was the starting point that Maxwell adopted to develop his contributions
to EM.

Despite the stories usually told in physics textbooks, Maxwell did
not considered electricity as a fluid, except for illustration purposes.
Following Faraday, Maxwell was certain that electricity (whatever
it is) had the ability of moving when associated to conductors, but
he restrained from further hypothesis \citep[{[552],}][]{maxw73}.
Then he proceeded to conceptualise $P$ (\ref{eq:P,A}) no longer
as a potential energy but rather as the \textsf{electrokinetic energy},
assimilating the currents $I_{i}$ with velocities. Then, by analogy
with Lagrange's formulation of mechanics, he would write
\[
A_{i}=\frac{\delta P}{\delta I_{i}}
\]
The variation $\delta I_{i}$ included the possibility of changing
the locus of the cable (additionally, he formally extended the expressions
to conductors not necessarily produced as cables). The quantity $A_{i}$
is then the electrical momentum following the standard wording in
Lagrange's mechanics. He then tried to determine experimentally whether
the electrical momentum had an associated mechanical momentum as well,
concluding that they were completely decoupled \citep[{[574-575],}][]{maxw73}.
Thus, the energy (and then the Hamiltonian), corresponding to two
pieces of matter is the addition of the energy of the ponderable matter
and the electrokinetic energy \citep[{[571],}][]{maxw73}, being the
electrokinetic energy independent of the motion of the ponderable
matter. The electromotive force (\ref{eq:fem}) is then the external
force required to balance the internal force that is necessary to
maintain the current \citep[{[576],}][]{maxw73}.

It is important to realise at this point that physical hypotheses
play two different roles. Some of them, like the fluid hypothesis
in Maxwell, are pedagogical devices directed to facilitate the perception
and explanation of abducted relations. These kind of hypothesis can
be changed without changing the abduction. For example, we can think
of massless electrical particles instead of fluids, as Weber did,
and nothing is changed in our expressions. Other kind of hypotheses
force us to modify the equations involved in the theory. In the present
situation, the decoupling of the kinetic energies (the electrokinetic
one and the one arising from the movement of matter) results in that
some energy terms have been considered to be identically zero\footnote{When current is involved, Maxwell's belief is that “electricity“ is
moving, some way or the other. Following the Lagrangian formulation
of mechanics, in {[}578{]} he calls $\dot{y}$ the variables associated
to currents, while $y$ is the conjugated variable (other authors
will relate $y$ to the transport of net charges). Further, he realises
that the mechanical variables $x,\dot{x}$ corresponding to ordinary
matter are not coupled to $\dot{y}$ and that the energy associated
to currents (the electrokinetic energy) is an homogeneous quadratic
function of the currents. Further, in {[}632-8{]} the relationship
between magnetic and electrokinetic energy is presented, thus closing
the circle with C. Neumann's formulation, since electrokinetic energy
relates to the vector potential $A$ and the magnetic field. Hence,
the whole electrodynamic energy of a material system is obtained as
the sum of electrostatic and electrokinetic energy.}. Such a hypothesis cannot be dropped or modified without a complete
reworking of the results. In particular, currents cannot be read as
$qv$, being $v$ a subjective velocity (a velocity that changes with
our choice of reference frame), maintaining Maxwell's ideation.

The second, and best known, contribution of Maxwell derives from his
conviction that forces of action at distance were not really needed
in order to describe EM phenomena \citep[{[59-60] [552],}][]{maxw73}.
In practice, Maxwell never went that far, large portions of his work
assume instantaneous action at a distance. Early in his treatise he
indicates that one of his goals is to pursue the hypothesis “... that
electric action is not a direct action between bodies at a distance,
but is exerted by means of the medium between the bodies...“ \citep[{[60],}][]{maxw73}.

Using Gauss formula, and the generalised expression corresponding
to $A$ (\ref{eq:P,A}) Maxwell produces 
\begin{eqnarray}
A(x,t) & = & \frac{\mu_{0}}{4\pi}\int_{V}\left(\frac{j_{1}(x_{1}(y,t),t)}{|x_{1}(y,t))-x|}\right)\left[\frac{\partial x_{1}}{\partial y}\right]d^{3}y\label{eq:Afinal}\\
\Delta A(x,t) & = & -\mu_{0}j_{1}(x,t)\nonumber 
\end{eqnarray}
here, $x_{1}(y,t)$ stands for the location of a small volume, $d^{3}y$,
of ponderable matter at time $t$ with initial condition $x_{1}(y,0)=y$,
and $\Delta$ is Laplace's operator\footnote{Actually, Maxwell did not write the integral form. He moved intuitively
from the expression in cables (wires) to the expression in three-dimensional
matter arrangements. The expression of the electromagnetic momentum
can be transformed by a coordinate change to
\begin{eqnarray*}
A(x,t) & = & \frac{\mu_{0}}{4\pi}\int_{V(t)}\frac{j_{1}(y,t)}{|y-x|}d^{3}y
\end{eqnarray*}
we have preferred the formulae (\ref{eq:Afinal}) because it displays
more readily the transformation properties of $A$.}. Maxwell proposed that a current corresponding to the variation of
the polarisation of the dielectric should be added to the equation,
so that $j_{1}=j_{1}^{g}+j_{1}^{d}$ would be \footnote{In standard textbooks this is called the \textsf{Ampère-Maxwell law.
}However, the contribution from $j_{1}^{g}$ is not from Ampère, but
rather from F. Neumann who inspired his developments in Ampère's force.} the sum of the galvanic current, $j_{1}^{g}$, and a current arising
from the variation of the induced polarisation. In the case of observer
(detector) and source at relative rest, $j_{1}^{d}=K\epsilon_{0}\frac{\partial E}{\partial t}$.
Here $K$ is a dimensionless constant characteristic of the dielectric,
the \textsf{specific inductive capacity}, that takes the minimum value
of $K=1$ in vacuum and it is also $1$ (or slightly above) in air
\citep[{[52], [60],[75]}][]{maxw73}. Further $E=-\frac{\partial A}{\partial t}-\nabla V$
(also for the case of relative rest) is the the electromotive force
($V$ stands for an undetermined potential function in \citep[{eq. B, [598],}][]{maxw73},
with $v=0$).  It is interesting to notice that Maxwell focused his
attention on the differential (derived) form $\Delta A(x,t)=-\mu_{0}\left(j_{1}(x,t)-\epsilon_{0}K\frac{\partial}{\partial t}\left(\frac{\partial A}{\partial t}+\nabla V\right)\right)$
which is not entirely equivalent to the integral form. The equivalence
holds only if the (unstated) boundary conditions are properly satisfied.
If we, alternatively, fix our attention on the integral form, we realise
that for the convergence of the integral expression it is required
that the current decays at infinity faster than $\frac{1}{|x|^{2}}$,
a condition that $\frac{\partial A}{\partial t}$ does not satisfy.
This means that the ideation is slightly inconsistent, it is actually
an improper form of proposing that the equation governing the vector
potential is
\begin{equation}
\Box A=-\mu_{0}(j_{1}+\epsilon_{0}K\nabla V)\label{eq:BoxDef}
\end{equation}
where $\Box=\Delta-(K\mu_{0}\epsilon_{0})\frac{\partial^{2}}{\partial t^{2}}$
is the D'Alembert operator. Later, Maxwell would seek solutions to
this equation.

\subsubsection{Maxwell on reference systems}

Maxwell's Treatise on Electromagnetism \citep{maxw73} is a complex
work. In particular, the equations are not self-contained. The reader
is expected to follow and understand the derivations. Quite often
the notation is simplified in excess, resulting in expressions that
do not apply to the general case. For example, dropping a gradient
contribution in the electromotive force renders several final equations
only apt for the use in the case of closed circuits. The handling
of the space is not transparent either. Partial derivatives with respect
to time occur and, unfortunately, they are indicated with the same
symbol ${\displaystyle \frac{d}{dt}}$ as the “total“ (convective,
following the body) derivative.

In Art. {[}600{]} Maxwell discusses the invariance in form of the
electromotive force with respect to the choice of reference system.
The problems of notation are particularly acute in {[}598-600{]}.
The general setup is that $A(x,t)$ is allowed to have a more general
(and undefined) form than in eq.(\ref{eq:P,A}). The location of the
secondary cable $x_{2}(l,t)$ is allowed to change in time and so
is the location and current of the primary, but since there is no
explicit expression for $A$ this will be reflected as the partial
derivative in time ${\displaystyle \frac{\partial A(x,t)}{\partial t}}$
that collects all the changes in time due to changes in the primary
circuit. Maxwell centred his attention in how ${\displaystyle \frac{\partial A(x,t)}{\partial t}}$
must change under a change of reference system (that could comprise
rotations and translations with respect to the original system). To
avoid complexities, and since this is sufficient for our discussion,
we will address only the translational case (i.e., where the coordinate
axes in a given inertial system are parallel to the corresponding
axes in another system). Instead of stating Maxwell's result for ${\displaystyle \frac{\partial A(x,t)}{\partial t}}$,
we collect all the hypotheses introduced in {[}600{]} and rewrite
his result in terms of invariance of form.

\paragraph{Maxwell's invariance theorem}

Let $A(x,t)$ be the electromagnetic momentum, in terms of the position
$x$ with respect to a coordinate system $L$, and $\mathcal{E}$
the electromotive force (introduced in {[}597{]}). Maxwell computes
in \citep[{eq.B, [598],}][]{maxw73}
\begin{eqnarray}
\mathcal{E} & = & v\times(\nabla\times A)-\frac{\partial A}{\partial t}-\nabla\psi\label{eq:E598}
\end{eqnarray}
where ${\displaystyle v=\frac{dx}{dt}}$ is the velocity of a point
on the (moving) secondary circuit with respect to $L$ and $\psi(x,t)$
is an undetermined scalar function. Maxwell explains that “We shall
find, however, that when we know all the circumstances of the problem,
we can assign a definite value to $\psi$, and that it represents,
according to a certain definition, the electric potential...“\footnote{In \citep[{[70],}][]{maxw73}Maxwell defines: ``The Potential at
a Point is the work which would be done on a unit of positive electricity
by the electric forces if it were placed at that point without disturbing
the electric distribution, and carried from that point to an infinite
distance''}

Consider coordinates $x^{\prime}$ on a system with motion given by
$\bar{x}(t)$ relative to a system with coordinates $x$, so that
$x^{\prime}=x-\bar{x}$ \footnote{Maxwell writes ``Let $x^{\prime}$, $y^{\prime}$, $z^{\prime}$
be the coordinates of a point referred to a system of rectangular
axes moving in space, and let $x$, $y$, $z$ be the coordinates
of the same point referred to fixed axes'', what makes evident he
is considering absolute space, since ``at rest'' does not refer
to any particular reference. However, the result is valid for any
pair of systems where the hypotheses are fulfilled.}. Then, the following result holds:

\textbf{Theorem} (\textbf{Maxwell's invariance theorem})\label{Theorem-Maxwell's-invariance}:

Let $x=x^{\prime}+\bar{x}(t)$, and correspondingly $v=v^{\prime}+\dot{\bar{x}}$.
Define ${\displaystyle A^{\prime}(x^{\prime},t)\equiv A(x,t)}$ \footnote{Maxwell refers to this expression as: “the theory of the motion of
a body of invariable form“. For any property of matter, this relation
is immediate.}, then the value of the electromotive force at a point $x$ does not
depend on the choice of reference system if and only if $\psi(x,t)$
transforms as $\psi^{\prime}(x^{\prime},t)\equiv\psi(x^{\prime}-\bar{x},t)-\dot{\bar{x}}\cdot A^{\prime}(x^{\prime},t)$.
In formulae, $\mathcal{E}^{\prime}(x^{\prime},t)=\mathcal{E}(x,t)$,
where

\[
\mathcal{E}^{\prime}(x^{\prime},t)=v^{\prime}\times(\nabla\times A^{\prime}(x^{\prime},t))-\frac{\partial A^{\prime}(x^{\prime},t)}{\partial t}-\nabla\psi^{\prime}(x^{\prime},t).
\]

\textbf{Proof}: First, according to the definition, we have ${\displaystyle A^{\prime}(x^{\prime},t)=A(x^{\prime}+\bar{x},t)}$.

Next, we note that by straightforward vector calculus identities,
Maxwell's electromotive force (eq. \ref{eq:E598}) can be restated
as 
\[
\mathcal{E}(x,t)=-\frac{\partial A(x,t)}{\partial t}-\left(v\cdot\nabla\right)A(x,t)-\nabla\left(\psi(x,t)-v\cdot A(x,t)\right)
\]
In the new coordinate system we compute: 
\begin{eqnarray*}
\mathcal{E^{\prime}}(x^{\prime},t) & = & v^{\prime}\times(\nabla\times A^{\prime}(x^{\prime},t))-\frac{\partial A^{\prime}(x^{\prime},t)}{\partial t}-\nabla\psi^{\prime}(x^{\prime},t)\\
 & = & -\frac{\partial A^{\prime}}{\partial t}(x^{\prime},t)-\left(v^{\prime}\cdot\nabla\right)A^{\prime}-\nabla\left(\psi^{\prime}(x^{\prime},t)-v^{\prime}\cdot A^{\prime}(x^{\prime},t)\right).
\end{eqnarray*}
Subsequently, under the present assumption ${\displaystyle A^{\prime}(x^{\prime},t)\equiv A(x,t)}$
we may rewrite 
\[
\frac{\partial A^{\prime}(x^{\prime},t)}{\partial t}=\left.\frac{\partial A(x,t)}{\partial t}\right|_{x^{\prime}+\bar{x}}+\left(\dot{\bar{x}}\cdot\nabla\right)A
\]
leading to 
\begin{eqnarray*}
\mathcal{E^{\prime}}(x^{\prime},t) & = & -\frac{\partial A(x,t)}{\partial t}-\left(\dot{\bar{x}}\cdot\nabla\right)A-\left(v^{\prime}\cdot\nabla\right)A^{\prime}-\nabla\left(\psi^{\prime}(x^{\prime},t)-v^{\prime}\cdot A^{\prime}(x^{\prime},t)\right)\\
 & = & -\frac{\partial A(x,t)}{\partial t}-\left(v\cdot\nabla\right)A-\nabla\left(\psi^{\prime}(x^{\prime},t)-v^{\prime}\cdot A^{\prime}(x^{\prime},t)\right)\\
 & = & -\frac{\partial A(x,t)}{\partial t}-\left(v\cdot\nabla\right)A-\nabla\left(\psi(x,t)-\dot{\bar{x}}\cdot A(x,t)-v^{\prime}\cdot A^{\prime}(x^{\prime},t)\right)\\
 &  & -\frac{\partial A(x,t)}{\partial t}-\left(v\cdot\nabla\right)A-\nabla\left(\psi(x,t)-v\cdot A(x,t)\right)=\mathcal{E}(x,t)
\end{eqnarray*}

where we used the condition $\psi^{\prime}(x^{\prime},t)\equiv\psi(x,t)-\dot{\bar{x}}\cdot A(x,t)$,
to proceed from the second line to the third.$\boxempty$

Maxwell's Art. {[}601{]} states “It appears from this that the electromotive
intensity is expressed by a formula of the same type, whether the
motions of the conductors be referred to fixed axes or to axes moving
in space, the only difference between the formulae being that in the
case of moving axes the electric potential $\psi$ must be changed
into $\psi+\psi^{\prime}$.“ (here being $\psi^{\prime}=-A^{\prime}\cdot\dot{\bar{x}}$).
$\psi$'s numeric value depends on the choice of reference system.
In the appendix to Chapter IX, J. J. Thomson --the curator of the
third edition-- indicates: “...It does not appear legitimate to assume
that $\psi$ in equations (B) represents the electrostatic potential
when the conductors are moving, for in deducing those equations Maxwell
leaves out a term $-\frac{d}{ds}(A\cdot v)$ since it vanishes when
integrated round a closed circuit...“.

While Thomson's concern is reasonable and follows from the calculations
on the Treatise in {[}598{]}, Maxwell did not pursue the issue in
depth, only suggesting (in {[}598{]} and {[}630{]}) that $\psi$ is
to be identified with $V$. We will see in the next Section that Lorentz
adopts the same identification, thus leading to the standard formulae
for the electromagnetic force found in textbooks. The exact nature
of the potential $\psi$ is an issue to be decided through experiments.

\subsubsection{Electromagnetic Energy}

The Treatise considers in Art. {[}85{]} the potential energy associated
to building up a given arrangement of charges, relating it to the
electrostatic potential. It returns to this topic in {[}630-31{]},
restating it as properties belonging to the electric field. An important
difference is that the computation in {[}85{]} was intended to be
performed only in the regions of space occupied by charged matter.
It can be formally extended to all space, assuming that the density
of charge outside the (bounded) region of charged matter is zero.
However, the computation in {[}630-31{]} \textsf{must} extend to all
space, otherwise a compensating term on the boundary of the integration
region is required. In formulae and using modern notation 
\begin{eqnarray*}
\frac{1}{2}\int_{U}\rho(x)V(x)d^{3}x & = & \frac{1}{2}\int_{U}\epsilon_{0}(\nabla\cdot E)V\,d^{3}x\\
 & = & \frac{1}{2}\int_{U}\epsilon_{0}\left(\nabla\cdot(EV)+|E|^{2}\right)\,d^{3}x\\
 & = & \frac{1}{2}\int_{U}\epsilon_{0}|E|^{2}d^{3}x+\frac{1}{2}\int_{U}\epsilon_{0}\nabla\cdot(EV)\,d^{3}x\\
 & = & \frac{1}{2}\int_{U}\epsilon_{0}|E|^{2}d^{3}x+\frac{1}{2}\int_{\partial U}\epsilon_{0}(EV)\cdot\hat{n}\,d^{2}s
\end{eqnarray*}
where $U$ is the region of space where the charge density $\rho$
is supported. We have used the relation ${\displaystyle \nabla\cdot E=\frac{\rho}{\epsilon_{0}}}$.
Using Gauss' theorem, the last term in the third equation is restated
as a surface integral over the boundary $\partial U$ of $U$. Only
when this surface integral is zero, we can identify the $\rho V$
integral with the $|E|^{2}$integral. For bounded charge distributions
the potential and field behave in such a way that the surface integral
vanishes at infinity.

Maxwell extends in Art. {[}631{]} the validity of this result to any
electric field. Thomson indicates in a footnote \citep[v.2 p.271,][]{maxw73}
that the deduction holds only in the electrostatic case. He adds that
in the general case the energy should be considered as that contained
in the polarisation of all the dielectric (ether included). In such
a form it extends to all sources of polarisation.

Other forms of energy such as the magnetic energy are put by Maxwell
in the same form, as integrals of the square of the fields over all
space. A step that would have pleased Faraday.

\section{The ether\label{sec:The-ether}}

At the very end of the Treatise, in the last Article, {[}866{]}, and
after discussing action at a distance very much following Clausius,
Maxwell states:
\begin{quotation}
But in all of these theories the question naturally occurs : --If
something is transmitted from one particle to another at a distance,
what is its condition after it has left the one particle and before
it has reached the other? If this something is the potential energy
of the two particles, as in Neumann's theory, how are we to conceive
this energy as existing in a point of space, coinciding neither with
the one particle nor with the other? In fact, whenever energy is transmitted
from one body to another in time, there must be a medium or substance
in which the energy exists after it leaves one body and before it
reaches the other, for energy, as Torricelli remarked, ' is a quintessence
of so subtile a nature that it cannot be contained in any vessel except
the inmost substance of material things.' Hence all these theories
lead to the conception of a medium in which the propagation takes
place, and if we admit this medium as an hypothesis, I think it ought
to occupy a prominent place in our investigations, and that we ought
to endeavour to construct a mental representation of all the details
of its action, and this has been my constant aim in this treatise.
\end{quotation}
Far from proving the existence of the ether, the paragraph establishes
a promising program of study that rests on Maxwell's substantialism,
for energy must leave one body to reach the other as if it were matter,
which indeed, appears to be almost the case if we are to believe that
energy is contained in matter, this is, localised in the space occupied
by matter. This conception of energy and matter is a far cry from
Faraday's leading vision, but it must be conceded to Maxwell that
his intuition deserved to be studied.

The arguments in favour of the ether or against action at distance
never went further that those by Maxwell. These arguments are the
consequence of an axiom of the proposed ideation: we must conceive
the unknown by means of analogies with matter accessible to the intuition.
Cf. with Kelvin's mechanical intuition \citep[p.235]{thompson2011life}
discussed in Subsection \ref{subsec:Real+NAP}.

In all, Maxwell's equations are supported by action at a distance
(present in Faraday's induction law and Coulomb's electrostatic force),
F. Neumann's energy of a system of currents (which also rests in action
at a distance) and in the polarisation current, an ingredient that
operates even in empty space, where there are no charge or current
carriers to support it. It is only here that the ether enters in Maxwell's
theory, quite early in the Treatise and basically because Maxwell
could not conceive an alternative explanation. Even when Faraday opened
to alternative possibilities, Maxwell felt forced to rectify him (Cf.
Subsection \ref{subsec:Maxwell's-epistemic}). In the next Subsection
we will se how Hertz tried to rectify what he regarded as faulty in
Maxwell's conception.

\subsection{Hertz as a theoretician\label{subsec:Hertz}}

While Hertz' most famous contributions come from his experiments,
he wrote two papers where he exposed his views on Maxwell equations.
In \citep[pp- 19-21][]{hert93} he explains his motivations (referring
to Maxwell's theory) as follows:
\begin{quotation}
(p. 19) “Casting now a glance backwards we see that by the experiments
above sketched the propagation in time of a supposed action-at-a-distance
is for the first time proved. This fact forms the philosophic result
of the experiments and, indeed, in a certain sense the most important
result. The proof includes a recognition of the fact that the electric
forces; can disentangle themselves from material bodies, and can continue
to subsist as conditions or changes in the state of space. The details
of the experiments further prove that the particular manner in which
the electric force is propagated exhibits the closest analogy with
the propagation of light; indeed, that it corresponds almost completely
to it. {[}...{]} Since the year 1861 science has been in possession
of a theory which Maxwell constructed upon Faraday's views, and which
we therefore call the Faraday-Maxwell theory. This theory affirms
the possibility of the class of phenomena here discovered just as
positively as the remaining electrical theories are compelled to deny
it.“

(p. 20) “I have not always felt quite certain myself of having grasped
the physical significance of his statements. Hence it was not possible
for me to be guided in my experiments directly by Maxwell's book.
I have rather been guided by Helmholtz's work, as indeed may plainly
be seen from the manner in which the experiments are set forth. But
unfortunately, in the special limiting case of Helmholtz's theory
which leads to Maxwell's equations, and to which the experiments pointed,
the physical basis of Helmholtz's theory disappears, as indeed it
always does, as soon as action-at-a-distance is disregarded. I therefore
endeavoured to form for myself in a consistent manner the necessary
physical conceptions, starting from Maxwell's equations, but otherwise
simplifying Maxwell's theory as far as possible by eliminating or
simply leaving out of consideration those portions which could be
dispensed with inasmuch as they could not affect any possible phenomena.
This explains how the two theoretical papers (forming the conclusion
of this collection) came to be written.“

(p. 21) “To the question: 'What is Maxwell's theory?'. I know of no
shorter or more definite answer than the following:-- Maxwell's theory
is Maxwell's system of equations. Every theory which leads to the
same system of equations, and therefore comprises the same possible
phenomena, I would consider as being a form or special case of Maxwell's
theory;{[}...{]}“
\end{quotation}
Hertz' operation is of an epistemological character. Actually, his
reading of Maxwell-Faraday contradicts Faraday's writing, for in Faraday
there was a duality between action and the (naive) view of matter
while in Hertz actions disentangle themselves from the bodies to meet
the requirements of Hertz epistemology. Also, the last quoted sentence
from p. 19 implies to ignore Lorenz' theory \citep{lore67} (which
is mentioned by Maxwell in \citep[{[805]}][]{maxw73}).

Hertz confesses not to \textsf{understand} the physical significance
of Maxwell's statements. This raises the question about Hertz' idea
of \textsf{understanding} which is greatly influenced by Helmholtz
(who was his Ph.D. advisor and life-long protector). Hertz needed
to see “pictures“ of the real based upon the elements he has learned
to trust: “a mechanistic conception within which natural phenomena
were to be explained by the action of mechanically moved matter“ \citep{schiemann1998loss}.
This view represents an extraordinary departure from Faraday's philosophical
attitude, since Faraday was willing to change his most basic building
blocks if this were necessary, while in Hertz we see the opposite
attitude: the persistence of beliefs.

The general idea that transpires from the quoted pages is that it
is possible (p. 21) to separate the process of construction of a theory,
what we have called the “rules of correspondence“ used in the construction,
from the theory's mathematical content. Instead, the theory has to
be provided with what Hertz called \textsf{interpretation}, a prescription
about how to apply the theory to experiments. Along this lines, Hertz
offers four interpretations of Maxwell's equations, including Maxwell's,
Helmholtz' and his own and preferred one, the most consistent with
the idea of the ether. Indeed, he attempts to recast Maxwell's work
in terms of his views about the ether.

\begin{figure}[h]
\resizebox{0.45\textwidth}{!}{%
\begin{tikzpicture}
\filldraw[color=blue!60, fill=blue!20, very thick](0,0) circle (1);
\filldraw[color=blue!60, fill=blue!20, very thick](4,0) circle (1);
\draw[->,red] (3,-0.75) -- (1,-0.75); \filldraw[color=red, fill=red](3,-0.75) circle (0.1);
\draw[->,red] (0.5,-1.25) -- (0.5,-2.75); \filldraw[color=red, fill=red](0.5,-1.25) circle (0.1);
\draw[->,red] (3.25,-2.75) -- (3.25,-1.25); \filldraw[color=red, fill=red](3.25,-2.75) circle (0.1);
\node[star,star points=7,star point ratio=0.6, draw, fill=yellow] at (0.5,-3) {\ };
\filldraw[color=blue!60, fill=blue!20, very thick](-1,-5) rectangle(1,-3);
\filldraw[color=blue!60, fill=blue!20, very thick](3,-5) rectangle(5,-3);
\draw[->,blue!20, ultra thick] (1.2,0) -- (2.8,0); 
\draw[->,blue!20, ultra thick] (0,-2.75) -- (0.,-1.25); 
\draw[->,blue!20, ultra thick] (4,-1.25) -- (4,-2.75); 
\node[black] at (-0.5,-2) {$\Pi$}; 
\node[black] at (2,0.5) {$\phi$}; 
\node[black] at (4.5,-2) {$\Gamma$}; 
\node[black] at (0,0) {Theory};  
\node[black] at (4,0.25){Theoretical}; 
\node[black] at (4,-0.25){conclusions};
\node[black] at (0,-4) {Observations}; 
\node[black] at (4,-3.75) {New};
\node[black] at (4,-4.25) {observations};
\node[star,star points=7,star point ratio=0.6, draw, fill=yellow] at (0.5,-3) {\ };
\end{tikzpicture} 
}\hspace{1cm}\resizebox{0.45\textwidth}{!}{%
\begin{tikzpicture} 
\filldraw[color=blue!60, fill=blue!20, very thick](0,0) circle (1);
\filldraw[color=blue!60, fill=blue!20, very thick](4,0) circle (1);
\draw[->,red] (3,-0.75) -- (1,-0.75); \filldraw[color=red, fill=red](3,-0.75) circle (0.1);
\draw[->,red] (3.25,-2.75) -- (3.25,-1.25); \filldraw[color=red, fill=red](3.25,-2.75) circle (0.1);
\filldraw[color=blue!60, fill=blue!20, very thick](3,-5) rectangle(5,-3);
\draw[->,blue!20, ultra thick] (1.2,0) -- (2.8,0); 
\draw[->,blue!20, ultra thick] (4,-1.25) -- (4,-2.75); 
\node[black] at (2,0.5) {$\phi$}; 
\node[black] at (4.5,-2) {$\Gamma$};
\node[black] at (0,0.25) {Theory $\equiv$}; 
\node[black] at (0,-0.25) {Equations};
\node[black] at (4,0.25){Theoretical}; 
\node[black] at (4,-0.25){conclusions};
\node[black] at (4,-3.75) {New};
\node[black] at (4,-4.25) {observations};
\end{tikzpicture} 
}

\caption{The epistemological change introduced by Hertz. \label{fig:Epistemological-change-introduce}}
\end{figure}
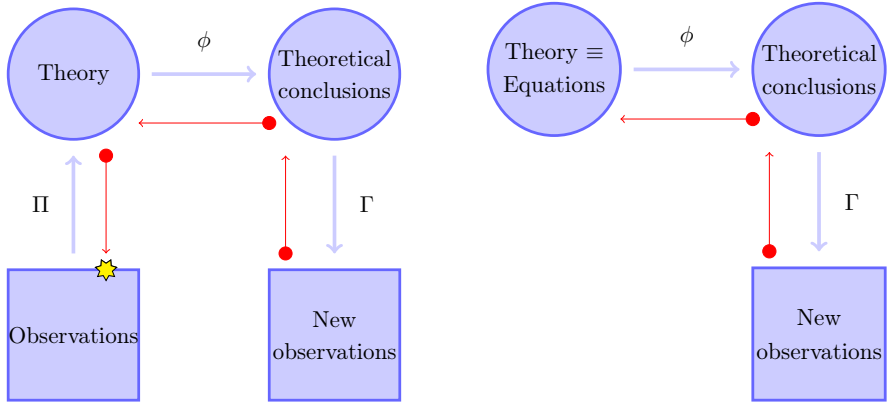

Hertz' departure from the traditional scientific epistemology is illustrated
in Figure \ref{fig:Epistemological-change-introduce}. Let $\Pi$
be a projection of the observable into a symbolic language (say mathematics),
$\phi$ the logical elaboration and $\Gamma$ the interpretation.
The left panel describes the traditional scientific construction,
where it is required that $\Pi\circ\Gamma=Id$ and $\Gamma\circ\Pi=Id^{\prime}$,
in other words that no distortions on both theory and observations
is introduced by the interpretation $\Gamma$ with respect to the
correspondence $\Pi$ applied in the construction. The thin red arrows
show the \textsf{flow of falsity}\citep{lakatos_1978}. In the traditional
epistemology of science (say that of Faraday), inconsistencies between
predictions and new observations flow back all the way through the
diagram triggering an improvement in the construction: Falsity can
force us to change our view of the world. The right panel describes
Hertz' view, where the flux of falsity cannot affect the ideation
since the ideation is suppressed. This conception represents a weakening
of the conditions for a theory to be acceptable, allowing the scientist
to change interpretations constrained only by their predictive success.

Hertz' deep epistemological change has been highlighted by D'Agostino:
“...by separating the mathematical structure of a theory from its
modes of representation he {[}Hertz{]} has profoundly challenged the
conception of a physical theory as an indivisible unity of the two
-- a conception accepted by Maxwell and other nineteenth century
mathematical physicists.“ \citep{dago68,dago04}.

Maxwell exposed his programme to address Faraday's ideas and experiments
in the first few pages of \citep{maxw56}, also reviewed in \citep{dago68}.
He advocates care in formulating models, so as not to fall in “...that
blindness to facts and rashness in assumption which a partial explanation
encourages.“ \citep[p. 155-156,][]{maxw56}. Further, Maxwell introduces
the idea of \textsf{physical analogies}, (p. 156), similar to what
we may call “working hypotheses“, finally declaring his goal: “...to
present the mathematical ideas to the mind in an embodied form“ (p.
187), namely integrating the mathematics with its physical significance.

The discussion about the four interpretations of Maxwell's equations
aims to retain the equations while removing their foundations. Action
at a distance was regarded as an obstacle. Since Maxwell's and Helmholtz'
views are supported by it, a new view was required. The equations
(which were good enough since they led to the wave equation) had to
be detached from the scaffolding used in their construction. Hertz
believed that Maxwell started from mechanics and ended with the ether
(actually an oversimplification), and sought a fully ether-based interpretation
instead.

For Hertz, action at a distance was not to be taken into account as
a possibility and hence the potentials $A,V$ were undesirable. Instead,
he regarded the electric and magnetic fields $(E,B)$ as the state
of the ether (p. 251), therefore deserving a central importance in
physics. Moreover, in Hertz' interpretation Maxwell did not consider
bodies in movement (p. 247) (despite any average reader of Maxwell's
Treatise can verify that this statement is false) and proceeds to
incorporate the movement under his famous hypothesis that the ponderable
matter drags the ether (p. 242). Forcing his system to correspond
with the traditional (Galilean) view, he incorporates terms representing
this effect (eq. 3, p. 251). The equations obtained with the new interpretation
are regarded as new, but they are actually identical to what is called
equation (B) in \citep{maxw64,maxw73}, now under a new interpretation.\footnote{It is remarkable that this mistake has persisted for centuries in
standard textbooks. It might have originated in the fact that Heaviside
was the first to write Maxwell's equations in their modern form (around
1885, \citep[p. 429, ][]{heaviside2011electrical}), for the special
case of bodies at rest. However, the most likely origin is Maxwell
himself who wrote under the title “The propagation of electromagnetic
disturbance“: “Let us next determine the conditions of the propagation
of an electromagnetic disturbance through a uniform medium, which
we shall suppose to be at rest, that is, to have no motion except
that which may be involved in electromagnetic disturbances.“\citep[{[783]}][]{maxw73}.
Is Maxwell thinking in the possibility of motion with respect to absolute
space? He is certainly considering a non moving ether, where the reference
for the movement remained in his brain.}

\subsection{H. A. Lorentz\label{subsec:Lorentz}}

The dutch physicist Hendrik Lorentz introduced a number of ideas that
have propagated to the present day. In a work from 1895 \citep[p. 2,][]{lorentz1895attempt},
he explains part of his scientific trajectory, that first developed
along action at a distance for some time. He informs us that “This
I have shown in a previous paper {[}13{]}, in which I admittedly have
derived the equations of motion from actions at a distance, and not,
what I now consider to be much easier, from Maxwell’s expressions.“
And later (p.3) he states “The influence that was suffered by a particle
B due to the vicinity of a second one A, indeed depends on the motion
of the latter, but not on its instantaneous motion. Much more relevant
is the motion of A some time earlier, and the adopted law corresponds
to the requirement for the theory of electrodynamics, that was presented
by Gauss in 1845 in his known letter to Weber {[}18{]}“\footnote{Lorentz' reference {[}18{]} corresponds to \citep[bd.5 p. 627-629,][]{gaus70}.}.

The views of Lorentz constitute a progression. In 1892 Lorentz considered
the advantages of Hertz approach:
\begin{quote}
M. Hertz ne s'occupe guère d'un rapprochement entre les actions électromagnétiques
et les lois de là mécanique ordinaire. Il se contente d'une description
succincte et claire, indépendante de toute idée préconçue sur ce qui
se passe dans le champ électromagnétique. Inutile de dire que cette
méthode a ses avantages.\citep[p. 368,][]{lorentz1892CorpsMouvants}
\end{quote}
In this work he considers both the case of the ether being dragged
by the ponderable bodies (Chapter II) and the case in which the bodies
move in the ether without dragging it (Chapter IV). He also considers
the hypothesis of the electric fluid (§31). Following Hertz and Fizeau,
Lorentz assumes \citep[p. 1,][]{lorentz1895attempt}:
\begin{quote}
...that ponderable matter is absolutely permeable, namely that at
the location of an atom, also the aether exists at the same time,
which would be understandable if we were allowed to see the atoms
as local modifications of the aether.
\end{quote}
However, Lorentz conceives an ether at “rest“. He later adds:
\begin{quote}
That we cannot speak about an absolute rest of the aether, is self-evident;
this expression would not even make sense. When I say for the sake
of brevity, that the aether would be at rest, then this only means
that one part of this medium does not move against the other one and
that all perceptible motions are relative motions of the celestial
bodies in relation to the aether.\citep[p. 2,][]{lorentz1895attempt}
\end{quote}
We have to observe that the notion of “perceptible motion“ is not
reconcilable with an imperceptible ether. The struggle to conciliate
the concept of space with electromagnetism continues in Lorentz. It
has been asserted that Lorentz' ether is some form of (absolute) space
(see e.g., \citep[p. 172,][]{ritz1908recherches}). Lorentz' ether
hypothesis contradicts that of Hertz (and also that of Stokes \citep{stokes1845iii})
in which the ether was dragged by ponderable bodies. Indeed, in \citep{lorentz1886influence}
Lorentz criticises different alternatives of ether drag.

While he admits that the Michelson and Morley experiment \citep{michelson1887relative}
(failing to detect a velocity of the earth relative to the ether)
represents an objection to his view, he proceeds to reformulate his
view incorporating the now famous contraction in the direction of
movement \citep[§89--§92,][]{lorentz1895attempt}. Lorentz attitude
corresponds to the “persistence of belief“ in the terms of Peirce:
The existence of the ether is never questioned, the propagation of
light cannot be conceived without an emission theory and a propagating
medium. In this respect, Lorentz retains the view of Maxwell and Hertz
and in so doing he opposes Faraday's view of vibrating rays. However,
Lorentz' ether cannot be reconciled with an ether that is dragged
by matter. In this issue concerning the ether “at rest“ or not at
rest, Lorentz remains close to Maxwell's Art. {[}783{]} of the Treatise,
while Hertz advanced a different view.

In his version of Maxwell's theory Lorentz leaves behind Maxwell's
practice of the \emph{epojé}. Thus, where Maxwell sustained the doubts
resulting from several undecidable possibilities, Lorentz adopts the
hypotheses.

The writing of Lorentz presents an evolution towards an authoritative
form in contrast with that of Ampère, Faraday, Maxwell and Hertz,
to mention just a few. Lorentz late writing is much closer to current
textbooks where a doctrine is being transmitted without providing
clues on how the author has established her/his results.\footnote{Actually, Maxwell would disagree with us, since this is Maxwell's
criticism of Ampère\citep[{[528],}][]{maxw73}, but in comparison
to Lorentz, Ampère is transparent.}

\subsubsection{The current}

As a central issue in his career, \citep{lorentz1892CorpsMouvants,lorentz1895attempt,lorentz1899simplified},
Lorentz adopted the early view of electrical particles. It is remarkable
that this idea of Gauss relates to the ether-free formulation and
is central to, for example, Weber's theory \citep{webe46}, considered
in Section \ref{relational}. In paragraphs §75 to §80 in \citep{lorentz1892CorpsMouvants},
he considers the force on a ponderable body that moves through the
ether with velocity $v$ relative to it (for the precedents and history
of the force see \citep[p. 143,][]{assi94}). In §75 the total current
is introduced,
\begin{equation}
J=j+\frac{\partial\!D}{\partial t}\label{eq:TotCurr}
\end{equation}
(in modern notation), being ${\displaystyle \frac{\partial D}{\partial t}}$
Maxwell's displacement current (in vacuum $D=\epsilon_{0}E$). Lorentz
proposes $j=\rho v$, this is the product of the density of charge
times the velocity relative to the ether, an expression present in
the current theory of electromagnetism and now denominated \textsf{Lorentz'
current}. He then proceeds to find the force following Maxwell's idea
regarding the variation of the energy with respect to a virtual displacement
of the particle. Because of the relevance of the expression of the
Lorentz' force we will address its derivation in a dedicated subsection.

The Lorentz current is introduced in all textbooks of physics that
we know about. However, it deserves to be questioned from an experimental
point of view. A bounded density of charge implies that charges can
be obtained in infinitesimally small amounts (just by integrating
this density in appropriate small domains), this is, in an amount
smaller than any given one, for example, smaller than the charge of
the electron. Nevertheless, the work by Millikan \citep{millikan1913elementary}
persuaded physicists that the charge of an electron is the minimum
value for electric charge, the quantum of electricity. If the charge
density must integrate in all domains to a multiple of the electron
charge, it cannot be continuous. Following this line of thought to
arbitrarily small distances, it follows that charges and the electron
in particular must be point-like objects and that charge density is
unbounded. That charged particles are not point-like is an experimental
observation that is in itself inconsistent with the assumptions made
by Lorentz in his deduction, but furthermore, Lorentz' construction
is inconsistent with our understanding of conductors and electron
states in conductors after quantum theory. As Einstein perceived \citep{eins40},
present field theories and quantum theory are not compatible, the
efforts made to forcibly put them together added insurmountable difficulties
that can only be dealt with by destroying the mathematical building
\citep{nati15} on which they rest.

\subsubsection{Lorentz' force}

The changes introduced by Hertz, Lorentz and others to Maxwell equations\footnote{The modification consists in accepting Maxwell equations in the cases
of bodies without relative motion and then proposing different extensions
based upon different ideas regarding the ether for the case of bodies
in motion (no longer relative motion, but with respect to the ether).
Motion is then considered with respect to the ether although no experimental
basis exists for it. All the motions considered in Faraday's experiments
and Maxwell formulae are relative motions.} (despite still naming the modified equations after Maxwell) make
Maxwell's (modified) equations unsuited to describe cases of induction
that appear in the foundational experimental background of the theory.
Experiments such as Arago's disk \citep[{[81],}][]{fara39} or the
rotating magnet of Faraday \citep[{[217--], }][]{fara39,assi94,munl04}
cannot be understood using Maxwell's (modified) equations. This matter
constitutes today the so-called “exceptions to the flux rule“ (referring
to the Faraday-Maxwell induction law) \citep[Ch. 17-2, Vol. ii,][]{feynman1965}.
We are instructed then that “the correct physics is always given by
the two basic laws ...“ being the first of the two “basic laws“ the
expression of Lorentz force. The resolution of this misunderstanding
has been given in \citep{munl04}. Using Faraday-Maxwell's flux law
and following Faraday's and Maxwell's original insight, rather than
the modified view proposed by Lorentz to accommodate for the ether,
Munley accounts for the rotating magnet and shows how the discrepancy
was built when the signification of the original flux law was altered
to its present status.

The force experienced by a charged point-like particle by the influence
of the electric and magnetic fields still plays today a central role
in electrodynamics. However, its origin is seldom discussed (cf.,
e.g., \citep{jackson1962classical}, where the Lorentz' force is mentioned
on p. 3, 260, 553, etc. (3rd. edition), but it is never derived from
basic principles). It comes as a piece of received knowledge without
a sound insertion in the theory.

The force is presented in \citep[§74-§80 (Ch. IV),][]{lorentz1892CorpsMouvants}
under the following assumptions:
\begin{itemize}
\item The state of the ether is defined by the associated fields created
by matter. There is a distinguished body on which the force will be
computed. The body is a rigid solid. The charge density is nonzero
only at the location of each body and it adds up to a smooth function
$\rho$, with its associated Lorentz current in the case of the moving
body. No other current in the distinguished body is considered. The
ether acts through the fields $E$ and $B$, satisfying Maxwell's
equations.
\item It is assumed that each point that takes part in the electromagnetic
movements is known and determined by our cognition of all the particles
of the system and the electric field in all the points of the space
\citep[§75 f. (Ch. IV),][]{lorentz1892CorpsMouvants}.
\item The force with which the ether acts on the distinguished body (with
charge density $\rho_{0})$ within this electrically active matter
is computed by an application of the variational method (with roots
in Lagrange and D'Alembert), inspired in Maxwell. A \textsf{virtual
displacement} with respect to the ether is proposed for the particle,
leaving the rest of the matter without modification (hence, the displacement
can be thought of both as relative to the ether as well as relative
to the rest of the bodies). The variation is not completely calculated
mathematically (despite it being possible). Rather, some steps are
circumvented with arguments pertaining to the ether, as e.g., the
restricted variation in §78 and its subsequent effects in §80 b .
The final result reads
\[
F_{L}=\int d^{3}x\,\rho_{0}\left(E+v\times B\right)
\]
\end{itemize}
There are a number of epistemological issues around this result. First,
Lorentz departs from Maxwell in the proposal of a current. While Maxwell
leaves as an undecided question the fundamental nature of the current,
Lorentz postulates that the only possible current for a charged particle
is given by its velocity relative to the ether at rest. Second, Lagrange's
method rests in allowing arbitrary variations compatible with the
boundary conditions. The peculiar variation proposed by Lorentz, automatically
restricts its validity. The force can be such only in a situation
where the proposed variation would be the most general possible. Indeed,
the mechanical force computed by Maxwell \citep[{eq. C, [619],}][]{maxw73}
is more general than Lorentz' expression. But perhaps the more striking
innovation is to put and end to the separation between guessing and
conjectures and mathematics. In Newton, Maxwell and all the precedent
physics, there is a time for synthetic propositions, usually inspired
in experimental observations, while the theory develops further through
analytic propositions performed with the correctness and consistency
of mathematics. Not only Lorentz engages in interpretations as Hertz
but he goes beyond them by substituting mathematical steps with his
imagination on the behaviour of the ether, see \citep[§80,][]{lorentz1892CorpsMouvants}

Lorentz Lagrangian reads 
\begin{eqnarray*}
{\cal L} & = & \frac{1}{2}\int\left(\frac{1}{\mu_{0}}|B|^{2}-\epsilon_{0}|E|^{2}\right)\,d^{3}x\\
 & = & \frac{1}{2}\int\left[A\cdot J\right]\,d^{3}x-\frac{1}{2}\int\rho V\,d^{3}x
\end{eqnarray*}
where $(A,V)$ are the potentials, $(j,\rho)$ the measurable current
and charge densities, $J$ as in eq.\eqref{eq:TotCurr}, while $(E,B)$
stand for the electric and magnetic fields. The integral extends over
the region occupied by the system. These quantities are related by
Maxwell's modified equations
\begin{eqnarray}
\nabla\cdot B & = & 0\nonumber \\
\nabla\cdot E & = & \frac{\rho}{\epsilon_{0}}\nonumber \\
\nabla\times B & = & \mu_{0}\left(j+\frac{\partial E}{\partial t}\right)\label{Maxwell eq.}\\
\nabla\times E & = & -\frac{\partial B}{\partial t}\nonumber \\
\nabla\times A & = & B\nonumber \\
-\frac{\partial A}{\partial t}-\nabla V & = & E\nonumber 
\end{eqnarray}

The fourth and sixth equations differ from Maxwell's original formulation
in that the present version holds only for stationary problems and
rigid circuits. In this way, the equations can no longer describe
Arago's problem and the like (see above). From the expression of the
electromagnetic Lagrangian an action, ${\cal A}$ can be formally
written as
\begin{eqnarray*}
{\cal A} & = & \int dt{\cal L}=\frac{1}{2}\int dt\left[\int\left(\frac{1}{\mu_{0}}|B|^{2}-\epsilon_{0}|E|^{2}\right)\,d^{3}x\right]\\
 & = & \frac{1}{2}\int dt\left[\int d^{3}x\,\left[A\cdot J\right]-\int d^{3}x\,\rho V\right]
\end{eqnarray*}
where in all cases the limits of integration have been absorbed in
the support of the integrands. We defer to Appendix \ref{sec:lagrangian-appendix}
a more technical derivation of the electromagnetic force from the
Lagrangian stated above, rendering explicit the restrictions that
lead to the Lorentz' force. In short, Lorentz proposal leads to
\[
F_{L}=\int d^{3}x\left[(\rho v)\times B+\rho E\right]
\]
if we consider the material current to be 
\[
j=\rho v
\]
where $v$ is a velocity relative to the ether and we use by habit
the Lagrangian formulation of mechanics.

In Lorentz' original context the Lagrangian is not invariant under
changes of reference systems and the velocity involved is the velocity
with respect to the frame of reference or the ether. We can derive
by Lorentz' method the force exerted by the ether on all the system
as 
\[
F_{L}=\int d^{3}x\left[(\rho v)\times B+\rho E\right]
\]
and for the force exerted by the ether on the particle $0$
\[
F_{L}^{0}=\int d^{3}x\left[(\rho_{0}v_{0})\times B+\rho_{0}E\right]
\]
Thus, in Lorentz, the ether (or the space, after suppressing the material
ether) exerts a force on every electric body which adds up to some
non necessarily zero amount. It is also well known that Lorentz' force
does not conform the action-reaction principle \citep[II 26-2,][]{feynman1965}\citep[Ch. 6,][]{assi94}.

Lorentz will later \citep[eqs. Ia-Va,][]{lorentz1899simplified} summarise
electrodynamics with five equations, namely the four Maxwell's equations
described from a reference system fixed to the ether at rest plus
Lorentz' force. Basically, this idea rests on the assumption that
the electromagnetic fields are given: they are properties of the ether
that suffer no influence from the particle subject to $F_{L}$.

The action above, expressed as $\mathcal{A}=\frac{1}{2}\int dt\left[\int d^{3}x\,\left[A\cdot J\right]-\int d^{3}x\,\rho V\right]$
has an interesting symmetry related to the now famous \textsf{Lorentz
transformations.} Let $X=(a,b)$ (with $a\in R^{3}$and $b\in R$)
be a four vector such as $(A,{\displaystyle \frac{V}{C^{2}}})$ or
$(j,\rho)$ or $(x,t)$, where $C$ is Weber's constant (later known
as the “speed of light“). Further, let $T$ be the transformation
\begin{eqnarray}
T_{u}X & = & \left(\gamma(a-uCb)+(1-\gamma)\hat{u}\times(a\times\hat{u}),\gamma(b-\frac{u\cdot a}{C})\right)\label{eq:TL}
\end{eqnarray}
where $u$ is a dimensionless parameter with $|u|<1$, ${\displaystyle \gamma=\frac{1}{\sqrt{1-u^{2}}}}$
and $\hat{u}=\frac{u}{|u|}$ a unit vector. The Lorentz transformation
$TL_{u}$ of a four vector valued function $F$ of the four vector
space-time, $X=(x,t)$, reads
\begin{eqnarray*}
TL_{u}F & = & T_{u}F(\xi,\tau)\\
(\xi,\tau) & = & T_{-u}X
\end{eqnarray*}
and leaves invariant the action ${\cal A}$. Hence, it represents
a symmetry of the action. The discussion of this symmetry and its
possible meaning corresponds to the contribution of Einstein. Further,
notice that if $j$ would represent the current associated to a particle
at rest in some reference frame, (which is zero in Lorentz' conception)
its representation in another frame as given by a Lorentz transformation,
would be 
\begin{eqnarray*}
TL_{u}(j,\rho) & = & T_{u}(j,\rho)(\xi,\tau)\\
(\xi,\tau) & = & T_{-u}X
\end{eqnarray*}
which disagrees with the Lorentz current (there is a factor $\gamma$
in excess). In other words, the construction is inconsistent with
special relativity and only acceptable under the new epistemology
that has no place for constructive consistence.

\subsection{Mathematical picture of DAD\label{subsec:DAD}}

Having reached the point of experimental refutation of the idea of
the ether, we feel obliged to collapse what originates from this idea.
Thus, from the competing theories of propagation, ether and DAD, we
must turn to the latter again and see whether Lorentz' current and
force belong to it. Especially in the case of the force since it was
``derived'' assuming at the same time displacements with respect
to the complement of the probe in the universe and with respect to
the ether. The work is imminently mathematical and is presented in
Appendix \ref{sec:lagrangian-appendix} We show in the Appendix that
the following elements fit harmoniously in unity: Faraday's insight
regarding field interactions, Gauss' insight on retarded action, Maxwell's
Lagrangian approach, equations and transformation, Lorenz's delays,
C. Neumann's minimal action approach, Lorentz' Lagrangian and force,
the principle of action and reaction and Newton's equations. But above
all, in the epistemological side, Newton's approach and motto: \emph{hypotheses
non fingo} appears as superior, for the production of a theory than
Helmholtz-Hertz pictorial method which needs the support of physical
hypotheses.

\subsection{Einstein contribution and his view of the ether\label{subsec:Albert}}

Einstein's 1905 work \citep{eins05} represents the final metamorphosis
of the electromagnetism initiated by Faraday and Maxwell. As Dingle
\citep{ding60} observes, (accepted) electromagnetism was not revised
later and continues to be the same since then. Einstein's equations
for electromagnetism are directly taken from Lorentz \citep{lorentz1904electromagnetic}
although the velocity participating in the equations will change its
meaning to “velocity with respect to a frame of reference“. In so
doing he adheres to the epistemological view of Hertz: to keep the
equations while producing a new interpretation. The motivation for
a new interpretation is clearly written as:
\begin{quotation}
It is known that Maxwell’s electrodynamics---as usually understood
at the present time---when applied to moving bodies, leads to asymmetries
which do not appear to be inherent in the phenomena. Take, for example,
the reciprocal electrodynamic action of a magnet and a conductor.
The observable phenomenon here depends only on the relative motion
of the conductor and the magnet, whereas the customary view draws
a sharp distinction between the two cases in which either the one
or the other of these bodies is in motion. For if the magnet is in
motion and the conductor at rest, there arises in the neighbourhood
of the magnet an electric field with a certain definite energy, producing
a current at the places where parts of the conductor are situated.
But if the magnet is stationary and the conductor in motion, no electric
field arises in the neighbourhood of the magnet. In the conductor,
however, we find an electromotive force, to which in itself there
is no corresponding energy, but which gives rise---assuming equality
of relative motion in the two cases discussed---to electric currents
of the same path and intensity as those produced by the electric forces
in the former case.
\end{quotation}
which is the opening paragraph of the 1905 work\footnote{Einstein's discussion of the symmetry is not clear. First, we cannot
know what it is meant by “electrodynamics---as usually understood
at the present time---“, but most likely it refers to the electrodynamics
as described by Lorentz. Certainly, there is no asymmetry in Faraday's
experimental work and in Maxwell's original formulation. The movement
they considered is relative motion and the results were checked experimentally.
However, a description of the situation using subjective coordinates
with reference to either the magnet or the conductor and breaking
down the interaction by dissociating the source from the detector
(despite the fact that there can be no induction without both conductor
and magnet) may have inspired the expectation by Einstein of a “field
arising {[}or not, depending on the choice of subjective reference{]}
in the neighbourhood of the magnet“ although there is no reason for
it since the terms of the relation are not equivalent. Einstein will
not come back to this matter in the paper but only to the issue of
relative motion. In Appendix \ref{sec:lagrangian-appendix} we show
that the sort of symmetry of form Einstein was seeking is present
in the DAD, i.e., in electromagnetism after suppressing the ether.
Further, also the assertion “there is no corresponding energy“ cannot
make sense outside the subjective description since the phenomenon
is rooted in the relative motion and the energy involved in the electromotive
force and induced current is always there, regardless of the observer
and his/her viewpoint.}. In the second paragraph, after rejecting absolute motion and the
idea of the ether as reference, he will introduce “The principle of
relativity“ (without mentioning Poincaré), a conjecture that is raised
to the level of principle. He also introduces the second (explicit)
postulate of the theory: “that light is always propagated in empty
space with a definite velocity $c$\footnote{In this work we use $C$ throughout for the associated electromagnetic
quantity, i.e., $C^{2}=(\mu_{0}\epsilon_{0})^{-1}$ .} which is independent of the state of motion of the emitting body“.

As it frequently occurs in physics, Einstein's explicit postulates
have to be complemented with implicit postulates, that are taken for
granted without even mentioning them. Among them we count in this
case “there exists something real we call space“, “there is something
real that we call relative velocity“ --this one in particular after
the opening paragraph-- and “light is something emitted“ (and in
this respect it is body-like). These three postulates are by 1905
hidden in the habits and other forms of irrational acting of scientists.
We have already explained \citep{sola18b} that space emerges from
the construction of the child and it is an auxiliary concept in mechanics.
Much later, Einstein will become more explicit about his view of the
ether
\begin{quotation}
When we speak here of aether, we are, of course, not referring to
the corporeal aether of mechanical wave-theory that underlies Newtonian
mechanics, whose individual points each have a velocity assigned to
them. This theoretical construct has, in my opinion, been superseded
by the special theory of relativity. Rather the discussion concerns,
much more generally, those things thought of as physically real which,
besides ponderable matter consisting of electrical elementary particles,
play a role in the causal nexus of physics. Instead of ‘aether’, one
could equally well speak of ‘the physical qualities of space’. Now,
it might be claimed that this concept covers all objects of physics,
for according to consistent field theory, even ponderable matter,
or its constituent elementary particles, are to be understood as fields
of some kind or particular ‘states of space’. \citep{einstein24}
\end{quotation}
Rather than description tools, the fields become entities of ``real''
existence, particular 'states of space'.

This is a remarkable paragraph since it goes back to Hertz' idea of
electromagnetic fields as the “state of the ether“, just changing
the word ether by space. Actually, despite the usual claims about
Special Relativity --for example \citep[p.354, ][]{torr07}--, only
the materiality of the ether has been suppressed, Poincaré's vision
of the ether, already presented (Section \ref{subsec:Poincar=0000E9}),
lives in this paragraph. It is the fundamental need, introduced by
the new epistemology, of sustaining a mechanistic (material) view
of actions what leads to consider the space as an immaterial ether.

Einstein postulates that the Lorentz transformations (named in honour
of Lorentz but introduced as such by Einstein), which leave the electromagnetic
action invariant (and correspondingly, are associated to equivariant
transformations of Maxwell equations), correspond to changes of reference
frames in relative motion. Since neither Einstein nor Poincaré discussed
the fundamentals of the Relativity principle (there is no critical
motion towards the fundamentals, the principle is a conjecture, an
intuition in them) there are no restrictions imposed to the transformations.
Einstein will then postulate the Lorentz transformation as a proper
substitute for the Galileo transformations.

Let us now study Einstein postulates from the point of view of the
No Arbitrariness Principle.

\subsubsection{A metaphysical symmetry}

We shall first make explicit the equivariance (too often called covariance)
in Maxwell's equation. To simplify the exposition we will introduce
the vector and scalar potential in the Lorentz gauge
\begin{eqnarray*}
E & = & -\frac{\partial A}{\partial t}-\nabla V\\
B & = & \nabla\times A\\
\Box(A,\frac{V}{C^{2}}) & = & -\mu_{0}(j,\rho)
\end{eqnarray*}
Where $\Box$ is D'Alembert operator, defined in eq.(\ref{eq:BoxDef}).
The third equation establishes a relation between two four-vectors
and since
\[
TL_{v}\circ\Box=\Box\circ TL_{v}
\]
the four-vector ${\displaystyle (A,\frac{V}{C^{2}})}$ changes under
the Lorentz transformation in the same form than the four vector $(j,\rho)$,
hence $(A,{\displaystyle \frac{V}{C^{2}}})$ and $(j,\rho)$ are \texttt{equivariant}
under a Lorentz transformation\footnote{We notice that the wave equation is an incomplete statement of a Physical
law. To be complete, a law expressed by differential equations needs
(generalised) boundary conditions to be expressed. Then, the equivariance
requires the differential equation to be used with boundary conditions
that preserve the form under the change of coordinates proposed. This
matter is completely de-emphasised in physics textbooks.}. This is the sort of invariance of form sought by Einstein. We shall
recall that a change of reference frame corresponds to a frame that
moves relative to the original one. Let us now suppose there is a
neutral current (no accumulation of static charges is present) going
through a cable, and the cable is not moving in the frame where we
have measured the current (let's say that the galvanometer is part
of the circuit, as it usually is). Consider now a second observer
in motion with respect to the galvanometer and the circuit, with some
velocity $v$. According to Einstein's view, the charge and current
corresponding to the second observer can be obtained by transforming
the original four vector $(j,0)$ with a Lorentz transformation. However,
no method of measurement has ever been offered for moving observers
(independent of the fields). How do we measure a current while flying
by? We are forced to conclude at this time that the charge and current
so obtained are metaphysical. The symmetry does not exist except in
our construction. They are only tautologies obtained through formulae
in the theory that cannot be physically verified. On the contrary,
DAD is not metaphysical, it relates measurable quantities in two systems
(source and detector, even if in relative motion) among each other.

\subsubsection{Collapsing Einstein's Special Relativity theory.}

In section \ref{sec:The-scientific-attitude} we have discussed the
collapse of a reasoning that is produced when a contradiction is found.
The idea is present in the form of \emph{reductio ad absurdum} in
mathematics. As we have seen, some of the postulates of Einstein are
not explicit. We will focus our attention in the concept of “relative
velocity“ which is mentioned on intuitive grounds at several places
of the 1905 work. The concept makes sense in relational mechanics
as a result of the invariance of relative position and time differences
\citep{sola18b}, however, if we change the structure of space-time,
we open for the possibility of destroying the concept. The real character
of the relative velocity is expressed in mathematical terms by its
invariance with respect to reference frames. As Peirce would have
put it: the real does not depend on opinions. This is to say, that
there should be an operation “$\ominus"$ that produces the relative
velocity $v_{ab}$ from the observed velocities of the two members
(say $v_{a}^{s}$ and $v_{b}^{s}$). This is
\[
v_{ab}=v_{a}^{s}\ominus v_{b}^{s}
\]
where the superscript $s$ runs over the equivalent observers. This
is, the relative velocity, $v_{ab}$, is invariant under changes of
observers (reference frames). It corresponds to the foundation of
the concept of group of transformations that all those operations
that leave an object invariant (be it equation, figure or any other)
form a group.

The concept of relative velocity is assumed and discussed as a part
of Special Relativity. However, the Lorentz transformations do not
form a group and the velocity addition by Einstein does not form a
group either.\footnote{A quick demonstration is as follows. The generators in the Lie algebra
of the Lorentz transformation do not close an algebra but rather have
Lie-products in the algebra of rotations. Equivalently, the Lorentz
transformations correspond with the cosets in the Poincaré-Lorentz,
$SO(3,1)$, group where the set of rotations, $SO(3)$, is a subgroup
but not a normal subgroup. Therefore, there is no “quotient group“
for $SO(3,1)$ relative to $SO(3)$, but just a quotient manifold.
The addition in the coset manifold corresponds to Einstein's velocity
addition and therefore it does not form a group. The structure of
the Poincaré group in relation with its subgroups and cosets is well
presented in \citep{gilm74}. } Thus, the theory collapses since there is no conceivable invariant
quantity $v_{ab}$ in it (thus contradicting one of the implicit assumptions)
but there are at most \textsf{opinions} about it depending on the
choice of reference frame. We offer more mathematical details in Appendix
\ref{sec:SR}. The problem with special relativity is then that it
does not speak about reality (as otherwise understood) since it fails
to satisfy the first postulate of logical realism: there exists a
reality independent of the observers.

Many objections to Special Relativity have appeared throughout the
decades, all of them meeting the same compact rejection despite being
intrinsically different. In 1974 Thomas E Phipps published a manuscript
discussing whether the Lorentz contraction had any real existence
\citep{phipps1974kinematics}. The answer was negative, after setting
up a simple measurement device consisting of a rotating a disc with
radial scratches. The relevant issue is that about 70 years had been
necessary to set up an experiment to refute (or not) a basic prediction
of the theory in the standard terms of Natural Science theories. On
the contrary, many article pages and textbooks had been written in
the mean time discussing the consequences of Lorentz contractions
from the point of view of “thought experiments“. \footnote{Thought experiments cannot challenge beliefs. At most, they can check
consistency issues between conflicting beliefs. As such, they can
be used as a method for indoctrination, to let go “false“ beliefs
retaining the “proper“ ones.}

\section{Discussion}

In physics, the XIX century was signed by the quest of understanding
problems such as: What is light and how does it propagate? How does
light propagation relate to the propagation of electricity in a conductor?

As stated before, Maxwell admits in the Treatise that he cannot conceive
light propagation outside the emission theory, with a source that
emits light, a medium on which light propagates and eventually a detector.
Moreover, the emitted light should behave like a wave. Kirchhoff \citep{kirchhoff1857liv}
had already advanced a wave-like model for electricity propagation
in wires. Unfortunately, F. Neumann's energy suggests a form for the
vector potential $A$ that cannot host waves. Starting from the persuasion
that light is an electromagnetic phenomenon as suggested by the experiments
of Faraday, Lorenz and others and from the abundant experimental evidence
of light as a wave phenomenon, the quest becomes how to incorporate
it to the electromagnetic phenomena evidenced by experiments with
circuits and magnets. This is the point of divergence. For Gauss,
Lorenz and later C. Neumann and others a delay in the propagation
of electromagnetic disturbances is what is needed and, especially
in Lorenz, no other justification needs to be added. In other terms,
this is inference or abduction in its pure form. The reformulation
is possible because it does not contradict experiments and achieves
a higher level of conceptual unity.

Pure reason appears not to be sufficiently convincing for others,
including Maxwell. Maxwell introduces an equivalent term in a differential
equation sustaining his view of the (possible) existence of the ether,
a matter worth to investigate. Thus, Maxwell's introduces the displacement
current. The next step in this quest is here represented by Hertz,
who advanced in the physical properties of the ether and adjusted
Maxwell's equations to them. If Maxwell's equations and the ether
were weakly linked in the original, now they became a unity. Further,
Hertz opens for the possibility of separating the equations from the
ideas that originated them. The next turn by Lorentz was to conceive
an ether at rest, adapting likewise Maxwell's equations to the situation.

Along this process, the explanatory power of Maxwell's theory (restricted
to the four equations that are in use still today) was reduced to
particular cases, leaving out important foundational experiments.
Electromagnetic systems such as Arago's disk and Faraday's unipolar
generator where the intervening parts are in relative motion were
instead addressed with the introduction of the Lorentz' force. It
is interesting to consider the introduction of Lorentz' force in this
context. In Maxwell, the electromotive force $e.m.f.$ is the force
that moves electricity (be it fluid, particles or any other thing)
and reads $e.{\displaystyle m.f.\equiv\phi=v\times B-\nabla\psi-\frac{\partial A}{\partial t}}$
\citep[{eq. B, [619],}][]{maxw73}, where $\psi$ is undetermined.
Lorentz' force is then Maxwell's electromotive force exerted over
a free-to-move charge. We get the following correspondence of symbols
\[
E=-\frac{\partial A}{\partial t}-\nabla\psi
\]
with the additional differences that Lorentz interprets the symbol
$v$ (originally a relational velocity) as the velocity with respect
to the ether and $\psi$ is identified with $V$.

Taking the rotor in the last equation we get the “Maxwell equation“
(not present in Maxwell's Treatise)
\[
\nabla\times E=-\frac{\partial B}{\partial t}
\]
In this way, Faraday's induction law has been split into two parts
and the same can be said for the electromotive force. The operation
reveals why the “new“ Faraday law (the equation above) no longer explains
some experiments by Faraday, that now require the Lorentz' force.
Thus, Lorentz goes in the reverse direction than Maxwell: while the
former seeks unity the later separates equivalent electrical effects.
To confuse things further, names and symbols are kept but their mathematical
meaning is changed.

At last, the contradictions and refutations related to the ether being
dragged or not with the movement of the earth became insurmountable.
It is clear in Lorentz that the ether is just an instrument, the relevant
dynamical quantities are the fields, expressing the state of the ether.
It was a small step then to conceive the fields as properties of space,
thus getting rid of the ether in the wording of the theory, while
in practice Lorentz' formulation is used still today, with the additions
of Einstein's special relativity. Before Einstein the electromagnetic
theory of light was untenable in view of its internal conflicts. After
Einstein the problem shifted to a more subtle one, since it is based
on concepts that the theory cannot sustain (such as relative velocity
and the group property of coordinate transformations) and on assumptions
(the constancy of the velocity of light) that were not put to experimental
test for many decades.

As previously discussed in Section \ref{sec:The-scientific-attitude},
only reason is the supporting pillar of scientific theories, in particular
the absence of internal contradictions and the ability to honestly
resist refutation attempts. On the contrary, a characteristic of the
construction of EM has been that failures, drawbacks, contradictions
and refutations had little effect. Since there appeared to be no alternative
to the emission/propagation theory of light (once the alternative
lines of thought were suppressed or ignored), no mishap whatsoever
could weaken this substantialist belief.\textcolor{brown}{{} }However,
alternatives did exist, only that they were never sufficiently studied,
and still aren't.

Thus, the historic construction of current electromagnetism reveals
a much larger and decisive underlying process: it represents a breaking
point that changes the goals of science from seeking the harmony of
the cosmos to the much limited search for some sort of usefulness
or success. The scientific goal becomes vague or poorly defined and
it substantially broadens the meaning of science. When Einstein states
“The justification (truth content) of the system rests in the proof
of usefulness of the resulting theorems on the basis of sense experiences“,
we do not know the service or purpose that has been stated for science:
what does useful mean in this context? The meaning will soon be provided
by the society at large, true will then become “useful for production“,
and \emph{Scientia} will be spelled \emph{techné}.

When science is being redefined, the scientist producing successful
predictions will later be appreciated as a “seer“ (Oxford dictionary:
“A person of supposed supernatural insight who sees visions of the
future.“). Thus, prediction (“vision“) is what is to be commended
and not the consistency implicit in harmony. It is no longer a surprise
that authors as L. Smolin and J. Harnad contrast “seers“ and “craftsmen“
(in their assessment of string theory \citep[Ch. 18; ][]{smol08})
as the theoretical physicist subtypes. For Smolin and Harnad, Einstein
is the prototype of the seer.

The new wisdom may have ancestral roots, since sorcerers were the
guardians of accumulated knowledge in most primitive cultures and
technology is a fundamental element in defining a culture. We observe
that the new scientist, the theoretical physicist, occupied an empty
niche in the society emerging from the second industrial revolution
and the decline of the Enlightenment culture. The new scientists become
the priests that interpret the oracle of “science“ relevant for technology.

Departing from the philosophical and mathematical realm that was the
place of physics until the middle of the XIX century, the new epistemic
view puts intuition above reason when they come in conflict. But since
“We have no power of Intuition, but every cognition is determined
logically by previous cognitions“ \citep[pp. 230; ][]{peir55}, intuition
sets us back to the constructs of the child and in particular, the
notion of space, time and of reality \citep{piag99}\footnote{Apparently Science and scientists are unaware of this epistemic operation,
that occurs completely behind the scenes, guided by habit and training.}. The construction of the child is elevated to a dogma: to understand
is to put the observed in terms of the real (as constructed by the
child). The transition goes from a philosophical ideation to simple
ideation \citep{huss83}. The inferred or abducted is rejected when
it does not conform the intuitive form: Not only instantaneous action
at distance is rejected, but delayed action at distance is rejected
as well. It is not the case that Hertz cannot understand Maxwell,
it is a conscious decision of enforcing his own view of what it means
to understand but taking benefit of the previously produced equations.
This means to give no consideration to the inferences of Faraday and
Maxwell (and of a large list of mathematicians); they are rejected
\emph{in limine.} A free game of interpretations is then entertained,
the equations become runes that have to be deciphered, interpreted.
In a last move Einstein observed that a symmetry that must be present
is not present in the inherited wisdom. The symmetry (see Appendix
\ref{sec:lagrangian-appendix}) certainly was present in Faraday's
perceptions and experiments as well as in Maxwell's original equations,
what was needed was (is) to restore what had been destroyed by free
interpretation. Instead, we get a new interpretation, being Einstein
immersed in the new culture of theoretical physics.

The new approach adopted by theoretical physics has reproductive advantages
as well since it builds on top of students' intuitions. Physics textbooks
do not discuss notions of space or time, they just construct physics
supported on the children's intuitions for such concepts and build
new “intuitions“ in the form of habits by repetitive exercising. In
contrast, the critical approach requires the development of confidence
in the own forces of the student, it rests upon the \emph{bildung}
\citep{sorkin1983wilhelm} including its tense relation with teaching
as it emerges from W. von Humboldt words: “Whatever man is inclined
to, without the free exercise of his own choice, or whatever only
implies instruction and guidance, does not enter into his very being,
but still remains alien to his true nature, and is, indeed, effected
by him, not so much with human agency, as with the mere exactness
of mechanical routine“\citep{humb92}.

\section{Conclusions}

For the construction of mechanics we have an ample experience of the
notions of body, space, time and motion, that were produced in our
infancy \citep{sola18b}. Thanks to this early organisation of our
perceptions, mechanics can be built with two pillars: simple intuition
(experience) and reason. In contrast, our direct relation with electromagnetism
is not part of our early intuitive construction. Experience is scarce,
it is provided by experiments limited in several forms and is subject
to a conscious interpretation in terms of our pre-existent beliefs.
Ultimately, experiments challenged beliefs such as: the principle
of action and reaction, the material existence of the ether and what
is matter and how matter acts onto (relates with) matter.

Ideally, electromagnetism can be constructed with these two pillars,
acting both as requisites on equal footing, and that was Maxwell's
original attempt. In replacement of an insufficient experience he
(and others) introduced analogies, although he (following Faraday's
teachings and Newton's tradition) was very reluctant to formulate
physical hypothesis and kept them to a minimum. In contrast, the Göttingen
school appears as not needing intuition, or at least, as keeping always
intuition under rational supervision (``philosophical intuition''
in the terminology of Husserl), intuition was not on an equal footing
than reason but just an aid.

While Maxwell's epistemology is rooted in British empiricism, in continental
Europe the dictum of Hegel ``What is rational is real and what is
real is rational'' appears to have influenced the scientific minds.
Thus, reality is the rational organisation of the electromagnetic
phenomena that gives unity to the observed but exists only in relation
with the particular (observed) realisations. Much later, Jean-Paul
Sartre would beautifully write: ``an electrical current does not
have a secret reverse side; it is nothing but the totality of the
physical-chemical actions which manifest it (electrolysis, the incandescence
of a carbon filament, the displacement of the needle of a galvanometer,
etc.). No one of this actions alone is sufficient to reveal it. But
no action indicates anything that is behind itself; it indicates itself
and the principle of the series.''\citep[Introduction, ][]{sart56}.
Both lines of thought, the continental and the insular, had a broad
coincidence when it comes to the final mathematical expressions, their
divergence lying in the foundations sustaining the actual development.

The epistemological grounds of Maxwell's approach were laid in 1856
\citep{maxw56}. In replacement of an insufficient experience he (and
others) introduced analogies, although Maxwell (following Faraday's
teaching and Newton's tradition) was cautious and regarded his analogies
as tentative. Maxwell in his treatise as well as in his 1865 paper,
follows a construction method that parallels the method of Lagrangian
mechanics. It can be said then that Lagrangian formulations transcend
(or surpass in Piaget-García's language) mechanics becoming a method
for the construction of mathematical physics. The beliefs obtained
by this method are accepted only because they organise the experience
in electromagnetism, i.e., the theory is abduced. As Lorenz insightfully
indicates, the ether in Maxwell, and to a lesser degree in Faraday,
``had only been useful to furnish a basis for our imagination''.

However, in one point Maxwell's insight came short of Faraday's, namely
in conceiving light as anything else than a wave emitted an propagated
through some propagation medium. He even attempted to ``correct''
Faraday when the latter advanced an alternative view\footnote{It is remarkable that Faraday's vibrating ray theory is compatible
with (and probably was influential to) the developments of the Göttingen
School at those times.}. We recognise that Maxwell was in front of a deep difficulty: how
to understand a reasoning that requires to change what we consider
understanding? If understanding that one is (or may be) wrong about
something is not difficult, understanding that what we call ``to
understand'' may not be correct appears almost impossible. 

Maxwell complained referring to the Göttingen school: ``There appears
to be, in the minds of these eminent men, some prejudice or \emph{à-priori}
objection, against the hypothesis of a medium in which the phenomena
of radiation of light and heat and the electric actions at a distance
take place. {[}...{]} Hence the undulatory theory of light has met
with much opposition, directed not against its failure to explain
the phenomena, but against its assumption of the existence of a medium
in which light is propagated.''\citep[{[865], }][]{maxw73}. It should
be noted that the ``undulatory theory of light'' did not meet any
opposition from the Göttingen school. They also provided a wave equation
and attempted different motivations for it, based in Gauss' suggestion
of delayed action at a distance. The opposition focused in the hypothesis
of a propagating medium and in the peculiar and contradictory properties
that were required for this medium.

In fact, in the same article {[}865{]} Maxwell acknowledges that the
Göttingen school had good reasons to be sceptical about the ether:
``...It is true that at one time those who speculated as to the causes
of physical phenomena were in the habit of accounting for each kind
of action at a distance by means of a special æthereal fluid, whose
function and property it was to produce these actions. They filled
all space three and four times over with æthers of different kinds,
the properties of which were invented merely to ' save appearances,'
so that more rational enquirers were willing rather to accept not
only Newton's definite law of attraction at a distance, but even the
dogma of Cotes, that action at a distance is one of the primary properties
of matter, and that no explanation can be more intelligible than this
fact...'' Nevertheless, if Maxwell's form of knowing (his epistemology)
was correct, the ether must be real, and hence it becomes necessary
to know more about it, he indeed closes the Treatise (last paragraph)
stating that ``all these theories lead to the conception of a medium
in which the propagation takes place, and if we admit this medium
as an hypothesis, I think it ought to occupy a prominent place in
our investigations, and that we ought to endeavour to construct a
mental representation of all the details of its action, and this has
been my constant aim in this treatise.''

Some years later a new epistemology spread through continental Europe
emanating from Berlin and in coincidence with the German unification
and the construction of the first Reich: the \textsf{\emph{bild}}
conception \citep{dago04}. This new epistemology required for understanding
the construction of mental images of the real. Whether this real was
accessible to the senses or not, it did not matter. Hence, under the
new epistemology, the ether became real without carrying the research
proposed by Maxwell.

Hertz' \emph{bild} conception appears as a militant epistemology for
it suppressed large portions of Maxwell's treatise arguing that it
reflected a mechanistic approach incompatible with the ether. Nevertheless,
he conveniently kept Maxwell's equations but without offering an alternative,
``non mechanistic'', derivation. In the same movement it achieved
three goals: first, to ignore Faraday-Maxwell's epistemology; second,
to suppress any mention of the old German electromagnetism in its
evolved form, probably feeling justified by the argument that the
equations were ``the same'' than those in Maxwell; and third, to
suppress the Lagrangian basis of the organisation of electromagnetism.
The meaning of science was changed by detaching the symbolic relations
produced, disregarding the construction and allowing for a free interpretation
of the equations. This movement began the transition from modern science
to techno-science, freeing the later of the rigidities imposed by
reason. Such an idea goes completely against Hegel's conceptions as
well, for Hegel ``Everything, other than the reality which is established
by the conception, is transient, surface existence, external attribute,
opinion, appearance void of essence, untruth, delusion, and so forth.
Through the actual shape {[}Gestaltung{]}, which it takes upon itself
in actuality, is the conception itself understood. This shape is the
other essential element of the idea, and is to be distinguished from
the form {[}Form{]}, which exists only as conception {[}Begriff{]}.''\citep{hege01b}
\footnote{Notice that the paragraph is more meaningful if we translate \emph{Gestaltung}
as: shaping, forming, manifesting, ...}

Schematically, Maxwell requirements to accept a belief as a (temporary)
truth were: to mathematically organise the subject (rationality) and
to be compatible with a substantialist (intuitive) view. The Göttingen
school put rationality over intuition, while the Berlin school put
intuition over rationality. Hertz must be recognised as the one who
understood that, unlike the experience with mechanics, in electromagnetism
intuition and reason had come to some degree of incompatibility. Maxwell's
goal was impossible to achieve.

Later experiments in the search of the ether showed that none of the
conceptions of the ether was compatible with experiences. The well
known Michelson-Morley experiment shows that if light is considered
as moving through space the way bodies move (i.e., if it can be regarded
in analogy with material bodies or perturbations of matter) then space-time
cannot be conceived in Galilean terms. When substantialism fails to
reach conscious state\footnote{We may indeed speak of ``hidden substantialism'', as in Maxwell
(and almost all of his followers) that could not conceive light propagation
otherwise.} and in so doing avoids the inspection of reason, the antecedent is
never questioned. This amounts to putting intuition above reason when
they come in conflict. Hence, the accepted conclusion of the experiment
became that the world was not as Galileo and Leibniz conceived it.
Thus, substantialism is not an interpretation of physics, it pre-exist
physics and forges it. It is difficult to put substantialism under
examination because we tend to believe (or we are indoctrinated to
believe) that we (the scientists) understand a world whose rules pre-exist
our understanding effort. The constructivist thinks/knows otherwise
and he/she is willing to offer his form of understanding for philosophical/rational
examination, this is, to criticism.

The historical construction of electromagnetism cannot be separated
from its place in history. The goals behind the attack against the
school of Göttingen carried out by Clausius do not appear to be only
academic. During the second part of the XIX century, Germany, which
was very active in science, was emerging to industrialisation on the
economic side and was reunified on the political side. The growth
of the universities (in terms of number of students) and a new emphasis
in research resulted in the installation of a second professor in
physics, this is, there was an important expansion in the number of
physicists working in the academia. The ``second physicist'' is
the origin of the ``theoretical physicist'' as we learn from  \citep{jung17}.
The first ``theoretical physicist'' they recognise as such is Rudolph
Clausius. In this social and political context, the expansion of the
academic positions should have resulted in the dissemination of the
theories and epistemology dominating in Berlin. \footnote{For example, Albert Einstein was born in the German Empire, Kingdom
of Württemberg, and his first advisor in Zürich was Heinrich Friedrich
Weber (not to be confused with Wilhelm Eduard Weber) who was for some
time the first assistant of Helmholtz at Berlin.}

Prisoner of his own doctrine\footnote{A quote of Chuang Tse in Ortega y Gasset's ``Mission of the university''
reads: ``How will I be able to speak about the sea with the frog
when she has never gone out of her pond? How will I be able to speak
about the ice with the summer bird when he is anchored in his season?
How will I be able to speak with the savant about Life when he is
prisoner of his doctrine?'' \citep{orte10} (translated by the authors).} and of his time, Einstein tried to restore some rationality to the
electromagnetism of his time, but he resorted to the same epistemological
approach that had left it devoid from reason: a new interpretation
of the equations separated from their conception. The principle of
relativity presented by Poincaré and used by Einstein impress us as
true because it builds upon the habits acquired in mechanics and it
has a reminiscence of the principle of no-arbitrariness. The absence
of a process to attain its conception undermines its truth; actually,
in a Hegelian view it is untrue. The principle is based on the hidden
proposition: every real, objective, interaction can be presented in
subjective form. However, this proposition is not necessarily true
and perpetuates Newton's concept of force. After assuming that electromagnetism
could be presented in subjective form, the new proposal failed to
incorporate Lorentz' space-time as a real concept and not just as
mere opinion. For a concept to be real, it must be intersubjective,
which in terms of mathematics demands the existence of a group structure.
The substantialist view of light and electromagnetic interactions
clashes with the inherited structure of the space-time. Since --as
for Maxwell before-- an alternative to this substantialist view could
not be conceived (or it passed unnoticed), the remaining option was
an attempt to change space-time. It is worth to emphasise that for
the relational point of view, the Michelson-Morley experiment of 1881
only gives the expected results, and actually, one would have considered
a waste of time to perform it \footnote{On the contrary, Michelson's experiment from 1925 in the relational
view reveals the relative velocity between source and detector.}.

The construction of Special Relativity begins by assuming the existence
of a relational velocity (real, unique, not just an opinion) that
at the end will find no place in the theory. This happens on the mathematical
side because of the lack of an appropriate group structure and on
the construction side because intuitions such as relative velocity
belong to a different conception of space-time. The solution to this
nightmare was a new epistemological reform, where theories arise from
``free thinking'' and are only vaguely related to the observed/measured
reality. This is, by insisting in that concepts have a life and a
reality by themselves and not in relation to the conceptualisation,
science becomes mere appearance and lacks reality (always thinking
in Hegel's key). It is only opinion, for every reference system in
special relativity is entitled to its opinion, but there is no common
point or consistent equivalence between the opinions, hence, there
is no reality.

The absence of the real, however, is no longer a problem when the
measure of truth is usefulness \citep{eins40}. This completes the
epistemological voyage towards vulgar pragmatism, transforming science
from understanding nature (an in so doing adapting to it and empathising
with it) into being a platform for technology, sharing its goals,
equating truth with success and allowing for technology (science)
to dominate nature, then playing the role wanted by the society at
large: the industrial society.

The accepted version of electromagnetic theory (as discussed e.g.,
in textbooks) is the outcome of both electromagnetism as a scientific
problem and of a series of epistemic decisions that impulsed the reformulation
of science in the second half of the XIX century. The Enlightenment
era with Rousseau's social contract and Kant's critical reason was
declining and giving way to a new era signed by industrial development
and a return to imperial thinking (and acting).

Electromagnetism develops symbiotically with the new form of ``savant''
adapted to the epoch, the theoretical physicist. Indeed, theoretical
physicists will adopt electromagnetism as a model of scientific construction.
The equations constitute ``the theory'' and as such will be placed
above any criticism, as occurred with Maxwell's equations and is the
case also for Relativity. Underneath the equation level a broad interpretation
game develops. Critical thinking, including its role of challenging
the foundations of theories upon mismatches with experiment, is suppressed.
Where the theory could be in fault it is patched with new \emph{ad
hoc} substantialist forms. In this way elements such as neutrinos,
dark matter or dark energy emerge. Their universal form is prescribed
by the epistemology and only the particular form depends on empirical
data.

S. Traweek expresses the ``common sense'' of \textsf{normal} (in
the sense of Kuhn \citep{kuhn62}) physicists in the terms ``it must
be true because it works'', reflecting the decisive motto of vulgar
pragmatism. She perceives that ``high energy physicists construct
their world and represent it to themselves as free of their own agency,
a description, as thick as I could make it, of an extreme culture
of objectivity: a culture of no culture'' \citep[Epilogue,][]{traw92}.
As we have seen, much of the physicist's conception of the world comes
from themselves, not from their knowledge, but rather from the ignorance
about their own limitations.

This state of things was mediated by changes in the society at large.
The foundations of physical science have ceased to be available to
the general reasoning and can only be handled by specialists. Like
medieval guilds, specialists become the guardians of a way of doing
things within a community. Such elitist practice avoids exposing the
basic dogma of substantialism and lesser practices such as ``mathematical
fetishism'' (the adoration of mathematical formulae) to the inspection
by the philosopher. Philosophy of science more often than not adopts
the form of ``praise of the scientist'', considering only issues
that science has left open, but laymen are not admitted in the discussion
of space, time, or the universe, no matter how philosophically solid
they could be.\footnote{We here agree with the critics of elitism in science made by I. Lakatos
\citep{laka78} He writes: ``In the demarcationist tradition, philosophy
of science is a watchdog of scientific standards.''. He asserts as
well: ``Among scientists the most influential tradition in the approach
to scientific theories is elitism.''} It is often said that the theoretical physicist has rescued cosmology
from religion but in so doing he/she has become dogmatic.

Enlightenment was signed by the supremacy of reason and the social
contract. The present postmodernism (understood as reason's loss of
supremacy and the abandonment of critical reason), that emerged in
science before pervading culture in general, it has now to be walked
through and negated (in Hegel's meaning). Our time is the epoch of
global warming, the fires consuming the Amazonas, the mass extinction
of species, etc., the milestones of an era we cannot escape and that
is in need of a second Enlightenment, a reconstruction of reason.
For this, a new social and environmental contract is required, regaining
the plenitude of reason and its right to express itself in all issues,
recovering its critical strengths, the search for foundations and
the unity of reason. We beg the reader to allow us to believe that
the elaboration of this manuscript, being possible now and not in
earlier times, is a sign that the transit to a new epoch has already
begun.

\bibliographystyle{spbasic}
\bibliography{nuevasreferencias,referencias}

{\vspace{10mm}

M. A. Natiello is professor of applied mathematics at the Centre for
Mathematical Sciences, Lund University. He has contributed research
in chemistry, physics, mathematics, biology. His current research
interests are Applied Philosophy of Sciences and Population Dynamics
with a focus on social, ecological and epidemiological aspects.

H. G. Solari is professor of physics at the School of Exact and Natural
Sciences, University of Buenos Aires and research fellow of the National
Research Council of Argentina (CONICET). He has contributed research
to physics, mathematics, biology (ecology and epidemiology) and more
recently has been engaged in Applied Philosophy of Sciences (defined
as a reflexive and critical practice of science grounded in philosophy).
Most of his work is of interdisciplinary type.

\appendix\newpage

\part*{Appendices}

\section{Delayed action at a distance\label{sec:Delayed}}

The origins of delayed action at a distance have been reproduced in
several works along the years. The initial impetus was apparently
given by Gauss around 1845, as mentioned in his letter to Weber \citep[bd.5 p. 627-629,][]{gaus70}.
The idea of delays connect with Faraday's ``ray vibrations'' \citep[p. 447]{fara55},
from 1846 as well. The first explicit mathematical formulation was
given by Riemann \citep{riem67}(presented in 1858 and published in
1867) and L. Lorenz \citep{lore67} in adjacent articles in the same
issue of Annalen der Physik. The idea was discussed in detail by Carl
Neumann \citep{neum68} soon after. Eventually, the idea of retarded
effects entered the textbooks. In more recent times, the issue has
been reconsidered in various contexts by Moon\&Spencer \citep{moon1989binary,moon1989universal,moon1994electrodynamics},
and Bilbao \citep{bilbao2016,bilb14} among others.

For the sake of the issue about the propagation of electromagnetic
effects, the common grounds are the use of electric and magnetic potentials.
Basically, eq.(\ref{eq:Afinal}) for the magnetic vector potential
with no delays is re-proposed as ${\displaystyle A(x,t)=\frac{\mu_{0}}{4\pi}\int_{U}\left(\frac{j(y,t)}{|x-y|}\right)\,d^{3}y}$.
Technically, the potentials are obtained by a \textsf{convolution},
i.e., the integral of the material support of the potential (a current
source $j$ occupying the region $U$) with the \textsf{kernel} 
\[
\mathbf{W}(x-y,t-s)=\frac{\mu_{0}}{4\pi}\frac{\delta(t-s)}{|x-y|}
\]

in the following way:
\[
A(x,t)=\int ds\,\int_{U}d^{3}y\,\mathbf{W}(x-y,t-s)j(y,s)
\]
where $\delta$ is the delta distribution.

Under the prerequisites of the No Arbitrariness Principle \citep{sola18b}
the introduction of the delays reconsiders the form of the vector
potential for a current source $j$ occupying the region $U$ acting
on a particle at position $x$, time $t$ in a situation of relative
rest between source and target. The delayed potential is 
\begin{equation}
A(x,t)=\frac{\mu_{0}}{4\pi}\int_{U}\left(\frac{j(y,t-\frac{1}{C}|x-y|)}{|x-y|}\right)\,d^{3}y\label{eq:Aret}
\end{equation}
A similar expression for the electric potential, generalising Poisson's
law reads
\begin{equation}
V(x,t)=\frac{1}{4\pi\epsilon_{0}}\int_{U}\frac{\rho(y,t-\frac{1}{C}|x-y|)}{|x-y|}\,d^{3}y=\frac{\mu_{0}C^{2}}{4\pi}\int_{U}\frac{\rho(y,t-\frac{1}{C}|x-y|)}{|x-y|}\,d^{3}y\label{eq:Vret}
\end{equation}
Note that $A$ and ${\displaystyle \frac{V}{C^{2}}}$ satisfy the
same constitutive equation in relation to $j$ and $\rho$ and that
these two quantities are assumed to be a property of matter, i.e.,
they are identically zero outside the region $U$ where matter exists
(in the sequel we skip indicating $U$ to lighten the notation). In
convolution terms the delayed kernel 
\[
\mathbf{W^{d}}(x-y,t-s)=\frac{\mu_{0}}{4\pi}\frac{\delta(t-s-\frac{1}{C}|x-y|)}{|x-y|}
\]
is now used.

This formulation follows the ideas of C. Neumann and Lorenz as expressed
in eq.(\ref{eq:univ}). All involved quantities are relational, independent
of the coordinate system and therefore universal. Different subjective
representations of the resulting interaction are equivalent, being
the Galilean transformations the underlying group.

Without delays, the vector potential from eq.(\ref{eq:Afinal}) satisfies
Poisson's equation $\Delta A=-\mu_{0}j$. Maxwell's path to the wave
equation was to incorporate the required time-derivatives in his instantaneous
action theory through the introduction of the displacement current
$j_{D}$, to be added to the galvanic current $j$. Basically, ${\displaystyle -\mu_{0}j_{D}=\frac{1}{C^{2}}\frac{\partial^{2}A}{\partial t^{2}}}$.
This ``current'' was to be present even in the absence of matter
and Maxwell related it to the ether.

The delayed potentials, on the other hand, satisfy a wave-equation
automatically. Indeed,starting from eq.(\ref{eq:Aret}) or, correspondingly
from eq.(\ref{eq:Vret}), we have (we do not write the arguments of
$j_{i}$ when it helps to simplify the notation, we also write $r$
for $|x-y|$):
\begin{eqnarray*}
\nabla_{x}A_{i} & = & \frac{\mu_{0}}{4\pi}\int d^{3}y\,\left(\frac{-\frac{\partial}{\partial t}j_{i}\nabla_{x}\frac{r}{C}}{r}+j_{i}\nabla_{x}\frac{1}{r}\right)\\
\Delta A_{i} & = & \nabla_{x}\cdot\nabla_{x}A_{i}\\
 & = & \frac{\mu_{0}}{4\pi}\int d^{3}y\,\left(j_{i}\Delta\frac{1}{r}-2\nabla_{x}\frac{1}{r}\cdot j_{i,t}\nabla_{x}\frac{r}{C}-\frac{\frac{\partial}{\partial t}j_{i}\Delta\frac{r}{C}}{r}+\frac{\frac{\partial^{2}}{\partial t^{2}}j_{i}}{r}|\nabla_{x}\frac{r}{C}|^{2}\right)
\end{eqnarray*}
Moreover, standard operations of vector calculus give
\begin{eqnarray*}
j_{i,t}\left(2\nabla\frac{1}{r}\cdot\nabla\frac{r}{C}+\frac{\Delta\frac{r}{C}}{r}\right) & = & 0\\
|\nabla\frac{r}{C}|^{2} & = & \frac{1}{C^{2}}
\end{eqnarray*}
and therefore 
\begin{eqnarray*}
\Delta A_{i}(x,t) & = & \frac{\mu_{0}}{4\pi}\int d^{3}y\,j_{i}(y,t-\frac{r}{C})\Delta\left(\frac{1}{r}\right)+\left(\frac{1}{C^{2}}\right)\frac{\mu_{0}}{4\pi}\int d^{3}y\,\frac{\partial^{2}}{\partial t^{2}}\frac{j_{i}(y,t-\frac{r}{C})}{r}
\end{eqnarray*}
The time derivative in the last term can be extracted outside the
integral, thus yielding,
\begin{eqnarray*}
\Box A_{i}(x,t) & \stackrel{def}{=} & \Delta A_{i}(x,t)-\left(\frac{1}{C^{2}}\right)\frac{\mu_{0}}{4\pi}\int d^{3}y\,\frac{\partial^{2}}{\partial t^{2}}\frac{j_{i}(y,t-\frac{r}{C})}{r}\\
 & = & \Delta A_{i}(x,t)-\left(\frac{1}{C^{2}}\right)\frac{\partial^{2}}{\partial t^{2}}A_{i}(x,t)\\
 & = & \frac{\mu_{0}}{4\pi}\int d^{3}y\,j_{i}(y,t-\frac{|x-y|}{C})\Delta\left(\frac{1}{r}\right)\\
 & = & -\mu_{0}j_{i}(x,t)
\end{eqnarray*}
Maxwell's and Lorenz' alternatives are the only two options starting
from static potentials satisfying Poisson equation. In order to arrive
to an inhomogeneous wave equation, either one adds a ``correction''
to an instantaneous current $j(y,t)$ (or charge $\rho(y,t)$) or
one builds up the sources following the general solutions of the wave
equation. In other words, the option is ether (or dependence of charge
and current with the frame of reference) or delays.

\subsection{Relative motion\label{subsec:App-VRTKernel}}

The assumption of delayed action opens up new possibilities that were
precluded by instantaneous action. The question arises whether two
systems in relative motion will have the same sort of interaction
as systems at relative rest whenever their relative distance at the
time $t$ in consideration is the same in both cases. For the sake
of clarity consider two identical sources, in relative motion that
for a given time $t$ ``coincide'' (ideally) in space, along with
a detector that is at rest relative to one of the sources. Instantaneous
action at time $t$ as given by eq.(\ref{eq:Afinal}) cannot distinguish
between the sources, having both the same current (or charge) and
being at the same distance relative to the detector at that given
time. However, delayed action conveys a clear distinction since the
current (or charge) of the source at an earlier time enters the description,
and at that earlier time it was possible to distinguish the sources:
their distances to the detector were not the same.

Recasted under the hypothesis of delayed action, the phenomenon of
induction originating in the relative motion of source and detector
reveals itself as a natural consequence. The delay hypothesis provides
the wave equation and it also reorganises previous knowledge in a
more integrated manner.

Let $(\Phi,\zeta)$ represent any of the pairs $(A_{i},j_{i})$ or
$({\displaystyle \frac{V}{C^{2}},\rho})$ connected to each other
by the same convolution kernel. Consider further that source and detector
move relative to each other with constant relative velocity $v$.
If $x$ represents a local coordinate of the detector, we are now
interested in computing the potentials $A$ and $V$ at a point $x^{\prime}=x+vt$
relative to the source. Without the delay hypothesis a natural choice
is:
\begin{equation}
\Phi_{v}(x,t)=\Phi(x+vt,t)=\frac{\mu_{0}}{4\pi}\int\frac{\zeta(y,t)}{|x+vt-y|}\,d^{3}y\label{eq:GralSol}
\end{equation}
where $\Phi_{v}$ expresses the value of the potential at the point
$x$ of the detector. This corresponds to the convolution kernel 
\[
\mathbf{W_{v}}(x-y,t-s)=\frac{\mu_{0}}{4\pi}\frac{\delta(t-s)}{|x+vt-y|}.
\]
For zero relative velocity, eq.(\ref{eq:Afinal}) is recovered. In
this expression motion is of no consequence other than that of altering
the relative distances.

We cannot expect the above equation to hold for low velocities, since
already the delay is missing in it. When taking into account the delay
hypothesis, whatever proposed generalisation of $\mathbf{W}$ must
have the correct limit for low velocities, but also satisfy the known
experimental behaviour of electromagnetic phenomena, namely the \textsf{Doppler
effect} \citep{ding60a,mand62,kaiv85}. This effect expresses the
observed fact that the frequencies associated to electromagnetic phenomena
perceived by a detector in motion relative to a source are different
from the frequencies of the source in a precisely determined way.
However, $\mathbf{W_{v}}$ does not comply with Doppler.

An option that has not been previously investigated is
\[
\mathbf{W_{v}^{d}}(x-y,t-s)=\frac{\mu_{0}}{4\pi}{\displaystyle \frac{\delta(t-s-{\displaystyle \frac{1}{g(v)C}}|x-y+vg(v)(t-s)|)}{|x-y+vg(v)(t-s)|}}
\]
 where ${\displaystyle g(v)}$ is a monotonically increasing function
of relative velocity such that $g(0)=1$. This function has to be
chosen in such a way that the delayed kernel $\mathbf{W_{v}^{d}}$
gives the correct expression for the Doppler effect. Further, 
\begin{eqnarray*}
\Phi_{v}(x,t) & = & \frac{\mu_{0}}{4\pi}\int ds\,\int d^{3}y\,\zeta(y,s)\frac{\delta(t-s-{\displaystyle \frac{1}{g(v)C}}|x-y+vg(v)(t-s)|)}{|x-y+vg(v)(t-s)|}\\
 & = & \frac{\mu_{0}}{4\pi}\int ds\,\int d^{3}z\,\zeta(z+g(v)v(t-s),s)\frac{\delta(t-s-{\displaystyle \frac{1}{g(v)C}}|x-z|)}{|x-z|}\\
\Phi_{v}(x,t) & = & \frac{\mu_{0}}{4\pi}\int d^{3}z\,\frac{\zeta(z+{\displaystyle \frac{v}{C}}|x-z|,t-{\displaystyle \frac{1}{g(v)C}}|x-z|)}{|x-z|}
\end{eqnarray*}

The second line arises from the variable substitution $z=y-vg(v)(t-s)$
and the last line from time-integration.

Relative motion influences also the perception of charge and current.
For low velocities the equation of continuity requires that $j_{v}=j-\rho v$.
In the general case, from the second line above, with $\rho$ in place
of $\zeta$ and for constant velocity, we obtain 
\[
\frac{\partial\rho}{\partial t}=\nabla\rho\cdot g(v)v=\nabla\cdot\left(g(v)\rho v\right)
\]
whence $j_{v}=j-g(v)\rho v$.

Turning to the Doppler effect, we attempt to compute the Fourier transform
of the kernel. For a situation of relative rest between source and
detector, we have 
\begin{eqnarray}
\mathcal{F}\left(\mathbf{W^{d}}\right)(k,w) & = & \frac{\mu_{0}}{4\pi}\int d\tau\,\int d^{3}z\,\exp(-ikz-iw\tau)\frac{\delta(\tau-\frac{1}{C}|z|)}{|z|}\nonumber \\
 & = & \frac{\mu_{0}}{4\pi}\int d^{3}z\,{\displaystyle \frac{\exp({\displaystyle -ikz-i\frac{w}{C}|z|})}{|z|}=\frac{\mu_{0}}{|k|^{2}-\left({\displaystyle \frac{w}{C}}\right)^{2}}}\label{eq:FT}
\end{eqnarray}
where the substitutions $z=x-y$ and $\tau=t-s$ were used. Alternative
to the present computation by integration in the distributional sense,
we may consider the Fourier transform of $\Phi(x,t)$ recovering the
same result:
\begin{eqnarray*}
\mathcal{F}\left(\Phi\right) & = & \mathcal{F}\left(\mathbf{W^{d}}\right)\mathcal{F}\left(\zeta\right)=\mathcal{F}\left(\mathbf{W^{d}}\right)\left(-\frac{1}{\mu_{0}}\right)\mathcal{F}\left(\Box\Phi\right)\\
 & = & \mathcal{F}\left(\mathbf{W^{d}}\right)\left(-\frac{1}{\mu_{0}}\right)\left(-|k|^{2}+\left(\frac{w}{C}\right)^{2}\right)\mathcal{F}\left(\Phi\right)
\end{eqnarray*}
Repeating the process and substitutions for the case of relative motion,
we obtain a similar integral:

\begin{eqnarray*}
\mathcal{F}\left(\mathbf{W_{v}^{d}}\right)(k,w) & = & \int d\tau\,\int d^{3}z\,\exp(-ikz-iw\tau)\frac{\delta(\tau-\frac{1}{g(v)C}|z+g(v)v\tau|)}{|z+g(v)v\tau|}
\end{eqnarray*}
Letting $u=z+g(v)v\tau$ and $\sigma=g(v)\tau$, it follows that

\begin{eqnarray*}
\mathcal{F}\left(\mathbf{W_{v}^{d}}\right)(k,w) & = & \int\frac{d\sigma}{g(v)}\,\int d^{3}u\,\exp(-iku-i\frac{w}{g(v)}\sigma)\exp(ik\cdot v\sigma)\frac{\delta(\sigma-\frac{1}{C}|u|)}{|u|}\\
 & = & \frac{1}{g(v)}\int d^{3}u\,{\displaystyle \frac{\exp(-iku-i{\displaystyle \frac{w}{g(v)C}}|u|)\exp(ik\cdot{\displaystyle \frac{v}{C}}|u|)}{|u|}}\\
 & = & \frac{1}{g(v)}\frac{\mu_{0}}{|k|^{2}-\left({\displaystyle \frac{w}{g(v)C}-k\cdot\frac{v}{C}}\right)^{2}}
\end{eqnarray*}
The observed Doppler effect is recovered letting 
\[
g(v)=\frac{1}{\sqrt{1-\left({\displaystyle \frac{v}{C}}\right)^{2}}}
\]
since for $k\cdot v=|k|\,|v|$ the frequency shifts to $w=g(v)\left(vk\pm kC\right)=g(v)k(v\pm C)$
and for $k\cdot v=0$ it gives $w=\pm g(v)|kC|.$

Doppler experiments are accurate up to $O(\frac{v}{C})^{2}$, so experimental
observations are compatible with any delay factor such that $g(v)=1+\frac{1}{2}(\frac{v}{C})^{2}+O(\frac{v}{C})^{3}$.

\subsubsection{An alternative view}

Let us consider the change of variables $u=g(v)v$, for $g(v)^{-1}=\sqrt{1-\left({\displaystyle \frac{v}{C}}\right)^{2}}$.
Then we may recast $\mathbf{W_{v}^{d}}$ and the previous developments
in terms of $u$, using that

\[
v={\displaystyle \frac{u}{\sqrt{1+{\displaystyle \left(\frac{u}{C}\right)^{2}}}}}\quad\mathrm{and\quad}g(v)=\sqrt{1+\left(\frac{u}{C}\right)^{2}}.
\]
Letting then $\gamma=\sqrt{{\displaystyle 1+\left(\frac{u}{C}\right)^{2}}}$,
we find that 
\begin{eqnarray*}
\mathbf{W_{u}^{d}}(x-y,t-s) & = & \frac{\mu_{0}}{4\pi}{\displaystyle \frac{\delta(t-s-{\displaystyle \frac{1}{\gamma C}}|x-y+u(t-s)|)}{|x-y+u(t-s)|}}\\
\Phi_{u}(x,t) & = & \frac{\mu_{0}}{4\pi}\int ds\,\int d^{3}y\,\zeta(y,s)\frac{\delta(t-s-{\displaystyle \frac{1}{\gamma C}}|x-y+u(t-s)|)}{|x-y+u(t-s)|}\\
 & = & \frac{\mu_{0}}{4\pi}\int d^{3}z\,\frac{\zeta(z+{\displaystyle \frac{u}{C}}|x-y|,t-{\displaystyle \frac{1}{\gamma C}}|x-z|)}{|x-z|}
\end{eqnarray*}
This suggests a different connection between mechanics and electrodynamics.
We may regard $u$ as the mechanical velocity (the one that is determined
with rods and chronometers) and $v$ as a sort of ``electromagnetic
velocity'', bounded by $C$\footnote{This connects nicely with the fact that $u$ is an unbounded quantity
that can be associated to an additive group (the usual Galilean group
connected to mechanical velocities), with a nonlinear counterpart
for $v$, compatible with $v<C$\citep[Theorem 2]{sola18b}.}. This electromagnetic velocity usually is computed in indirect form,
by energy measurements in order to determine $\gamma=g(v)$ and thereafter
obtain $v$. For $v\ll C$, the difference between $u$ and $v$ is
negligible while for comparatively large velocities $u$ cannot be
measured independently.

Under this assumptions, $\mathbf{W_{v}^{d}}$ differs from $\mathbf{W_{v}}$
only in one $\gamma$ factor. The ``role'' of the factor is to adjust
the delay time. Electromagnetic interactions evolve ``more slowly''
when source and detector are in relative motion as compared with the
corresponding interactions at relative rest. It is not space-time
that undergoes deformations but only the properties of electromagnetic
interactions.

\subsection{Relationism and the speed of light\label{subsec:App-RecycledKernel}}

The idea that a relational view is necessarily associated to a ``speed
of light'' of the form $(C\pm v)=|\frac{w}{|k|}|$ has been so deeply
inculcated that it needs to be addressed. The question we want to
raise is: Are there relational kernels that have as low velocity limit
Maxwell's electromagnetism and also provide the correct Doppler shift
with $C=|\frac{w}{|k|}|$? The answer is affirmative. We provide here
an example. We first introduce the vector decomposition:

\begin{eqnarray*}
a & = & a_{\parallel}+a_{\perp}\\
a_{\parallel} & = & \left(a\cdot\hat{v}\right)\hat{v}\\
a_{\perp} & = & a-a_{\parallel}=\hat{v}\times(a\times\hat{v})
\end{eqnarray*}

Next, consider the convolution kernel
\begin{eqnarray*}
\mathbf{K}(\tau,X) & = & \frac{\mu_{0}}{4\pi}\delta(\tau^{\prime}-\frac{1}{C}|X^{\prime}|)\frac{1}{|X^{\prime}|}\\
\tau^{\prime} & = & \gamma(\tau+\frac{v}{C^{2}}X_{\parallel})\\
X^{\prime} & = & \gamma(X_{\parallel}+v\tau)+X_{\perp}
\end{eqnarray*}
where $\gamma=\frac{1}{\sqrt{1-\left({\displaystyle \frac{v}{C}}\right)^{2}}}$
and $v$ is the relative velocity between source and detector. Further,
let the propagation of an electromagnetic current or charge density
be of the same form as above, namely
\[
\Psi_{v}(x,t)=\int ds\,\int d^{3}y\,\mathbf{K}(t-s,x-y)\zeta(y,s)
\]
$\mathbf{K}$ is a relational kernel that does not depend on the choice
of origin of time or the reference for the space and coincides with
$\mathbf{W^{d}}$ for $v=0$. The Fourier transform, ${\cal F},$of
the potential $\Psi_{v}$ is now
\[
{\cal F}(\Psi_{v})={\cal F}(\mathbf{K}){\cal F}(\zeta)
\]
where
\[
{\cal F}(\mathbf{K})=\frac{\mu_{0}}{4\pi}\int d^{3}X\,\int d\tau\,\exp(-i(k\cdot X+wt))\delta(\tau^{\prime}-\frac{1}{C}|X^{\prime}|)\frac{1}{|X^{\prime}|}
\]
Letting $L_{v}$ be the Lorentz' transformation, we have
\begin{eqnarray*}
(k\cdot X+wt) & = & (k,w)^{\dagger}(L_{-v}L_{v})(X,\tau)\\
 & = & \left(L_{-v}(k,w)\right)^{\dagger}\left(L_{v}(X,\tau)\right)\\
 & = & \left(L_{-v}(k,w)\right)^{\dagger}(X^{\prime},\tau^{\prime})
\end{eqnarray*}
after a change of variables in the integral we obtain the expression
already considered in eq.\eqref{eq:FT}, the result being
\begin{eqnarray*}
{\cal F}(\Psi_{v}) & = & {\cal F}(\rho)\left(\frac{\mu_{0}}{|k^{\prime}|^{2}-\left({\displaystyle \frac{w^{\prime}}{C}}\right)^{2}}\right)\\
k^{\prime} & = & k_{\perp}+\gamma(k_{\parallel}-\frac{v}{C^{2}}w)\\
w^{\prime} & = & \gamma(w-k_{\parallel}v)
\end{eqnarray*}
It can now be verified that if ${\cal F}(\rho)=\delta(w-w_{0})$ the
observed frequency $w^{\prime}$ presents the correct Doppler shift
and that $\frac{|w^{\prime}|}{|k^{\prime}|}=C=\frac{|w|}{|k|}$ is
the relation for the emerging light.

\section{Electromagnetic Force and Lorentz' Lagrangian\label{sec:lagrangian-appendix}}

Maxwell introduced Lagrangian methods in electrodynamics transcending
(or surpassing) their mechanical origin. We compute here the electromagnetic
force $F_{em}$ exerted on a probe system by electric and magnetic
fields $E,B$ following Maxwell's method. We advance that specialising
Lorentz' hypothesis about current, the Lorentz force is recovered.

From the electric and magnetic potentials the corresponding fields
can be obtained via $B=\nabla\times A$ and ${\displaystyle E=-\nabla V-\frac{\partial A}{\partial t}}$.
Maxwell's equations read

\begin{eqnarray*}
\nabla\cdot B & = & 0\\
\nabla\cdot E & = & \frac{\rho}{\epsilon_{0}}\\
\nabla\times B & = & \mu_{0}\left(j+\frac{\partial E}{\partial t}\right)\\
\nabla\times E & = & -\frac{\partial B}{\partial t}
\end{eqnarray*}
Where the fields (and the equations) are given from a reference system
at rest with the source. Maxwell's total current reads ${\displaystyle J=j+\frac{\partial E}{\partial t}}.$
The Lagrangian introduced by Lorentz \citep[Chapter I and IV; ][]{lorentz1892CorpsMouvants}
reads 
\begin{eqnarray*}
{\cal L} & = & \frac{1}{2}\int dt\int\left(\frac{1}{\mu_{0}}|B|^{2}-\epsilon_{0}|E|^{2}\right)d^{3}x
\end{eqnarray*}
 and the variation to be considered is that of the probe, namely $\delta E=\delta E_{2}$,
$\delta B=\delta B_{2}$. The subscript $2$ stands for ``secondary''
electrical body.

Before proceeding further, we need to introduce the action, following
an insight that can be found in C. Neumann \citep{neum68}, thus departing
from Lorentz's approach that is based on a non-rigorous ``physical
argumentation'' supported upon his idea of a material ether. Let
the electromagnetic contribution to the mechanical action be
\[
{\cal A}=\int dt{\cal L}
\]
and accept, following Hamilton's principle, that $\delta E$ and $\delta B$
are zero in the extremes of the interval of time-integration. This
mechanical action may be recast in different ways by performing partial
integrations along with Gauss' relation 
\[
\int_{U}d^{3}x(\nabla\cdot F)=\int_{\partial U}d\sigma\cdot F
\]
and expressions derived from it for some electromagnetic field $F$.
Further, we accept Maxwell's hypothesis that the fields decay at infinity
in a such a form that the surface integral can be neglected. Maxwell's
equations were obtained under these conditions, which means that they
cannot be used if this restriction is lifted. \footnote{Notice that this implies that electromagnetism cannot be used to describe
an infinite universe. Only a finite universe is compatible with our
tools, an infinite universe is beyond our capabilities of explanation.
The finite universe of cosmology is an hypothesis required by our
limitations and not a conclusion reached from our knowledge.} After introducing the vector potential, $A$, and the scalar potential,
$V$ with the relations
\begin{eqnarray*}
E & = & -(\frac{\partial A}{\partial t}+\nabla V)\\
B & = & \nabla\times A
\end{eqnarray*}
and ``integrating by parts'' \citep[{e.g. [631] p. 270,}][]{maxw73}
in space and time, the following correspondence is found
\begin{equation}
\delta{\cal A}=\int dt\int d^{3}x\left[A\cdot\delta j-V\delta\rho\right]\label{eq:variation}
\end{equation}

In what follows, we will consider only the low velocity case, setting
$g(v)=1$, which is the case considered by Maxwell and Lorentz. We
need to introduce more details in the calculation. Let $\bar{x}(t)$
be the relative position between the primary and secondary electrical
bodies (the primary being everything else in consideration except
the probe). We define currents and potentials through
\begin{eqnarray}
\bar{\rho}_{2}(x,t) & = & \rho_{2}(x-\bar{x}(t),t)\nonumber \\
\bar{j}_{2}(x,t) & = & j_{2}(x-\bar{x}(t),t)+\dot{\bar{x}}\rho_{2}(x-\bar{x}(t),t)\label{eq:bar}\\
\Box A_{2} & = & \frac{\mu_{0}}{4\pi}\bar{j}_{2}\nonumber \\
\Box V_{2} & = & \frac{1}{4\pi\epsilon_{0}}\bar{\rho}_{2}\nonumber 
\end{eqnarray}
The potentials can be found using eq.\eqref{eq:GralSol}, but this
latter property will not used in the present discussion.

In such conditions, the variation of the current and of the charge
distribution due to the motion of the probe are: 

\begin{eqnarray*}
\delta\bar{j}_{2} & = & -(\delta\bar{x}\cdot\nabla)\bar{j_{2}}+\bar{\rho}_{2}\delta\dot{\bar{x}}\\
\delta\bar{\rho}_{2} & = & -(\delta\bar{x}\cdot\nabla)\bar{\rho}_{2}
\end{eqnarray*}
We have then
\begin{eqnarray}
\delta{\cal A} & = & \delta\int dt{\cal L}=\int dt\int d^{3}x\left[A\cdot\delta\bar{j}_{2}-V\delta\bar{\rho}_{2}\right]\nonumber \\
 & = & \int dt\int d^{3}x\left[\bar{j}_{2}\cdot\left(\delta\bar{x}\cdot\nabla\right)A-\bar{\rho}_{2}\left(\delta\bar{x}\cdot\nabla\right)V-\delta\bar{x}\cdot\frac{\partial}{\partial t}\left(A\bar{\rho}_{2}\right)\right]\label{eq:variada}
\end{eqnarray}
(the second line after some integrations by parts and the use of the
wave equation for the vector potential). Further transformation with
mathematical identities allows us to write

\[
\int dt\int d^{3}x\,\delta x\cdot\left[\bar{j}_{2}\times B+\bar{\rho}_{2}E\right].
\]
This is, following the standard use of Hamilton's principle in mechanics
we arrive to an electromagnetic contribution to the force on the probe
\[
F_{em}=\bar{j}_{2}\times B+\bar{\rho}_{2}E
\]

Lorentz considered only the case $j_{2}=0$, hence $\bar{j}_{2}=\bar{\rho}_{2}\dot{\bar{x}}$.
We have then derived from Hamilton's principle and the Lagrangian
base in Lorentz (which is actually based upon C. Neumann's Lagrangian
and action) the Lorentz' force, after assuming Lorentz' current. We
must emphasise however that the velocity in our deduction is relational
and in Lorentz' work is relative to the ether. The second difference
is that our presentation is fully mathematical while Lorentz' one
contains hand-waving arguments. The third difference is perhaps more
striking. Since $\bar{x}$ is a relational position the simultaneous
motion of primary and secondary with the same motion with respect
to a reference frame does not change the interaction, in other terms,
the principle of action and reaction holds and the conservation of
the total linear moment is assured by Noether's theorem.

Maxwell's derivation of the mechanical force (eq. C on the Treatise,
art. {[}619{]} p. 258), arrives to a similar result. For Maxwell,
$j_{2}$ is the ``total'' current, i.e., the sum of the galvanic
current and the displacement current. Remarkably, Maxwell's derivation
of this force is performed for the galvanic current only and therefore
it coincides with the present one. The displacement current was added
to the galvanic one by analogical thinking, without further justification.

\subsection{Maxwell's transformation and the Lorentz's force}

Whenever we have an integral expression like eq.\eqref{eq:variation}
it is possible to change the spatial variable of integration without
affecting the result. We intend to change from $x$ to $z,$with $x=z+\bar{x}(t)$,
the integration variable. But instead of performing the change in
eq.\eqref{eq:variation} we will save effort and perform it in eq.\eqref{eq:variada},
prior to the partial integration in time, namely 
\begin{eqnarray*}
\delta{\cal A} & = & \int dt\int d^{3}x\left[\bar{j}_{2}\cdot\left(\delta\bar{x}\cdot\nabla\right)A-\bar{\rho}_{2}\left(\delta\bar{x}\cdot\nabla\right)V+\delta\dot{\bar{x}}\cdot\left(A\bar{\rho}_{2}\right)\right]
\end{eqnarray*}
(with $\bar{j}_{2}(x,t),\quad\bar{\rho}_{2}(x,t)$ given by eq.\ref{eq:bar}).
We introduce the following notation

\begin{eqnarray}
z & = & x-\bar{x}(t)\nonumber \\
\underline{V}(z,t) & = & V(z+\bar{x}(t),t)\label{eq:underbar}\\
\underline{A}(z,t) & = & A(z+\bar{x}(t),t)\nonumber 
\end{eqnarray}

Hence, the variation reads now

\begin{eqnarray*}
\delta\int\!{\cal L}dt & = & \int dt\int d^{3}z\left[\left(j_{2}(z,t)+\dot{\bar{x}}\rho_{2}(z,t)\right)\cdot\left(\delta\bar{x}\cdot\nabla\right)\underline{A}-\rho_{2}(z,t)\left(\delta\bar{x}\cdot\nabla\right)\underline{V}\right]\\
 &  & +\int dt\int d^{3}z\left[\delta\dot{\bar{x}}\cdot\left(\underline{A}\rho_{2}(z,t)\right)\right]
\end{eqnarray*}
integrating by parts in time the last term and using the relations

\begin{eqnarray*}
\int dt\left[\delta\dot{\bar{x}}\cdot\left(\underline{A}\rho_{2}\right)\right] & = & \int dt\left[-\rho_{2}\delta\bar{x}\cdot\left[\frac{\partial\underline{A}}{\partial t}\right]-\delta\bar{x}\cdot\underline{A}\frac{\partial\rho_{2}}{\partial t}\right]\\
\left[\frac{\partial\underline{A}}{\partial t}\right] & \equiv & \frac{\partial}{\partial t}A(z+\bar{x}(t),t)=\left.\frac{\partial}{\partial t}A(x,t)\right|_{x=z+\bar{x}(t)}+\left(\dot{\bar{x}}\cdot\nabla\right)A(z+\bar{x}(t),t)\\
-\nabla\underline{V}-\left[\frac{\partial\underline{A}}{\partial t}\right] & = & -\nabla\underline{V}-\left.\frac{\partial}{\partial t}A(x,t)\right|_{x=z+\bar{x}(t)}-\left(\dot{\bar{x}}\cdot\nabla\right)\underline{A}\\
{\displaystyle \frac{\partial\rho}{\partial t}+\nabla\cdot j} & = & 0
\end{eqnarray*}
(the last equation is the continuity equation, valid in all setups),
we arrive after some algebra to

\begin{eqnarray*}
\delta\int\!{\cal L}dt & = & \int dt\int d^{3}z\left[j_{2}\cdot\left(\delta\bar{x}\cdot\nabla\right)\underline{A}-j_{2}\cdot\nabla\left(\delta\bar{x}\cdot\underline{A}\right)\right]\\
 &  & +\int dt\int d^{3}z\left[(\rho_{2}\dot{\bar{x}})\cdot\left(\delta\bar{x}\cdot\nabla\right)\underline{A}+\rho_{2}\delta\bar{x}\cdot(-\nabla\underline{V}-\frac{\partial\underline{A}}{\partial t})\right]
\end{eqnarray*}

Making use of the following relations,
\begin{eqnarray*}
\dot{x}\cdot\left(\delta\bar{x}\cdot\nabla\right)\underline{A} & = & \delta\bar{x}\cdot\nabla(\dot{x}\cdot\underline{A})\\
(\delta\bar{x}\cdot\nabla)\Phi(x,t) & = & -\delta\bar{x}\times(\nabla\times\Phi)+\nabla(\delta\bar{x}\cdot\Phi)\\
j_{2}\cdot\left(\delta\bar{x}\cdot\nabla\right)\underline{A}-j_{2}\cdot\nabla(\delta\bar{x}\cdot\underline{A}) & = & j_{2}\cdot\left(-\delta\bar{x}\times(\nabla\times\underline{A}\right)
\end{eqnarray*}
the result for the electromagnetic force is:

\begin{eqnarray*}
\delta\int{\cal L}dt & = & \int dt\int d^{3}z\left[j_{2}\cdot\left(-\delta\bar{x}\times(\nabla\times\underline{A})\right)\right]\\
 &  & +\int dt\int d^{3}z\left[\delta\bar{x}\cdot\rho_{2}\left(-\left[\frac{\partial\underline{A}}{\partial t}\right]-\nabla\left(\underline{V}-\dot{\bar{x}}\cdot\underline{A}\right)\right)\right]\\
 & = & \int dt\int d^{3}z\,\delta\bar{x}\cdot\left[j_{2}\times\underline{B}+\rho_{2}\left(-{\displaystyle \left[\frac{\partial\underline{A}}{\partial t}\right]}-\nabla(\underline{V}-\dot{\bar{x}}\cdot\underline{A})\right)\right]
\end{eqnarray*}

Hence we have two expressions for the mechanical contribution of the
electromagnetic force: The one obtained from eq.\eqref{eq:variada}
above and the present one, i.e.,
\[
F_{em}=\!\int\!d^{3}x\left[\bar{j}_{2}\times B+\bar{\rho}_{2}E\right]=\!\int\!d^{3}z\left[j_{2}\times\underline{B}+\rho_{2}\left(-{\displaystyle \left[\frac{\partial\underline{A}}{\partial t}\right]}-\nabla(\underline{V}-\dot{\bar{x}}\cdot\underline{A})\right)\right]
\]
(recall the relation among functions defined in \ref{eq:bar} and
\ref{eq:underbar}). This corresponds to the transformation proposed
by Maxwell and discussed in Theorem \ref{Theorem-Maxwell's-invariance}.

\subsection{The symmetry Einstein failed to see.\label{subsec:blindness}}

Let us consider a general quantity of the form 
\[
(\zeta_{1}|\zeta_{2})=\frac{\mu_{0}}{4\pi}\int\!ds\int\!dt\int\!d^{3}x\int\!d^{3}y\,\zeta_{1}(y,s)\zeta_{2}(x,t)\frac{\delta(t-s-{\displaystyle \frac{1}{C}}|x-y+u(t-s)|)}{|x-y+u(t-s)|}
\]
e.g., the mechanical action corresponding to the electromagnetic interaction
of $\zeta_{1}$ and $\zeta_{2}$. We write a first form (using $\tau=t-s$
and also $z=x+ut$ ) 
\begin{eqnarray*}
(\zeta_{1}|\zeta_{2}) & = & \frac{\mu_{0}}{4\pi}\int\!ds\int\!d\tau\int\!d^{3}x\int\!d^{3}y\,\zeta_{1}(y,s)\zeta_{2}(x,\tau+s)\frac{\delta(\tau-{\displaystyle \frac{1}{C}}|x-y+u\tau|)}{|x-y+u\tau|}\\
 & = & \frac{\mu_{0}}{4\pi}\int ds\,\int d^{3}z\,\int d^{3}y\,\left[\zeta_{1}(y,s)\zeta_{2}(z-u\tau,\tau+s)\right]_{\tau={\displaystyle \frac{1}{C}}|z-y|}\frac{1}{|z-y|}
\end{eqnarray*}

And a second (equivalent) form now using $z=y-u\tau$ 
\begin{eqnarray*}
(\zeta_{1}|\zeta_{2}) & = & \frac{\mu_{0}}{4\pi}\int\!dt\int\!d\tau\int\!d^{3}x\int\!d^{3}y\,\zeta_{1}(y,t-\tau)\zeta_{2}(x,t)\frac{\delta(\tau-{\displaystyle \frac{1}{C}}|x-y+u\tau|)}{|x-y+u\tau|}\\
 & = & \frac{\mu_{0}}{4\pi}\int dt\,\int d^{3}z\,\int d^{3}x\,\left[\zeta_{1}(z+u\tau,t-\tau)\zeta_{2}(x,t)\right]_{\tau={\displaystyle \frac{1}{C}}|x-z|}\frac{1}{|x-z|}
\end{eqnarray*}

The first form uses a forward potential and second a backward potential,
the different accounts for the difference between source and detector.
But except for this difference, forward fields can be defined following
the same relations that backward fields. Despite the well-known objection
of Einstein \citep[first paragraph]{eins05} it is possible to make
the reading in terms of the ``electric field arising in the neighbourhood
of the magnet'', although we are not used to think in these terms.

\section{Is there a relative velocity in special relativity?\label{sec:SR}}

Let us examine the standard view in special relativity (SR) that we
take from an authoritative source, the Feynman lectures of physics\citep[Ch. 16, Vol. i,][]{feynman1965}.
It starts stating that the \textbf{correct} transformations between
systems moving with relative velocity $v$ are Lorentz transformations
(emphasis added). We stated its general expression in eq.\eqref{eq:TL},
letting ${\displaystyle u=\frac{v}{C}}$ (we drop the index $C$ in
$T_{\frac{v}{C}}$ for simplicity):
\begin{eqnarray*}
T_{v}X & = & \left(\gamma_{v}(x-vt)+(1-\gamma_{v})\hat{v}\times(x\times\hat{v}),\gamma_{v}(t-\frac{v\cdot x}{C^{2}})\right)\\
 & = & \left(\gamma_{v}\left(\left(x\cdot\hat{v}\right)\hat{v}-vt\right)+\hat{v}\times(x\times\hat{v}),\gamma_{v}(t-\frac{v\cdot x}{C^{2}})\right)
\end{eqnarray*}
where $x\in\mathbf{R}^{3}$ is a spatial coordinate and $\gamma_{v}=(1-{\displaystyle \frac{v^{2}}{C^{2}}})^{-\frac{1}{2}}$.
As in Appendix \eqref{subsec:App-RecycledKernel}, we use that $x=\left(x\cdot\hat{v}\right)\hat{v}+\hat{v}\times(x\times\hat{v})$.
The book presents only the special case where the velocity $v$ between
systems is parallel to the $\hat{x}_{1}$-axis of the $S_{1}$ system,
in components: $(x_{1}^{\prime},x_{2}^{\prime},x_{3}^{\prime},t^{\prime})=T_{v}(x_{1},x_{2},x_{3},t)=({\displaystyle \gamma_{v}(x_{1}-vt),x_{2},x_{3},\gamma_{v}(t-\frac{vx_{1}}{C^{2}}))}$,
arguing that the general case is ``rather complicated''. Here $(x_{1}^{\prime},x_{2}^{\prime},x_{3}^{\prime},t^{\prime})\equiv(x^{\prime},t^{\prime})$
is the position and time in a system $S_{2}$. That the inverse transformation
corresponds to $T_{-v}$ is explicitly highlighted, for otherwise
``we would have a real cause to worry!''.

Further, Einstein's velocity transformation is presented by transforming
the line $(x,t)=(ut,t)$ with $t\in[-\infty,\infty]$ as points using
$T_{v}$, resulting in the set 
\[
(x^{\prime},t^{\prime})=t\left(\gamma_{v}\left(\left(u\cdot\hat{v}\right)\hat{v}-v\right)+\hat{v}\times(u\times\hat{v}),\gamma_{v}(1-\frac{v\cdot u}{C^{2}})\right)
\]
If $u$ is the (constant) velocity of a third system $S_{3}$ moving
with respect to $S_{1}$, we would like to describe its velocity as
seen by $S_{2}$ as $u^{\prime}=\frac{x^{\prime}}{t^{\prime}}$. We
note that the line $(x,t)=(vt,t)$ depicting the trajectory of $S_{2}$
according to $S_{1}$ transforms as $(x^{\prime},t^{\prime})=t({\displaystyle 0,\frac{1}{\gamma}})$
indicating that $S_{2}$ does not move according to $S_{2}$, as it
is always the case with a subjective vision. However, this reasoning
is arbitrary (it selects a preferred system $S_{1}$ for no reason)
unless the same calculation interchanging $S_{2}$ with $S_{3}$ could
give the relative velocity $-u^{\prime}$, and this for \textbf{any}
choice of system $S_{1}$. Only in this way, relative velocity between
$S_{2}$ and $S_{3}$ could be free from dependencies on arbitrary
choices. Computing the trajectory of $S_{2}$ as seen from $S_{3}$
as above, we obtain
\begin{eqnarray*}
(x^{\prime\prime},t^{\prime\prime}) & = & t\left(\gamma_{u}((\hat{u}\cdot v)\hat{u}-u)+\hat{u}\times(v\times\hat{u}),\gamma_{u}(1-\frac{v\cdot u}{C^{2}})\right)
\end{eqnarray*}
The two resulting trajectories, $(x^{\prime\prime},t^{\prime\prime})$
and $(x^{\prime},t^{\prime})$ are not parallel unless $u\parallel v$,
and similarly ${\displaystyle \frac{x^{\prime}}{t^{\prime}}\ne-\frac{x^{\prime\prime}}{t^{\prime\prime}}}$,
or equivalently ${\displaystyle \frac{x^{\prime}}{t^{\prime}}+\frac{x^{\prime\prime}}{t^{\prime\prime}}}\ne0$.
Therefore, the relative velocity depends on the arbitrary form we
choose to calculate it, i.e., there is no genuine concept of relative
velocity, it is a mere opinion that depends on arbitrary decisions.
The reader may want to verify the statement using $v=|v|\hat{x}_{1}$
and $u=|u|\hat{x}_{2}$. In short, we have:\\
\textbf{Theorem C.1:} The Lorentz transformation does not define an
equivalence relation.\\
\textbf{Proof:} The reflexive and symmetric properties of equivalences
are satisfied, but the transitive property is not.$\Box$\textbf{}\\
\textbf{Corollary C.1:} It is not possible to define the \textsf{inertial
class} in terms of the Lorentz transformation.

This problem emerges because of the lack of a group structure in Einstein's
velocity addition (the same problem arises for the Lorentz transformations
that by themselves do not constitute a group). In the velocity transformations,
the associative property is missing. If we have an operation that
we call velocity addition and symbolise it by $\oplus$, having the
basic property that $v\oplus(-v)=0$, then we would expect the relative
velocity to be 
\[
u\oplus(-v)
\]
and to satisfy the no arbitrariness relation
\[
u\oplus(-v)=-\left(v\oplus(-u)\right)
\]
or what is the same, that the law of transformation of reference systems
is transitive. If the operation $\oplus$ were to define a group,
we could apply the associative property of the group in the form
\[
\left[v\oplus(-u)\right]\oplus\left[u\oplus(-v)\right]=v\oplus\left[\left[(-u)\oplus u\right]\oplus(-v)\right]=v\oplus(-v)=0
\]
but this is not the case for relativistic velocity addition, since
the lhs of this equation corresponds to ${\displaystyle \frac{x^{\prime}}{t^{\prime}}+\frac{x^{\prime\prime}}{t^{\prime\prime}}}$.
All sorts of contradictions can be obtained in SR when three reference
systems moving without restrictions are considered.

If points are transformed from $S_{1}$ to $S_{2}$ ($S_{3})$ as
\begin{eqnarray*}
(x_{2},t_{2}) & = & T_{v}(x_{1},t_{1})\\
(x_{3},t_{3}) & = & T_{u}(x_{1},t_{1})
\end{eqnarray*}
for any set of points $B=\{(x_{1},t_{1})\}$ it follows that the transformation
from $S_{2}$ to $S_{3}$ should be
\[
(x_{3},t_{3})=T_{u}(x_{1},t_{1})=\left(T_{u}T_{-v}\right)(x_{2},t_{2})
\]
However, since the Lorentz transformations do not commute in general,
and they are symmetric, $T_{u}T_{-v}$ is not a Lorentz transformation
but rather an element in the Poincaré group. Any element $P$ of the
Poincaré group is written as:
\[
P=T_{w}R
\]
where $R$ is a rotation. Thus, $T_{u}T_{-v}=T_{w}R$. If we further
care to apply this transformation to $B^{\prime}=\left\{ (0,s),s\in[-\infty,\infty]\right\} $,
being $R(0,s)=(0,s)$ we get $T_{u}T_{-v}(0,s)=T_{w}(0,s)=\gamma_{v}sT_{u}(-v,1)$
which is Einstein's velocity addition law $u\oplus(-v)=w$ in its
general form. This law is neither associative nor commutative.

The rotation $R$ is around the vector $u\times v$ and is known in
SR as Thomas' rotation. It is a spurious rotation that appears as
a result of operating with two successive Lorentz transformations
in different directions and it has nothing to do with normal rotations.
It is worth noticing that if $u=0$ or $v=0$ the rotation is the
identity. Thus, the angle of rotation depends not only on the directions
of $u,v$ but also on their absolute value. Hence, it depends on the
intermediate system $S_{1}$ chosen to perform the transformation.
It follows that the transformation from $S_{2}$ into $S_{3}$ is
not unique, it depends on which system we have privileged to transform
into others with $R=Id$. Arbitrariness has a price: SR has no room
for objective relations, despite every observer being entitled to
her/his own opinion. In the case of electromagnetism the natural choice
would be the system where the electromagnetic disturbance takes place,
but at this point, the pretension of all systems being equivalent
is completely lost. In other words, we have proven the following \\
\textbf{Theorem C.2:} Given a system $S$ and two bodies, $\{a,b\}$,
moving uniformly following a straight line through the origin, for
the general case in which the respective velocities are not parallel,
it is not possible to define consistently a relative velocity.\\
\textbf{Proof:} $v_{a}\ominus v_{b}\ne-\left(v_{b}\ominus v_{a}\right)$$\Box$

Putting together Theorems C.1 and C.2, the inertial class that corresponds
to isolated systems does not exist in SR.

\end{document}